\documentclass[12pt, draftclsnofoot, onecolumn]{IEEEtran}
\IEEEoverridecommandlockouts
\usepackage{bbm}
\usepackage{dsfont}
\usepackage{lineno}
\usepackage{amssymb}
\usepackage{amsmath}
\usepackage{amsthm}
\usepackage{epsfig}
\usepackage{graphicx}
\usepackage{graphics}
\usepackage{epstopdf}
\usepackage{float}
\usepackage{subcaption}
\usepackage{multirow}
\usepackage{color}
\usepackage{tcolorbox}
\usepackage{makeidx}
\usepackage{xspace}
\usepackage{wrapfig}
\usepackage{lipsum}
\usepackage{mathtools,amsmath}
\usepackage{cuted}
\usepackage{cite}
\usepackage{amsfonts}
\usepackage{textcomp}
\usepackage{bm}

\usepackage{textcomp}
\usepackage{enumerate}
\usepackage{mathtools}
\usepackage{relsize}
\usepackage{url}
\usepackage{perpage}
\usepackage{setspace}
\usepackage{extpfeil}
\usepackage[linesnumbered,ruled,vlined]{algorithm2e}
\SetKwInput{KwInput}{Input}  
\SetKwInput{KwOutput}{Output}
\SetKw{KwBy}{by}
\makeatletter
\newcommand{\nosemic}{\renewcommand{\@endalgocfline}{\relax}}
\newcommand{\dosemic}{\renewcommand{\@endalgocfline}{\algocf@endline}}
\let\oldnl\nl
\newcommand{\nonl}{\renewcommand{\nl}{\let\nl\oldnl}}
\makeatother
\DontPrintSemicolon

\SetKwProg{Fn}{Function}{:}{}
\SetKwProg{Proc}{Procedure}{:}{}

\SetKwFunction{HK}{Hopcroft-Karp}
\SetKwFunction{RecursiveAssign}{RecursiveAssign}

\usepackage{hyperref} 

\newcommand{\ahmad}[1]{\textcolor{black}{ #1}}

\usepackage[scr=boondoxo,scrscaled=1.05]{mathalfa} 

\usepackage{enumitem}

\newenvironment{parletters}[1][]%
  {\begin{enumerate}[label=\alph*), #1]
   
   \renewcommand{\paragraph}[1]{\item \text{##1}.}}
  {\end{enumerate}}

\makeindex
\theoremstyle{plain}
\newtheorem{theo}{Theorem}
\newtheorem{defi}{Definition}

\newtheorem{prop}{Proposition}
\newtheorem{remark}{Remark}

\newtheorem{lem}{Lemma}
\newtheorem{ex}{Example}

\makeatletter

\renewcommand\@endtheorem{\vvv@endmarker\endtrivlist\@endpefalse}
\newcommand\vvv@endmarker{%
  {\unskip\nobreak\hfil\penalty50
  \hskip2em\vadjust{}\nobreak\hfil\openbox
  \parfillskip=0pt \finalhyphendemerits=0 \par
  \penalty 10000 \parskip=0pt\noindent}\ignorespaces}
\makeatother

\def \A {\mathcal{A}}
\def \mod {\text{{\normalfont mod}}}

\def \supp {\text{supp}}
\def \Supp {\text{Supp}}

\def \T {\mathcal{T}}

\def \z {\mathbf{z}}
\def \f {\mathbf{f}}
\def \pu {\underline{\mathbf{p}}}
\def \j {\underline{\mathbf{j}}}
\def \TensorJ {\bar{\mathcal{J}}}
\def \TensorF {\bar{\mathcal{F}}}

\def \TensorE {\bar{\mathcal{E}}}
\def \TensorS {\bar{\mathcal{S}}}
\def \TensorT {\bar{\mathcal{T}}}
\def \TensorR {\bar{\mathcal{R}}}
\def \TensorW {\bar{\mathcal{W}}}
\def \one \mathbf{1}

\def \min {\mathrm{min}}
\def \rank {\mathrm{rank}}

\MakePerPage{footnote}
\allowdisplaybreaks

\def\BibTeX{{\rm B\kern-.05em{\sc i\kern-.025em b}\kern-.08em
    T\kern-.1667em\lower.7ex\hbox{E}\kern-.125emX}}

\IEEEaftertitletext{\vspace{-1\baselineskip}}

\begin{document}



\title{Multi-User Non-Linearly Separable\\ Distributed Computing}

\author{Ali Khalesi, Ahmad Tanha, Derya Malak, and Petros Elia
\thanks{This research was partially supported by European Research Council ERC-StG Project SENSIBILITÉ under Grant 101077361, by the Huawei France-Funded Chair Toward Future Wireless Networks, and by the Program “PEPR Networks of the Future” of France 2030.}

\thanks{Ali Khalesi is with the Institut Polytechnique des Sciences Avancées (IPSA) and LINCS Lab, Paris, France (Email: ali.khalesi@ipsa.fr).

Ahmad Tanha is a PhD candidate at Sorbonne University, working with the Communication Systems Department at EURECOM Research Center, France (Email: ahmad.tanha@eurecom.fr).

Derya Malak and Petros Elia are with the Communication Systems Department at EURECOM, 450 Route des Chappes, 06410 Sophia Antipolis, France (Emails:\{malak, elia\}@eurecom.fr).}

\thanks{This paper will be presented in part \ahmad{at} the 2026 IEEE International Symposium on Information Theory (ISIT)~\cite{khalesi2026non}.}

}

\maketitle
\begin{abstract} 

This paper considers an $N$-server distributed computing setting with $K$ users requesting functions that are arbitrary multivariable polynomial evaluations of $L$ real (potentially non-linear) basis subfunctions, where each function output is raised to a bounded power. Our aim is to seek efficient task allocation and data communication techniques that reduce computation and communication costs. To this end, we take a tensor-theoretic approach, in which we represent the requested non-linearly decomposable functions using a properly designed tensor $\TensorF$, whose sparse decomposition into a tensor $\TensorE$ and a matrix $\mathbf{D}$ directly defines the task assignment, connectivity, and communication patterns.

We design a \emph{lossless} achievable scheme that integrates fixed-support SVD-based tensor factorization with multi-dimensional tiling of $\TensorE$ and $\mathbf{D}$, 
\ahmad{followed by} a bipartite graph matching-based recursive assignment of tiles.
This step transforms an overlapping decomposition into a disjoint one and reduces the resulting sum rank of the tiles, thereby decreasing the number of required servers. 
Under mild dimensionality conditions, we derive an explicit zero-error characterization of the achievable system rate $K/N$. 
Numerical simulations demonstrate the computational and communication savings
over existing state-of-the-art matrix factorization approaches
across a wide range of system parameters.

\end{abstract}


\section{Introduction}
\label{sec:intro}

As computational workloads continue to grow in size and complexity, distributed computing has become essential~\cite{verbraeken2020survey} for handling these workloads, particularly with the emergence of massive machine learning systems. While frameworks such as MapReduce~\cite{dean2008mapreduce} and Spark~\cite{zaharia2010spark} enable large-scale parallel processing across distributed servers to support a wide range of computation tasks (see, e.g.,~\cite{Brunero1,AVEST1,tandon2017gradient,raviv2020gradient,ye2018communication,wang2018fundamental,Charalambides25,Vithana23tit,Yan22,dutta2016short,ordentlich2025quant,ramamoorthy2019universally,Khalesi2025Tessellated,maheri2026universal}), limited computation and communication resources still impose a fundamental bottleneck that governs the design of distributed computing schemes.

Motivated by these intertwined bottlenecks and by the increasing complexity of the computed functions, we study the problem of distributed computation of \emph{non-linearly separable functions} in multi-user computing systems with bounded computational and communication resources. In particular, we consider $N$ distributed servers with limited computational and communication capabilities and $K$ users, each requesting the evaluation of their own 
polynomial of $L$ real (potentially non-linear) basis subfunctions. In this setting, a coordinator assigns computation tasks across the servers, aiming for lossless recovery of the user demands while also minimizing the required number of servers $N$, or equivalently, maximizing the \textit{achievable system rate} $K/N$. 
For this novel setting, we seek to provide new task allocation and data communication schemes, and to present bounds on the achievable system rate. 

Before detailing our setting, consider the following simple example, which illustrates the key constraints and metrics involved.

\begin{ex}\label{motiv-ex}
Let us first focus on the nature of the requested functions. Consider a setting with
$L=4$ basis subfunctions
\begin{align*}
f_1(\cdot) \triangleq e^{x_1}\, ,\quad 
f_2(\cdot) \triangleq \log(x_2)\, ,\quad
f_3(\cdot) \triangleq \sqrt{x_2}\, ,\quad
f_4(\cdot) \triangleq \cos(x_3)
\end{align*}
operating on fixed input vectors $x_1,x_2,x_3\in\mathbb{R}^B$ and producing the output files
$\{W_\ell=f_\ell(\cdot)\}_{\ell\in[4]}$\footnote{Naturally here, all operations are component-wise.}.
Consider the requested function
\begin{align*}
F_1(\mathbf{W})
&= 7\,W_1^{2}W_2^{3} + 8\,W_1W_3^{2}W_4
+ 6\,W_3W_4^{4}
+ 4\,W_1^{4}W_4^{2}
\end{align*}
which is clearly non-linearly separable over the variables in $\mathbf{W}=(W_1,W_2,W_3,W_4)$.

\begin{figure}
  \centering
\includegraphics[scale=0.75]{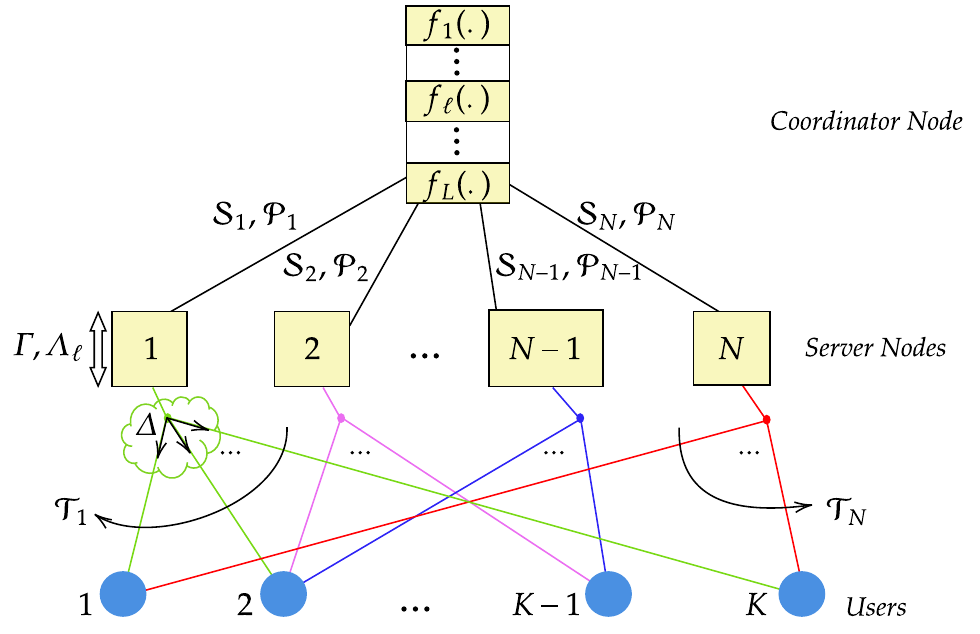}
  \vspace{-0.1cm}
\caption{The lossless $(K,N,L,\Gamma,\Delta, \{P_{\ell},\Lambda_{\ell}\}_{\ell\in [L]})$ distributed computing setting with a coordinator node, $N$ servers, and $K$ users.}
  \vspace{-0.2cm}
  \label{Fig: System Model}
\end{figure}

Our construction will be parameterized by the number of basis subfunctions $L$ (here, $L=4$),
and by \ahmad{per output file exponent bounds $\{P_\ell\}_{\ell\in[L]}$}, where the exponent of $W_\ell$
is restricted to lie in $(p_\ell-1) \in \{0,1,\ldots,P_\ell-1\}\, ,\ \ell\in[L]$.
Equivalently, $(P_\ell-1)$ is the maximum degree of $W_\ell$ that may appear in any requested
polynomial in $\mathbf{W}$.
In the above example, the maximum degrees are $P_1-1=4\, ,\ P_2-1=3\, ,\ P_3-1=2\, ,\ P_4-1=4$,
i.e., $(P_1,P_2,P_3,P_4)=(5,4,3,5)$.

For instance, $P_3=3$ implies that the exponent of $W_3$ is at most $2$, found in monomial $W_1W_3^2W_4$ in the second additive term of the requested function $F_1(\mathbf{W})$.

Let us assume a second requested function --- now by a second user --- of the form
\begin{align*}
F_2(\mathbf{W})
=  3\,W_2W_3^{3}
+ 2\,W_1^3W_3 
+ 11\,W_1^{2}W_2 
+13\,W_2^{2}W_4^{3}\, .
\end{align*}

We note that the declared maximum degrees $4,3,2,4$ --- which constrain every requested function --- are not respected, as this second polynomial entails maximum degrees $3,2,3,3$. 

Assuming $K=2$ users, respectively requesting $F_1(\mathbf{W})$ and $F_2(\mathbf{W})$, let us also assume 
$N = 3$ servers, and a network topology (see Figure~\ref{Fig: System Model}) that allows each server to communicate with up to $\Delta\le K$ users (let us assume in this example that $\Delta=1$), where each user must linearly combine the received signals to retrieve their own function. Another important parameter is $\Gamma\le L$ -- the computational constraint on the number of subfunctions that can be computed at each server. If we assumed $\Gamma=2$, we would not be able to compute $F_1(\mathbf{W})$ because its second term $8\,W_1W_3^{2}W_4$ could never be reproduced. We must have $\Gamma = 3$. There will finally be another computational constraint, on the ranges of powers $\Lambda_\ell$ of each output variable $W_\ell,\, \ell = 1,\ldots,L$, that can be computed at any one server, \ahmad{which determines the number of multiplications per output file required for each server to compute the desired exponent.} 
\end{ex}

The challenge here will be to assign subfunctions and specific powers of these subfunctions for specific servers to compute, and to determine the linear encoding and decoding processes at the servers and users, respectively. This same problem, it is worth noting, could conceivably be resolved by ``linearization'',  converting our problem to the linearly-separable setting found\footnote{For our above example, this would entail $L' = 8$ subfunctions, where the first subfunction would be 
$W_1^{2}W_2^{3}$, the second $W_1W_3^{2}W_4$, and so on.} in~\cite{Khalesi2025Tessellated}. Part of our contribution here is to present a more powerful approach, which involves converting the above problem to a sparse tensor decomposition problem. 
This new approach can have 
very substantial gains over the linearized approach, as detailed in Example~\ref{single-shot-example-simple} in~Section~\ref{Formulating} and also validated by numerical evaluations in Section~\ref{numerical} for various system parameters.

\subsection{Related Works}


There have been substantial efforts to attain the fundamental limits of distributed computing of linearly separable classes of functions, including those on 
gradient coding~\cite{Charalambides25,tandon2017gradient,raviv2020gradient,ye2018communication}, matrix multiplication~\cite{dutta2016short,ordentlich2025quant,tanhaCDMM26,ramamoorthy2019universally,jia2021capacity}, polynomial computing~\cite{wan2022secure,tanha24,Khalesi2025Tessellated,Khalesi25Tensor}.
%
Among existing models, the proposed setting is closest to that considered in~\cite{Khalesi2025Tessellated}, where the authors devised an equivalence between the distributed computing problem for linearly separable functions subject to sparsity constraints on the factor supports, and a longstanding open problem in sparse matrix factorization, for which they devised tessellation-based constructions accounting for support constraints. 
\ahmad{
The tessellation-based sparse matrix factorization scheme of~\cite{Khalesi2025Tessellated} partitions the demand matrix into submatrices of appropriate dimensions, whose sparsity patterns, and consequently their ranks\footnote{\ahmad{The rank of each $A \times B$ submatrix is explicitly defined as the maximum possible rank of a matrix of that size, namely $\min(A,B)$, where $A$ and $B$ denote its numbers of rows and columns, respectively.}}
determine the required computational resources. 
However, matrix factorization inherently operates over an expanded space of basis subfunctions, and therefore may become inefficient when the requested functions exhibit structures beyond linear separability. 
This limitation necessitates the development of a new framework tailored to a non-linearly separable class of functions. 
}
Our emphasis on 
\ahmad{this new framework}
is motivated by emerging machine learning applications~\cite{sefidgaran2023mdl,Vithana23tit,kavian2025heterogeneity}, most notably large language models, which require distributed computation of non-linear functions. Such functions can include activation functions~\cite{haziza2025sparsity} and attention mechanisms~\cite{hassani2025generalized} that involve higher-order terms of underlying basis subfunctions and their multiplications~\cite{Jayakumar2020Multiplicative}.

\ahmad{
Existing approaches, including the matrix factorization framework of~\cite{Khalesi2025Tessellated}
and coded-computing schemes such as~\cite{yu2019lagrange,reisizadeh2021coded,Moradi25,Abadi25},
typically rely on a linear separability assumption on the users' demands. 
However, non-linearly separable functions require an augmented representation that explicitly captures the per-basis subfunction degrees appearing in each monomial. 
These degrees introduce additional structural dimensions, naturally leading to a multi-dimensional representation, i.e., a tensor, that inherently encodes the non-linear coupling among subfunctions.
Moreover, existing schemes such as~\cite{Khalesi2025Tessellated} often rely on structural or disjoint-support assumptions on users' demands, effectively restricting them to configurations where the supports occupy non-overlapping regions of the demand matrix. 
However, in non-linearly separable settings, overlaps between supports naturally arise, 
leading to overlapping submatrices and, consequently, overlapping supports may yield inefficient decompositions. 
In particular, such constraints hinder the ability to attain low-rank representations, as they impose rigid matrix partitions whose ranks grow unfavorably with the underlying combinatorial structure of the demand tensor (or its matricized form\footnote{\ahmad{Matricization (also called unfolding or flattening) refers to reshaping a tensor into a matrix by grouping a subset of its modes into rows and the remaining modes into columns.}}). These challenges are consistent with known hardness results, including rank minimization of sparse or structured matrices, which is NP-hard~\cite{vavasis2009complexity,gillis2014complexity,Hanauer24}, and tensor rank minimization, which is NP-hard even to approximate~\cite{hillar2013most}.
As a result, existing approaches typically rely on heuristic constructions or simplified decompositions.
}

\subsection{Our Contributions}

In this work, we study the problem of lossless distributed computing of non-linearly separable functions. 
Our contributions are summarized as follows.
\begin{itemize}
\item \textbf{Tensor representation of user demands:}
We utilize the inherent structure of user demands to represent them using a full-rank tensor $\TensorF$. 
By embedding the range of exponents of a set of basis subfunctions into higher-order tensor modes, the proposed framework captures a broad class of non-linearly separable functions.

\item \textbf{Sparse tensor factorization framework:}
We introduce the sparse factorization of $\TensorF$ as  $\TensorF = \TensorE \times_1 \mathbf{D}$ (the operation $\times_1$ is detailed in Section~\ref{Formulating}).
The encoding tensor $\TensorE$ specifies, 
for each server~$n$, a subset of basis subfunctions and a subset of power terms to be computed by that server, 
%
while the decoding matrix $\mathbf{D}$ specifies a user set of size at most $\Delta$ to which each server transmits.
Naturally, the sparsity of $\TensorE$ and $\mathbf{D}$ respectively specify the computation and communication costs. 

\item \textbf{A new achievable scheme design:}
Focusing on lossless sparse tensor factorization, \ahmad{as detailed in Section~\ref{subsec:rate}}, we decompose $\TensorF$ into properly sized and carefully positioned subtensors, which are then factorized into subtensors of $\TensorE$ and $\mathbf{D}$, and lower bound the system rate, defined as $K/N$ (cf. Theorem~\ref{Achievability-Converse}), by leveraging novel tools from fixed-support sparse matrix factorization and multilinear singular value decomposition (SVD)~\cite{le2023spurious,de2000}. 
A naive approach to this problem is to first matricize the resulting demand tensor. 
However, such a transformation leads to a prohibitively large matrix representation, whose subfunction space grows exponentially with the number of basis subfunctions and combinatorially with their degree bounds, resulting in substantial rank and computational overhead. 
Our proposed approach yields an exponential reduction in \ahmad{total computation cost (cf. Proposition~\ref{comparison1}) and substantial savings in the required number of servers $N$ (cf. Example~\ref{single-shot-example-simple}, Case~$\mathrm{I}$), and communication resources (cf. Example~\ref{single-shot-example-simple}, Case~$\mathrm{II}$)} 
for lossless reconstruction of user demands compared to the sparse matrix factorization solution~\cite{Khalesi2025Tessellated}. 




\item \textbf{A novel assignment-based combinatorial reduction:}
We introduce a new \emph{tile assignment framework} (cf. Algorithm~\ref{Alg:GreedyAssignment}), \ahmad{as detailed in Section~\ref{subsec:refined_rate}}, that systematically resolves the overlap between subtensors (tile closures).  
Instead of allowing multiple assignments of the same index tuple, representing the demand coefficient across different tiles, the proposed method enforces a disjoint decomposition by assigning each tuple to exactly one feasible tile while respecting tile capacity constraints.  
This leads to a reduction in the induced \emph{sum rank} of the resulting subtensors, which in turn determines the number of required servers (cf. Theorem~\ref{Achievability-new}), over Theorem~\ref{Achievability-Converse}. By eliminating redundant contributions across overlapping supports, this step significantly lowers the sum rank and hence $N$, without imposing restrictive disjoint-support assumptions or incurring the complexity of general rank-minimization methods.
By combining structured tensor factorization with this assignment mechanism, the proposed scheme significantly outperforms approaches that rely solely on tensor decomposition and do not account for overlap (cf. Theorem~\ref{Achievability-Converse}).  
In particular, the assignment step 
allocates computational resources by transforming an overlapping tensor decomposition into a disjoint representation, thereby reducing the sum rank of the resulting subtensors. The observed gains stem from resolving overlap through a global tuple-to-tile assignment, which enables a more efficient utilization of the computational resources, as described in Section~\ref{numerical}.
Furthermore, the above results focusing on the single-shot scheme are extended to the multi-shot scenario with $T$ shots in Appendix~\ref{sec:multishot} (cf. Theorems~\ref{Achievability-Converse-multishot},~\ref{Achievability-new-multishot}).

\item \textbf{Numerical evaluation of the proposed assignment algorithm:}
We provide exact numerical evaluations of $N$ induced by Algorithm~\ref{Alg:GreedyAssignment} under various system configurations,~\ahmad{as detailed in Section~\ref{numerical}}.  
The results demonstrate the efficiency of the proposed assignment-based construction in reducing \(N\), compared to both the TDC scheme~\cite{Khalesi2025Tessellated} and the default tensor factorization approach with overlapping subtensors (cf. Theorem~\ref{Achievability-Converse}).  
\end{itemize}

\textbf{Notations.} For $n\in \mathbb{Z}^{+}$, we let $[n]
\triangleq \{1,\hdots, n\}$. For $a$,  $b \in \mathbb{Z}^{+}$ such that $a <b$, $[[a:b]]$ is an ordered set of integers, ranging from $a$ to $b$, and  $a \mid b$ denotes $a$ divides $b$. 
For any matrix $\mathbf{X} \in \mathbb{R}^{m \times n}$, then $\mathbf{X}(i,j)$, $\mathbf{X}(i,:)$, and $\mathbf{X}(:,j)$ represent its $(i,j)$-th entry, the $i$-th row, and $j$-th column, respectively, for $i \in [m],\: j \in [n]$, and $\Supp (\mathbf{X}) \in \{0,1\}^{m \times n}$ represents the locations of non-zero elements of $\mathbf{X}$. $[\mathbf{A},\mathbf{B}]$ indicates the horizontal concatenation of the two matrices. 
All the above matrix notations are extended to tensors. 
$\mathbbm{1}(.)$ is the indicator function. 
For a vector $\mathbf{x}\in \mathbb{R}^N$, $\| \mathbf{x} \|_0$ denotes the number of non-zero elements, and $\| \mathbf{x} \|
$ the Euclidean norm.
For a real number $x\in \mathbb{R}$, $\lceil x \rceil, \lfloor x\rfloor$ represent the ceiling and floor functions of $x$, respectively. $\mathbbm{1}(\ell \in \mathcal{Q})$ is the indicator function, returning 1 for $\ell \in \mathcal{Q}$. 
$\vee$ denotes a logical “OR” operator.


\section{System Model}
\label{System-Model} 
We consider a practical distributed computing framework with \(N\) servers and \(K\) users, where each user requests the evaluation of an arbitrary multivariate polynomial function. Each of these polynomial functions are composed of \(L\) real-valued, potentially non-linear basis subfunctions \(\{f_\ell(\cdot)\}_{\ell\in[L]}\). A coordinator node orchestrates the distributed computation of these functions through three phases, described next.

\paragraph{Demand Phase}
During the initial demand phase, each user $k \in [K]$ independently requests the computed output of a single real-valued function that takes the multivariate form
\begin{align} 
\label{nonlinearlySep1}
    F_{k}(W_1,\dots,W_L) 
    = \sum_{\substack{ \pu \in \prod_{\ell \in [L]} [P_{\ell}] }}
    c_{k,\pu}\: \prod_{\ell\in [L]} W_\ell^{p_{\ell}-1} 
\end{align}
where $\{W_\ell \triangleq f_\ell(x)\}_{\ell\in [L]}$ is the resulting real-valued \emph{output file}, representing evaluations of $\{f_\ell(\cdot)\}_{\ell\in[L]}$ on a common fixed input vector $x \in \mathbb{R}^d$, where $x$ denotes the underlying data instance (e.g., a feature vector or signal) on which all computations are performed. 
Furthermore, $c_{k,\pu}\in \mathbb{R}$ denotes the \emph{basis coefficient} corresponding to the monomial $\prod_{\ell\in [L]} W_\ell^{p_{\ell}-1}$, where $\pu \triangleq (p_1,\hdots,p_{L})$ is an index tuple corresponding to the exponent vector $(p_1-1,\hdots,p_{L}-1)$. 
We set $W'(\pu)=\prod_{\ell\in [L]}W_\ell^{p_{\ell}-1}$, so that we can transform~\eqref{nonlinearlySep1} into 
\begin{align}
 \label{eq:linearized}
 F_{k} = \sum_{\substack{ \pu \in \prod_{\ell \in [L]} [P_{\ell}] }} c_{k,\pu}\: W'(\pu) 
\end{align} 
which is the classical multi-user gradient coding problem. On the other hand, in this work, we explicitly account for multiplication costs in the high-dimensional problem~\eqref{nonlinearlySep1}, which introduces new algebraic challenges, as we will detail below.

\paragraph{Non-Linear Computing Phase}
Subsequently, the coordinator assigns to each server $n$, a set of basis subfunctions $\mathcal{S}_{n} \subseteq [L]$ to be computed locally. Each server $n$ then computes $\{W_\ell=f_\ell(.)\}_{\ell\in\mathcal{S}_n}$. We consider the computation cost
\begin{align}
\label{eq:Gamma}
\Gamma  &\triangleq  \max_{n \in [N]} |\mathcal{S}_{n}| 
\end{align}
representing the maximum number of basis subfunctions to be locally computed at any server. 
Each server operates under a computation constraint $|\mathcal{S}_n|\leq \Gamma\leq L$, which implies that up to $\Gamma$ basis subfunctions can appear in each demand, i.e., $c_{k,\pu} = 0$ in (\ref{nonlinearlySep1}) when the support of $\pu$ involves more than $\Gamma$ components, reflecting the per-server computation limit. 

The coordinator also assigns to each server $n \in [N]$ the corresponding set of exponent vectors $\mathcal{P}_n\subseteq \prod_{\ell\in [L]} [P_\ell]$. Each server $n$ then performs multiplications to compute the power terms $\{W_\ell^{p_{\ell}-1}\}_{\ell\in\mathcal{S}_n}$, and consequently, to obtain the corresponding monomial $\prod_{\ell\in \mathcal{S}_n} W_\ell^{p_{\ell}-1}$ for each $\pu\in \mathcal{P}_n$, thus rendering the computation phase non-linear.

To devise the multiplication cost, we assume that for any given $W_\ell,\,\ell\in \mathcal{S}_n$, each server $n$ computes the following ordered set of exponents with a cardinality\footnote{With heterogeneous server computational abilities, the cardinality would instead depend on both $\ell$ and $n$
.} $\Lambda_\ell,\, \ell\in \mathcal{S}_n$:
\begin{align}
\label{eq:range_of_exponents}
[[q\Lambda_\ell+1:(q+1)\Lambda_\ell]] \ ,\quad q\in\mathbb{N}\ .
\end{align} 

Alternatively, each server computes a power term $W_\ell^\alpha$ in a demanded function, where the exponent $\alpha$ is decomposed as 
\begin{align}
\label{eq:exponent_alpha_decompositiong}
\alpha\triangleq q\Lambda_\ell+r\ ,\quad q\in \mathbb{N}\ ,\quad r\in [[0:\Lambda_\ell-1]] \ .
\end{align}

Using the server's allowed range in (\ref{eq:range_of_exponents}), we decompose $W_\ell^\alpha=W_\ell^{q\Lambda_\ell+1}W_\ell^{(r-1)}$, and hence the cost of evaluating $W_\ell^\alpha$ is 
\begin{align}
\label{eq:cost_comp_power_term}
\lfloor \log_2 (q\Lambda_\ell+1) \rfloor +(r-1) \ 
\end{align}
where the logarithmic cost is due to repeated squaring, i.e., computing successive powers of two until reaching the desired anchor exponent, which is negligible compared to the $(r-1)$ multiplications required within the range of exponents. Hence, the overall complexity induced by self-multiplications of basis function $f_\ell(\cdot)$ per server 
is in order $\mathcal{O}(\Lambda_\ell)$,
versus $\mathcal{O}(\alpha)$ for naive repeated multiplications. For example, with $\alpha=851$ and $\Lambda_\ell=100$, 
from (\ref{eq:cost_comp_power_term}), the computation requires only $59$ multiplications, versus $\alpha-1=850$ multiplications for the naive approach. 
In general, $\Gamma\le L \ll \Lambda_\ell$ allows us to conclude that the cost of evaluating the monomial $\prod_{\ell\in \mathcal{S}_n} W_\ell^{p_{\ell}-1}$ from $\{W_\ell^{p_{\ell}-1}\}_{\ell\in \mathcal{S}_n}$ for $\pu\in \mathcal{P}_n$ at server $n$ is negligible compared to the cost of evaluating $\{W_\ell^{p_{\ell}-1}\}_{\ell\in \mathcal{S}_n}$. 
Thus, we consider the multiplication cost at each server
\begin{align}
\label{eq:Lambda}
\Lambda_{\ell} & \triangleq  \max_{\pu \in \mathcal{P}_{n}} {p}_\ell - \underset{\pu \in \mathcal{P}_{n}}{\min} p_\ell +1 \, , \quad \ell\in \mathcal{S}_n
\end{align}
representing the range of exponents of $\{W_\ell\}_{\ell\in \mathcal{S}_n}$ that must be computed, which in turn determines the maximum number of multiplications required locally, at each server, among all multiplicative terms in \eqref{nonlinearlySep1}. 
Each server also operates under a multiplication constraint $\Lambda_{\ell} \leq P_{\ell}$ for each $\ell\in [L]$.

\paragraph{Communication Phase}
Upon completing the local computations, server $n\in[N]$ forms signals
\begin{align}
\label{EncodedFiles}
  z_{n}\triangleq \sum\nolimits_{\pu \in \mathcal{P}_{n}} e_{n,\pu} \prod\nolimits_{\ell\in [L]} W_\ell^{p_{\ell}-1},\quad n\in [N]
\end{align} 
as dictated by the encoding coefficients $e_{n,\pu} \in \mathbb{R}, n\in [N],\,\pu \in \mathcal{P}_{n}$. 
Note that $p_\ell=1,\, \ell\notin \mathcal{S}_n$, reflects the absence of certain basis subfunctions and their powers in the multiplicative terms for computation by server $n$.
Subsequently, server $n$ proceeds to transmit $z_{n}$ 
to a subset of users $\T_{n} \subseteq [K]$, via an error-free shared link. Finally, during the decoding part of the last phase, each user $k\in[K]$ linearly combines its received signals to get
\begin{align}
    F'_{k} \triangleq \sum\nolimits_{n \in [N]} d_{k,n} z_{n} \label{DecedFiles}
\end{align}
dictated by the decoding coefficients $d_{k,n} \in \mathbb{R}, 
n\in [N]$, where $d_{k,n} =0,\, k \notin \mathcal{T}_{n}$. 
We consider the communication cost
\begin{align}  
\label{eq:Delta}
   \Delta \triangleq \max_{n \in  [N]} |\mathcal{T}_{n}|\ 
\end{align}
denoting the maximum number of users that each server can communicate to, where $\Delta\le K$. 


We highlight that the constraints \( \Gamma \), \( \Delta \), and \( \Lambda_\ell \) are strict, meaning that they must be satisfied for every instance of the problem. Thus, the costs from (\ref{eq:Gamma}), \eqref{eq:Delta}, and (\ref{eq:Lambda}) yield a system with normalized constraints 
\begin{align}
\label{eq:normalized_costs}
\gamma = \frac{\Gamma}{L} \ ,\ \  \delta = \frac{\Delta}{K} \ , \ \ \left\{\lambda_{\ell} = \frac{\Lambda_{\ell}}{P_\ell}\right\}_{\ell\in[L]} \ 
\end{align}
where $\gamma,\delta,\lambda_\ell\in[0,1],\, \ell\in[L]$. 
Accordingly, for each server $n$, we must specify the basis subfunctions $\mathcal{S}_n$, 
the exponent vectors $\mathcal{P}_n$,
and the users $\T_{n}$. 
Having to serve many users with fewer servers naturally places a burden on the system, bringing to the fore the concept of the \emph{system rate} 
\begin{align}
    R \triangleq \frac{K}{N} \ . \label{Rate1}
\end{align}

In a system parametrized by $(K,N,L,\Gamma,\Delta, \{P_{\ell},\Lambda_{\ell}\}_{\ell\in [L]})$, depicted in Figure~\ref{Fig: System Model}, our task is to devise an achievable scheme for the error-free recovery of any set of desired functions, subject to constraints on computation, communication, and multiplication loads. 
For general values of $\{\Lambda_{\ell},\, P_{\ell}\}_{\ell\in[L]}$ and for $\Gamma\geq 1$, a possible way to solve this problem is to exploit~\cite{Khalesi2025Tessellated}, and embed~\eqref{nonlinearlySep1} into the linearly-separable form in~\eqref{eq:linearized} with the corresponding output files $ W'(\pu)$ for each $\pu$, which requires $L'\triangleq\prod_{\ell \in [L]} P_{\ell}$ basis subfunctions 
to embed all multiplicative terms imposed by~\eqref{nonlinearlySep1}. 
    The required 
    \begin{align}\label{eq:N_TDC}
    N=\frac{K}{\Delta} \frac{L'}{\Gamma} \min(\Delta,\Gamma)
    \end{align}
    from~\cite[Theorem~1]{Khalesi2025Tessellated} for the single-shot scenario, directly implies that in this linearized alternative, the required 
    $N$ can grow exponentially with $L$.

\section{Problem Formulation in Tensor Form}
\label{Formulating}

In the $(K,N,L,\Gamma,\Delta, \{P_{\ell},\Lambda_{\ell}\}_{\ell\in [L]})$ framework, the desired functions in~\eqref{nonlinearlySep1} are fully represented by a tensor $\TensorF\in \mathbb{R}^{K \times P_1\times \hdots\times P_L}$ of all function coefficients $\{c_{k,\pu}\}$ across all the users, which is an order-$(L+1)$ tensor, i.e., a multi-way array with $L+1$ modes. With $\TensorF$ in place, we must decide on the computation assignment (encoding) and the communication protocol (decoding). 
For the error-free case, from~\cite{de2000}, this task is equivalent --- directly from \eqref{EncodedFiles},\eqref{DecedFiles}--- to solving a sparse tensor factorization problem subject to the sparsity constraints $\Gamma, \Delta$, and $\{\Lambda_\ell\}_{\ell=1}^L$, as specified in~\eqref{eq:Gamma},~\eqref{eq:Delta}, and ~\eqref{eq:Lambda}, respectively. This problem takes the form 
\begin{equation}
 \TensorF =\TensorE\times_1 \mathbf{D} \label{eq:DEF1}
\end{equation} 
where $\TensorE \in \mathbb{R}^{N\times P_1\times \hdots\times P_L}$ is the computing tensor, capturing the 
non-linear 
encoding tasks of servers that holds the coefficients $e_{n,\pu}$ from~\eqref{EncodedFiles}, $\mathbf{D}\in\mathbb{R}^{K\times N}$ is the communication matrix derived from the decoding coefficients $d_{k,n}$ in~\eqref{DecedFiles}, capturing the communication and linear decoding task done by each user, 
and finally, the operation $\times_1$ denotes the mode-$1$  product\footnote{The mode-$n$ product is defined similarly to (\ref{eq:mode-1-product}) with respect to the mode-$n$ unfolding of the tensor (cf. Definition~\ref{def:mode-n-unfold} in Appendix~\ref{tensor_basic}).
} of $\TensorE$ and $\mathbf{D}$ and is comprised of three consecutive operations:   
\begin{align}
\label{eq:mode-1-product}
    \TensorE \rightarrow \TensorE_{(1)}\, , \ 
    \TensorF_{(1)} = \mathbf{D}\TensorE_{(1)}\, , \  
    \TensorF_{(1)} \rightarrow \TensorF\ 
\end{align}
where the matrix $\TensorE_{(1)} \in \mathbb{R}^{N \times P_1P_2\hdots P_L}$ represents the mode-$1$ unfolding of the encoding tensor $\TensorE \in\mathbb{R}^{N\times P_1\times \hdots\times P_L}$.

To motivate the underlying idea behind the tensor construction, we next form tensors of desired function outputs, transmissions, and retrieved function outputs, as detailed below.


\paragraph{Desired Function Outputs in Vector Form}
We exploit the representation in~\eqref{nonlinearlySep1}, and consider the following vector of desired function outputs
\begin{align}
\label{function-vectors-0}
\f\triangleq [F_1,F_2,\hdots,F_K]^\intercal \in \mathbb{R}^K \ 
\end{align}
and for all $\pu \in \prod_{\ell \in [L]} [P_{\ell}]$, the tensor of basis coefficients
\begin{align}
\label{function-vectors-1}
\TensorF_k(\pu) \triangleq c_{k,\pu}\, ,\ \TensorF_k \in \mathbb{R}^{P_1\times P_2\times \hdots\times P_L},\, k \in [K]\  
\end{align}
as well as the tensor of multiplicative terms of $\{W_\ell\}_{\ell\in[L]}$
\begin{align}
\label{message-vectors-1}
    \TensorW(\pu) \triangleq \prod\nolimits_{\ell\in[L]} W_\ell^{p_\ell-1},\,\, \TensorW\in \mathbb{R}^{P_1 \times P_2 \times \hdots \times P_L}\ .
\end{align}
Next, using~\eqref{function-vectors-0}-\eqref{message-vectors-1}
leads us to the vector of function outputs
\begin{align}
    \f  =\textit{stack}_{1}(\TensorF_1,\hdots,\TensorF_K) \times^{[[1:L]]}_{[[2:L+1]]} \bar{\mathcal{W}}\label{Functions}
\end{align}
where $\textit{stack}_{1}(\TensorF_1,\hdots,\TensorF_K)$ forms an order-$(L+1)$ tensor $\TensorF\in\mathbb{R}^{K \times P_1\times \hdots\times P_L}$ 
composed by stacking the $K$ tensors $\{\TensorF_k\}_{k\in[K]}$ (each of order $L$) along a new first mode. 
In the above, the operation $\times^{[[1:L]]}_{[[2:L+1]]}$ simply generalizes the tensor contraction product\footnote{Tensor contraction product operation extends matrix multiplication to higher-order tensors by summing over shared modes.} (cf.~\cite{de2000}) by 
capturing an ordered set of common modes between two given tensors $\TensorF\in\mathbb{R}^{K\times P_1\times P_2\cdots\times P_L}$ and $\bar{\mathcal{W}}\in\mathbb{R}^{P_1\times P_2\cdots\times P_L}$. This generalized tensor contraction product yields a lower order tensor. Particularly, the $\times^{[[1:L]]}_{[[2:L+1]]}$-contraction product of $\TensorF$ and $\TensorW$ is obtained by contracting 
modes $[[2:L+1]]$ in $\TensorF$ and $[[1:L]]$ in $\TensorW$, corresponding to the indices $\pu \in \prod_{\ell\in [L]} [P_\ell]$, yielding an order-$1$ tensor (i.e., a vector) $\f\in\; \mathbb{R}^{K}$ with entries 
\begin{align}
\label{eq:contraction}
F_k=\sum_{\pu \in \prod_{\ell\in [L]} [P_\ell]} \bar{\mathcal{F}}(k,\pu)\, \bar{\mathcal{W}}(\pu)\, , \quad k\in [K]\, .
\end{align}


\paragraph{Transmissions in Vector Form}

In the communication phase, similarly to above, from~\eqref{EncodedFiles}, the transmission from server $n$ takes the following form $z_n = \TensorE_n \times^{[[1:L]]}_{[[1:L]]} \bar{\mathcal{W}}\in \mathbb{R}$, 
thus yielding the overall transmission vector
\begin{align}
    \mathbf{z}\triangleq [z_{1},z_{2},\hdots, z_{N}] = \bar{\mathcal{E}} \times^{[[1:L]]}_{[[2:L+1]]} \bar{\mathcal{W}}\, ,\ \ \mathbf{z}\in \mathbb{R}^{N} \
    \label{EncodedCashedData-1}
\end{align}
where for all $\pu \in \prod_{\ell \in [L]} [P_{\ell}]$,  the encoding coefficients $e_{n,\pu}$  from~\eqref{EncodedFiles} satisfy
\begin{align}
      \bar{\mathcal{E}}_{n}(\pu) & \triangleq e_{n,\pu},\,\, \bar{\mathcal{E}}_{n}\in \mathbb{R}^{P_1\times P_2\times \hdots \times P_L}\, . \label{encoding-tensor}
\end{align}

\paragraph{Retrieved Function Outputs in Vector Form}
In the decoding phase, from~\eqref{DecedFiles}, each retrieved function output takes the form $F'_k= \mathbf{d}_{k}^{\intercal}\mathbf{z}\in\mathbb{R}$, 
thus resulting in the vector of all outputs taking the form 
\begin{align}
\label{eq:received_functions}
\f' \triangleq [F'_{1},F'_{2},\hdots, F'_{K}]= [\mathbf{d}_1,\mathbf{d}_2,\hdots,\mathbf{d}_K]^{\intercal} \mathbf{z} \in\mathbb{R}^K
\end{align}
where the decoding coefficients $d_{k,n}$ from~\eqref{DecedFiles} satisfy
\begin{align}
        \mathbf{d}_k &\triangleq [d_{k,1},d_{k,2},\hdots, d_{k,N}]^\intercal \in \mathbb{R}^{N},\,\, k \in [K]\, .\label{decoding-vectors}
\end{align}

The tensors of basis coefficients in~\eqref{function-vectors-1}, the tensors of encoding coefficients \eqref{encoding-tensor}, and the vector of decoding coefficients in~\eqref{decoding-vectors} allow us to form the respective tensors 
\begin{align}
       \TensorF &\triangleq \textit{stack}_{1}(\TensorF_1,\hdots,\TensorF_K)\in \mathbb{R}^{K \times P_1 \times P_2 \times \hdots \times P_{L}},\label{Demand-Tensor-1}\\
        \TensorE &\triangleq \textit{stack}_{1}(\TensorE_1,\hdots,\TensorE_N)\in \mathbb{R}^{N\times P_1 \times P_2 \times \hdots \times P_{L}},\label{EncodingMatrix}\\
    \mathbf{D} &\triangleq [\mathbf{d}_1, \hdots , \mathbf{d}_K]^{\intercal} \in \mathbb{R}^{K \times N} \ .\label{DecodingMatrix}
\end{align}

We are now ready to establish the connection between the tensor forms in~\eqref{Demand-Tensor-1},~\eqref{EncodingMatrix}, and~\eqref{DecodingMatrix} and the tensor factorization problem. We note that employing the definitions of $\f$ and $f'$ from~\eqref{Functions} and~\eqref{eq:received_functions}, respectively, and setting the recovery error to zero, i.e., $\|\f' - \f \|^2  =0$,
we see that resolving our distributed computing problem requires that $\TensorF$ be decomposed as~\eqref{eq:DEF1}.

In terms of the corresponding connection to the sparsity of $\mathbf{D}$ and $\TensorE $, we recall from~\eqref{eq:Gamma}, our metric $\Gamma$, which directly from~\eqref{encoding-tensor} and~\eqref{EncodingMatrix}, implies the computation constraint as 
\[\max_{n \in [N]} \ \sum_{\ell\in [L]} \Bigl| \mathbbm{1} \Big(\supp\big(\bar{\mathcal{E}}(n, \underbrace{:,\hdots,:}_{\ell-1\,\,\text{terms}},p_\ell, 
 \underbrace{:, \hdots,:}_{L -\ell\,\,\text{terms}})\big) \neq \emptyset \Big) \Bigr| \leq \Gamma\, ,\quad p_\ell\in [2:P_{\ell}]\]

Furthermore, from~\eqref{eq:Delta}, we recall $\Delta$, which from~\eqref{decoding-vectors} and~\eqref{DecodingMatrix}, implies a communication constraint 
\[
\max_{n \in [N]} \ \bigl|\supp\big(\mathbf{D}(:,n)\big) \bigr| \leq \Delta \ .
\]

Finally, from~\eqref{eq:Lambda}, we recall $\Lambda_{\ell}$, which from~\eqref{encoding-tensor} and
\eqref{EncodingMatrix}, suggests, for all $\ell\in [L]\, ,\, p_\ell \in [P_\ell]$, the  multiplication constraints 
 \begin{align*}     
 \|\bar{\mathcal{E}}(n,p_1,p_2, \hdots, p_{\ell-1}, :,p_{\ell +1}, \hdots, p_L)\|_0 &\leq \Lambda_\ell\, .
 \end{align*}

\begin{figure}[t!]
      \centering
      \includegraphics[scale=0.75]{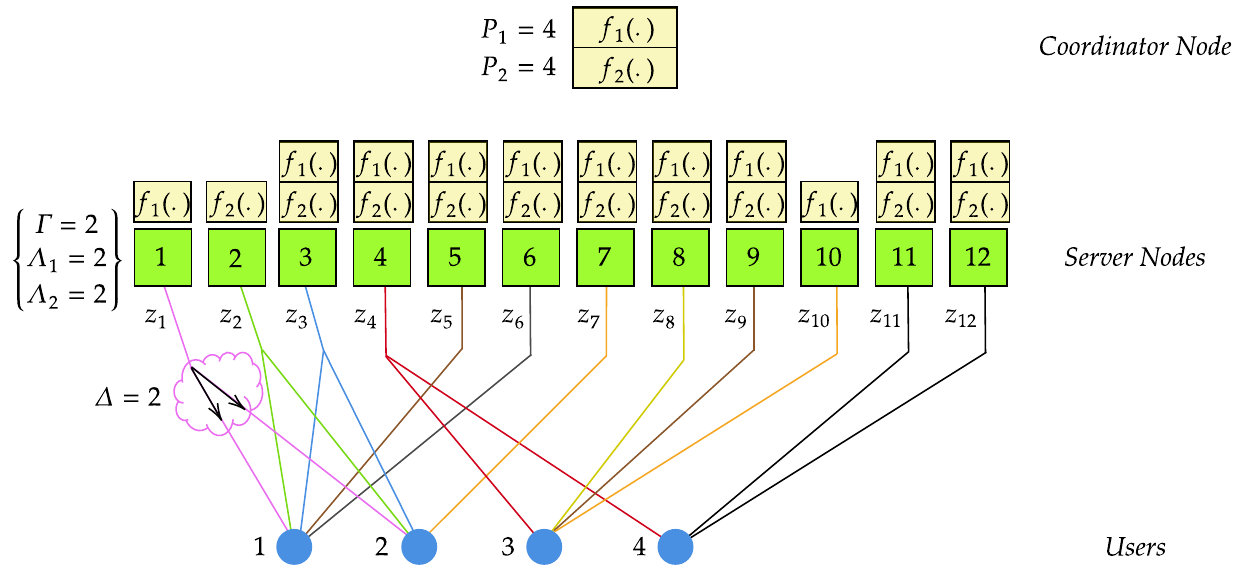}
      \caption{Corresponding to Example~\ref{detailed_ex}, the distributed computing system is illustrated for a set of system parameters, including computation cost $\Gamma=2$, communication cost $\Delta=2$, and multiplication costs $\Lambda_1=\Lambda_2=2$ for distributed computation of multivariate polynomial demands with the maximum degree ranges $P_1=P_2=4$.}\label{fig:det_ex}
\end{figure}

We next explain in detail a full example, motivating the proposed tensor factorization approach by the exact recovery of the user demands.
\begin{ex}\label{detailed_ex}
Let us consider the system model depicted in Figure~\ref{fig:det_ex}, where we have $N=12$ servers, $K=4$ users, $L=2$ basis subfunctions whose maximum range of degrees are specified by $P_1=P_2=4$. Let us assume each server is able to compute up to $\Gamma=2$ basis subfunctions, $\Lambda_1=\Lambda_2=2$ number of multiplications, and transmit to up to $\Delta =2$ users in one shot $T=1$.

Consider that in the demand phase, each user requests to compute a linear combination of the multiplicative terms of output files in $\mathbf{W}=(W_1,W_2)$, corresponding to 
\begin{align}\label{NLS_det_ex}
    F_1(\mathbf{W}) &= 
    2W_1  + 4 W_2 + 2W_1W_2 + 2W_1^3W_2+W_1^3W_2^2
    \, ,\nonumber\\
    F_2(\mathbf{W}) &=  
    2W_1W_2 + W_1  +W_2+2W_1^2W_2^2+5W^2_1  W^3_2 + W_1^3W_2^3
    \, ,\nonumber\\
    F_3(\mathbf{W}) &= 
    2 W_1W_2 + 4 W_1W_2^2 +  6W_1^2W_2^2 + 5 W_1^3 
    \, ,\nonumber\\
    F_4(\mathbf{W}) &= 
    3 W_1 W_2+ 2W_1W_2^2+5 W_1^3W_2^3
\end{align}
\ahmad{
where each user demand can be captured by $F_k(\mathbf{W})=\mathbf{F}_k \odot {\bf M},\, k=1,2,3,4$, using the Hadamard product of two matries $\mathbf{F}_k$ and ${\bf M}$, where
}
\begin{align*}
    {\bf M}(W_1,W_2) & =
\left[ \begin{array}{cccc}
   1 & W_1 & W_1^2 & W_1^3\\
   W_2 & W_1 W_2 & W_1^2 W_2 & W_1^3 W_2 \\
   W_2^2 & W_1 W_2^2 & W_1^2 W_2^2 & W_1^3 W_2^2\\
   W_2^3 & W_1 W_2^3 & W_1^2 W_2^3 & W_1^3 W_2^3\\
\end{array}\right],\, \quad {\bf M}\in \mathbb{R}^{P_2\times P_1}
\end{align*}
captures the computational tasks, whose elements are the multiplicative terms of the output files $ W_1$ and $ W_2$. \ahmad{Furthermore, the coefficients in~\eqref{NLS_det_ex} are described in matrix form as}
\begin{align*}
     \mathbf{F}_1 = 
\left[ \begin{array}{cccc}
   0 & 2 & 0& 0\\
   4 & 2 & 0 &2 \\
   0 & 0 & 0 &1\\
  0 & 0& 0 & 0\\
\end{array}\right],\
 \mathbf{F}_2 = 
\left[\begin{array}{cccc}
   0 & 1 & 0 &0\\
   1& 2 &0& 0 \\
   0 & 0 & 2 & 0\\
  0 & 0 & 5 &1\\
\end{array}\right],\
 \mathbf{F}_3 = 
\left[\begin{array}{cccc}
  0 & 0 & 0 & 5\\
   0 &2 &0 & 0 \\
   0 & 4 & 6 &0\\
  0 & 0& 0 & 0\\
\end{array}\right],\
 \mathbf{F}_4 = 
\left[ \begin{array}{cccc}
   0 & 0 & 0 & 0\\
   0 & 3 & 0& 0 \\
   0 & 2 & 0& 0\\
   0 &0 & 0&5 \\
\end{array}\right]
\end{align*}
where $\mathbf{F}_k=\TensorF(k,:,:)\in \mathbb{R}^{P_2\times P_1}$
is indeed a subtensor (here matrix) of the entire demand tensor $\TensorF\in \mathbb{R}^{K\times P_2\times P_1}$, \ahmad{which captures all coefficients in all users' demands in~\eqref{NLS_det_ex} with the first and second dimension reflecting $W_2$ and $W_1$, respectively}, as illustrated in Figure~\ref{TensorF_ex}.


\begin{figure}[t!]
\centering
\includegraphics[scale=0.9]{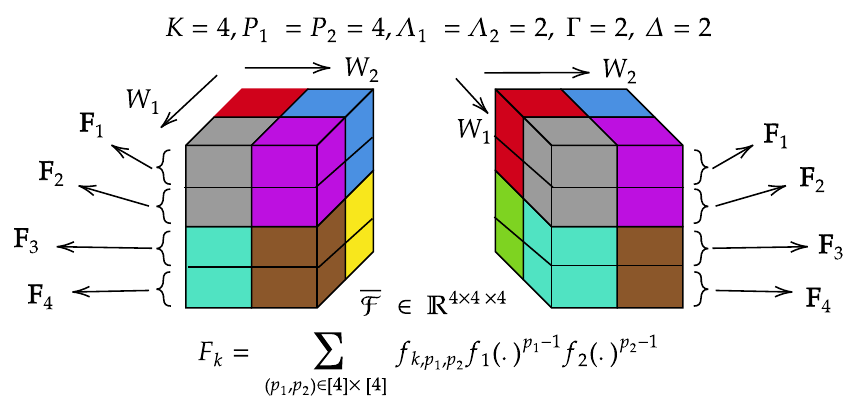}
\caption{Corresponding to Example~\ref{detailed_ex}, the tensor $\TensorF$ forming all user demands is illustrated here.}
\label{TensorF_ex}
\end{figure}

In the computation  phase, the coordinator allocates the set of basis subfunctions $\mathcal{S}_n$ to server $n=1,\ldots,12$,
determined as $\mathcal{S}_1 =\{1\},\, \mathcal{S}_2 = \{2\},\, \mathcal{S}_{10} = \{1\},\, \mathcal{S}_{[3:13] \backslash \{10\} } = \{1,2\}$, \ahmad{along with the set $\mathcal{P}_n$ of index tuples $\pu=(p_1,p_2)$ corresponding to the multiplicative terms $W_1^{p_1-1}W_2^{p_2-1}$ to be computed by server $n=1,\ldots,12$}, which is described as $\mathcal{P}_1=\{(2,1)\},\, \mathcal{P}_2=\{(1,2)\},\, \mathcal{P}_3=\mathcal{P}_4=\{(2,2)\},\, \mathcal{P}_5=\{(4,2)\},\, \mathcal{P}_6=\{(4,3)\},\, \mathcal{P}_7=\{(3,3),\, (3,4),\, (4,4)\},\, \mathcal{P}_8=\{(2,3)\},\, \mathcal{P}_9=\{(3,3)\},\, \mathcal{P}_{10}=\{(4,1)\},\, \mathcal{P}_{11}=\{(2,3)\},\, \mathcal{P}_{12}=\{(4,4)\}$.

Next, the coordinator node asks server $n=1,\ldots,12$ to compute ${z}_n$ according to
\begin{align*}
    z_{1} &= 
    2 W_1\, ,\ \  
       z_{2} = 4 W_2\, ,\ \ 
       z_{3} = W_1W_2\, ,\ \
       z_{4} = W_1W_2\, ,\ \
       z_{5} = 2W_1^3W_2\, ,\ \ 
       z_{6} = W_1^3W^2_2 \ ,\\
       z_{7} &=2W_1^2W_2^2 + 5W_1^2W_2^3 +W_1^3W_2^3\, ,\ \ 
       z_{8} = 4W_1W_2^2\, ,\ \ 
       z_{9} = 6W_1^2W_2^2\, ,\ \ 
       z_{10} = 5W_1^3 \ ,\\ 
       z_{11} &= 2W_1W_2^2\, ,\ \ 
       z_{12} = 5W_1^3W_2^3 
\end{align*}
where $z_n=\mathbf{E}_n \odot {\bf M}$ is the computational outputs that server $n$ evaluates, and the encoding coefficients above are described in the matrix form with the non-zero entries
\begin{align*}
\mathbf{E}_1(1,2)&=2\, ,\quad \mathbf{E}_2(2,1)=4\, ,\quad \mathbf{E}_3(2,2)=1\, ,\quad \mathbf{E}_4(2,2)=1\, ,\quad \mathbf{E}_5(2,4)=2\, ,\quad \mathbf{E}_6(3,4)=1\, ,\nonumber\\
\mathbf{E}_7(3,3)&=2\, ,\quad \mathbf{E}_7(4,3)=5\, ,\quad \mathbf{E}_7(4,4)=1\, ,\quad \mathbf{E}_8(3,2)=4\, ,\quad \mathbf{E}_9(3,3)=6\, ,\quad \mathbf{E}_{10}(1,4)=5\, ,\nonumber\\ 
\mathbf{E}_{11}(3,2)&=2\, ,\quad \mathbf{E}_{12}(4,4)=5
\end{align*}
where $\mathbf{E}_n=\TensorE(n,:,:)\in \mathbb{R}^{P_2\times P_1}$ denotes the encoding subtensor (here matrix) of the entire encoding tensor $\TensorE$, which captures the assignment of computational tasks to all the servers $n=1,\ldots,12$, and $\mathbf{E}_n(p_2,p_1)$ demonstrates the $(p_2,p_1)$-th entry of the encoding matrix $\mathbf{E}_n$, where $p_2\in [P_2],\, p_1\in [P_1]$.



In the communication phase, the coordinator requests server $n$ to send $z_{n}$ to the users in $\T_{n}$, where $\T_{1} = \{1\}, \: \T_{2} = \{1\},\: \T_{3} = \{1,2\}, \: \T_{4} =\{3,4\},\: \T_{5} =\{1\},\: \T_{6} = \{2\}, \T_{7} = \{2\}, \T_{8} = \{ 2\}, \T_{9} =\{3\},\: \T_{10} = \{3\}, \T_{11} = \{4\}, \T_{12} = \{ 4\}$.



Each user is then going to decode by linearly combining $z_n$'s according to
\begin{align*}
    F_1' =& z_{1} + z_{2} + 2 z_{3} +z_{5} +z_{6}\, , \quad 
    F_2' = 1/2 z_1+ 1/4 z_2 + 2 z_{3}  +  z_{7}\, ,\\
    F_3' =&   2 z_{4} + z_{8} + z_{9}+z_{10}\, , \quad
    F_4' = 3 z_{4} +  z_{11} + z_{12}\, .
\end{align*}

To decode, each user $k =1,2,3,4$ utilises a linear decoding algorithm according to the elements on the $k$-th row of the decoding matrix $\mathbf{D}$, where
\begin{align*}
    \mathbf{D}  \triangleq 
    \left[ \begin{array}{cccccccccccc}
  1& 1 & 2 &0 & 1& 1 & 0 & 0 & 0 & 0 & 0 & 0\\
1/2 & 1/4 & 2 & 0 & 0 & 0 & 1 & 0 & 0 & 0 & 0 & 0  \\
0 & 0 & 0 & 2 & 0& 0 & 0 &1 & 1 & 1 & 0 & 0 \\
0 & 0 & 0 & 3 & 0& 0 & 0 &0 & 0 & 0 & 1 & 1
\end{array}\right]\, .
\end{align*}

It can be easily verified that $F_k = F'_k,\, k =1,2,3,4$, confirming the lossless reconstruction of the demands of all users.
  
\end{ex}

As illustrated above, the design of the task assignment sets $\mathcal{S}_n$, $\mathcal{P}_n$ to the servers,~\ahmad{which} governs the sparsity of encoding tensor $\TensorE$, and the~\ahmad{design of} connectivity sets $\mathcal{T}_n$,~\ahmad{which} governs the sparsity of decoding matrix $\mathbf{D}$ is central to the proposed distributed computing framework. Our goal is to guarantee lossless recovery of all user demands under sparsity constraints while 
requiring significantly fewer servers than existing solutions~\cite{Khalesi2025Tessellated}.

We proceed to present the main results in the next section.
\section{Main Results}
\label{Results}

\ahmad{
In this section, we present the main findings of this paper, summarized below.}

\ahmad{
First in Section~\ref{subsec:rate}, we devise a sparse tensor factorization scheme for the proposed lossless
$(K,N,L,\Gamma,\Delta, \{P_{\ell},\Lambda_{\ell}\}_{\ell\in [L]})$ distributed computing setting, by upper bounding the required number of servers $N$, which is the common dimension between $\TensorE,\, \mathbf{D}$, directly impacting our sparse tensor factorization approach (cf. Theorem~\ref{Achievability-Converse}).
}
\ahmad{
We then examine Theorem~\ref{Achievability-Converse} numerically to illustrate the intuition and the performance of the proposed scheme (cf. Example~\ref{single-shot-example-simple}).}
\ahmad{
Next, we compare the proposed sparse tensor factorization approach with the sparse matrix factorization scheme of~\cite{Khalesi2025Tessellated} in terms of the total computation cost (cf. Proposition~\ref{comparison1}).}

\ahmad{
We then present a tighter achievable rate result in Section~\ref{subsec:refined_rate} by leveraging a combinatorial reduction algorithm (cf. Algorithm~\ref{Alg:GreedyAssignment} in Theorem~\ref{Achievability-new}).
Next, we give a numerical elaboration and illustration of the proposed Algorithm~\ref{Alg:GreedyAssignment} (cf. Example~\ref{ex:L3merged}).
Finally, we conclude this section by analyzing the complexity of the proposed Algorithm~\ref{Alg:GreedyAssignment} (cf. Remark~\ref{rem:complexity}).
}

\ahmad{
All the result holds without any restriction on the dimensions, provided that each subtensor of $\TensorF$ has full rank, a condition that is easily justified in our real-valued function settings.
}


\subsection{Achievable System Rate}\label{subsec:rate}

\ahmad{
We here characterize the achievable rate for the lossless $(K,N,L,\Gamma,\Delta, \{P_{\ell},\Lambda_{\ell}\}_{\ell\in [L]})$ distributed computing system under $P_{\ell}=P,\Lambda_{\ell}=\Lambda$ for all $\ell\in [L]$ and $(\Delta | K$, $\Lambda | P)$.
}

\begin{theo} 
\label{Achievability-Converse}
The achievable rate of the lossless $(K,N,L,\Gamma,\Delta, \{P_{\ell},\Lambda_{\ell}\}_{\ell\in [L]})$ distributed computing system, under $P_{\ell}=P,\Lambda_{\ell}=\Lambda$ for all $\ell\in [L]$ and $(\Delta | K$, $\Lambda | P)$, takes the form $R = K/N$, where 
\begin{align}\label{achiv-o}
N&\leq \frac{K}{\Delta}\, {L\choose{\Gamma}}\, \min \big(\Delta, \Lambda^\Gamma)\,
\big(\frac{P}{\Lambda}\big)^\Gamma\, .
\end{align}
\end{theo}

\begin{proof}

This proof has two parts. We first, in Part~I, define the $n$-th rank-one contribution support, representative rank-one support, or tile, along with their related parameters. These tiles represent equivalence classes of rank-one contribution supports, which show how any support constraint on $\mathbf{D} $ and $ \TensorE$ contributes to the supports on $\TensorE\times_1\mathbf{D}$. The correspondence above is shown via Lemma~\ref{Lemma-DE-Support}. 

Then in Part~II, to design the decoding matrix $\mathbf{D}$ and encoding tensor $\TensorE$ for lossless reconstruction of the demand tensor $\TensorF$, the following steps are involved.
\begin{enumerate}
    \item Designing the sizes of the tiles for the product tensor $\TensorF=\TensorE\times_1\mathbf{D}$.
  
    \item Creating and filling the non-zero elements of the tiles in the product tensor $\TensorF$.

    \item Placing the filled tiles in $\mathbf{D}$ and in $\TensorE$, accordingly.

    \item The number of servers required is obtained by associating the rank of each tile with the servers and then summing the ranks over all tiles\footnote{Quickly recall that for a $(L+1)$- dimensional tensor $\bar{\mathcal{E}}$, then $\bar{\mathcal{E}}(n,:,\hdots,:)$ represents a $L$ dimensional subtensor and $\mathbf{D}(:,n)$ for a matrix $\mathbf{D}$ is its $n$-th column, and $\Supp(\mathbf{D})$ ($ \Supp(\bar{\mathcal{E}})$) is a binary matrix (tensor), indicating the support of $\mathbf{D}$ ($\bar{\mathcal{E}}$). Also, when we refer to a support constraint, this will be in the form of a binary matrix (tensor) that indicates the support (the position of the allowed non-zero elements) of a matrix (tensor) of interest.}.
\end{enumerate}
 
Throughout the proof, the design borrows concepts and definitions from tensor theory in Appendix~\ref{tensor_basic}, and the blockwise SVD approach of~\cite{le2023spurious}, generalized to the multilinear scenario in~\cite{de2000} by utilizing tensors, as detailed in Appendix~\ref{MLSVD}.

We now proceed with Part I \ahmad{of the proof for Theorem~\ref{Achievability-Converse}}.

\textbf{Part I (Basic Concepts and Definitions).}
We here present the basic concepts and preliminaries of our proposed scheme. 
We first introduce the concept of \emph{rank-one contribution support}, which is essential in our tile design procedure by capturing the constitutive components of each tile.
\begin{defi}\label{def-r1}
Given two support constraints $\mathbf{I} \in \{0,1\}^{K \times N}$ and $\TensorJ \in \{0,1\}^{N\times P_1\times P_2\times\hdots\times P_L}$, then for any $n \in [N]$, we refer to 
\begin{align}
\TensorS_{n}(\mathbf{I}, \TensorJ) \triangleq \TensorJ(n, :,\hdots,:)\times_1 \mathbf{I}(:,n) 
\end{align}
as the $n$-th \emph{rank-one contribution support}. 
\end{defi}
We note that when the supports are implied, we may shorten $\TensorS_{n}(\mathbf{I}, \TensorJ)$ to just $\TensorS_n$.
The aforementioned $\mathbf{I}$ and $\TensorJ$ will generally represent the support of $\mathbf{D}$ and $\TensorE$ respectively, while $\TensorS_{n}(\mathbf{I}, \TensorJ)$ will generally capture some of the support of $\TensorE\times_1\mathbf{D}$ and thus of $\TensorF$. We have the following lemma for this. 
\begin{lem}\label{Lemma-DE-Support}
    For $\mathbf{I} \triangleq \supp(\mathbf{D}) $ and $\TensorJ \triangleq \supp(\TensorE)$, then
    \begin{align}
         \cup^{N}_{n=1} \TensorS_n(\mathbf{I}, \TensorJ) &= \cup^{N}_{n=1}\TensorJ(n,:,\hdots,:)\times_1\mathbf{I}(:,n)\nonumber\\
         &\supseteq \Supp(\TensorE\times_1 \mathbf{D})\, .\label{Equation-nultiplicative-contribution}
    \end{align}
\end{lem}
\begin{proof}[Proof of Lemma~\ref{Lemma-DE-Support} ]
The above follows from Definition~\ref{def-r1} and 
$\TensorE\times_1\mathbf{D} = \sum^{N}_{n=1}\TensorE(n,:,\hdots,:)\times_1 \mathbf{D}(:,n) $. 
\end{proof}

\ahmad{We continue with the proof of Theorem~\ref{Achievability-Converse} by introducing}
the concept of \emph{equivalence classes of rank-one supports}, which we will utilize later for enumerating the number of required servers.
\begin{defi}\label{Def2}
   Given two supports $\mathbf{I} \in \{0,1\}^{K \times N} $ and $\TensorJ \in \{0,1\}^{N\times P_1\times P_2\times\hdots\times P_L}$, the \emph{equivalence classes of rank-one supports} are defined by the equivalence relation $i \sim j$ on $[N]$, which holds if and only if $\TensorS_i = \TensorS_j$, as represented in Figure~\ref{fig:my_label1}. 
\end{defi}

The above splits the columns of $\mathbf{D}$ (and correspondingly the rows of $\TensorE$) into equivalence classes such that $i \sim j$ holds if and only if $ \TensorJ(i,:,\hdots,:)\times_1\mathbf{I}(:,i) = \TensorJ(j,:,\hdots,:)\times_1\mathbf{I}(:,j)$.

\ahmad{
We next introduce a basic assumption on $\mathbf{D}$ and $\TensorE$, termed the \emph{disjoint support assumption}, which will be used to derive
the result of Theorem~\ref{Achievability-Converse}.
\begin{defi}\label{tensor-disjointSupportAssumption}
(\emph{Disjoint Support Assumption}.)
We say that the matrix $\mathbf{D}\in \mathbb{R}^{K\times N}$ and the tensor $\TensorE \in \mathbb{R}^{N\times P_1\times \hdots \times P_{L}}$ accept the \emph{disjoint support assumption} if and only if for any distinct pair of columns $\mathbf{D}(:,i),\mathbf{D}(:,i'), i, i' \in [N]$ of $\mathbf{D}$ and the respective two subtensors  $\TensorE(i,\underbrace{:,\hdots,:}_{L\, \text{terms}}),\TensorE(i',\underbrace{:,\hdots,:}_{L\, \text{terms}})$ of $\TensorE $, then either we have 
$\text{supp}(\TensorE(i,:,\hdots,:) \times_{1} \mathbf{D}(:,i)) = \text{supp}(\TensorE(i',:,\hdots,:) \times_{1} \mathbf{D}(:,i') )$ 
or 
$\text{supp}(\TensorE(i,:,\hdots,:) \times_{1} \mathbf{D}(:,i)) \cap \text{supp}(\TensorE(i',:,\hdots,:) \times_{1} \mathbf{D}(:,i')) = \emptyset$. 
\end{defi}
}

Next, we describe the \emph{high-dimensional tiles} and their corresponding \emph{component columns and rows}, which are essential in our tensor factorization procedure.
\begin{defi} \label{Def2b}
    For two supports $\mathbf{I} \in \{0,1\}^{K \times N}, \TensorJ \in \{0,1\}^{N\times P_1\times P_2\times\hdots\times P_L}$, and for $\mathcal{C}$ being the collection of equivalence classes as in Definition~\ref{Def2}, then for each class $\mathcal{P} \in \mathcal{C}$, we call $\TensorS_{\mathcal{P}}$ to be the \emph{representative rank-one support} of class $\mathcal{P}$, which will also be called the \emph{tile} labeled\footnote{Note that $\TensorS_n = \TensorS_\mathcal{P}$ for any $n \in \mathcal{P}$. Furthermore, the term \emph{tile} will be used interchangeably to represent both $\TensorS_{\mathcal{P}}$ and $\mathcal{P}$.} by $\mathcal{P}$. Furthermore, for each tile $\TensorS_{\mathcal{P}}$, let $\mathbf{c}_{\mathcal{P}} \triangleq \mathbf{I}(:,n), n \in \mathcal{P}$ (resp. $\mathbf{r}_{\mathcal{P}} \triangleq \TensorJ(n,:,\hdots,:), n \in \mathcal{P}$) be the corresponding \emph{component column} (resp. \emph{component row}) of the representative rank-one support. Finally, $\mathcal{C}_{\mathcal{P}} \triangleq \text{supp}(\mathbf{c}_{\mathcal{P}} ) \subset [K] $ describes the set of the indices of the non-zero elements in $\mathbf{c}_{\mathcal{P}}$, while $\mathcal{R}_{\mathcal{P}} \triangleq \text{supp}(\mathbf{r}_{\mathcal{P}}) \subset \prod_{\ell\in [L]} [P_\ell]
    $ describes the set of indices of the non-zero elements in $\mathbf{r}_{\mathcal{P}}$. This is illustrated in Figure~\ref{fig:difference}.
\end{defi}

\begin{figure}[t!]
\centering
\[\begin{array}{cc} \includegraphics[scale=0.55]{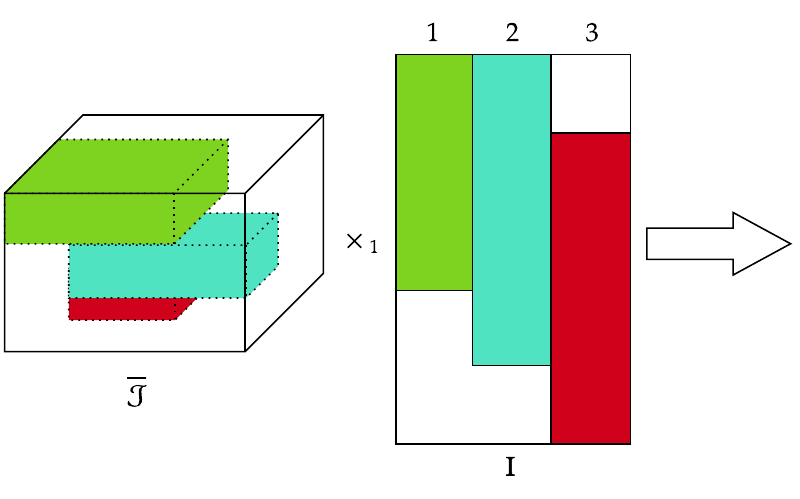} & \includegraphics[scale=0.45]{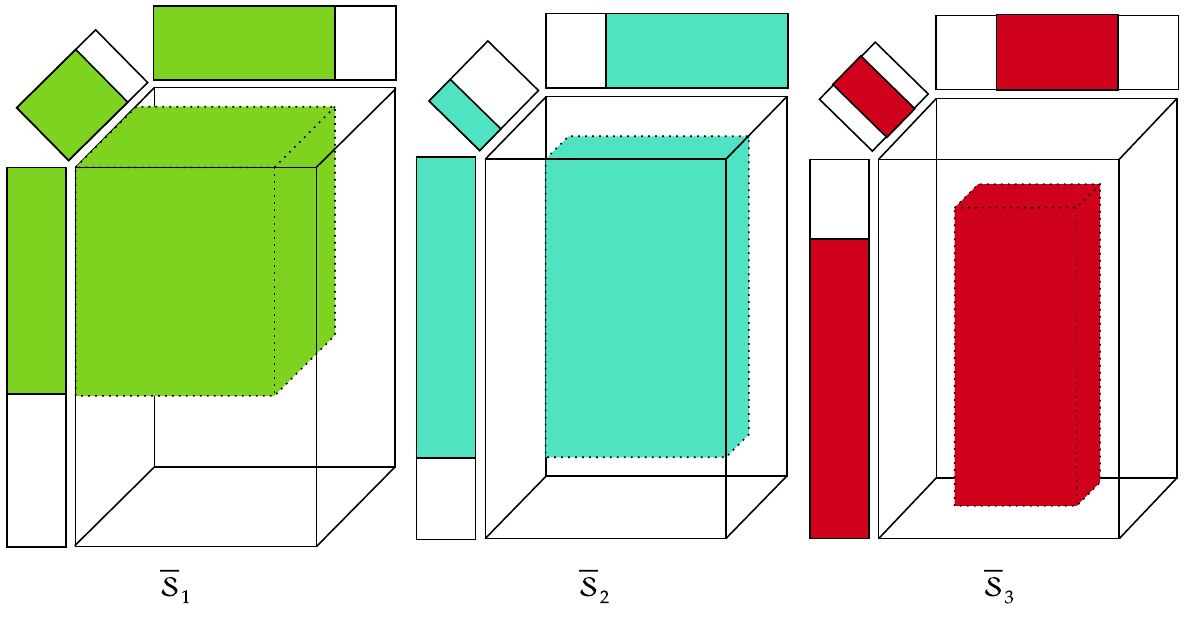} \end{array}\]
\caption{The figure on the left illustrates the support constraints $\mathbf{I}$ and $\TensorJ$ on $\mathbf{D}$ and $\TensorE $ respectively.  
The constraints $\mathbf{I}(:,1)$ and $\TensorJ(1,:,:)$ on the columns and rows of $\mathbf{D}$ and $\TensorE $ respectively are colored green, $\mathbf{I}(:,2)$ and $\TensorJ(2,:,:)$ are colored cyan and $\mathbf{I}(:,3)$ and $\TensorJ(3,:,:)$ are colored red. The product of a column with a row of the same color yields the corresponding rank-one contribution support $\TensorS_{n}(\mathbf{I},\TensorJ),n=1,2,3,$ as described in Definition~\ref{def-r1}, and as illustrated on the right side of the figure.}
\label{fig:my_label1}
\end{figure}

\begin{figure}[t!]
\centering
\includegraphics[scale=0.55]{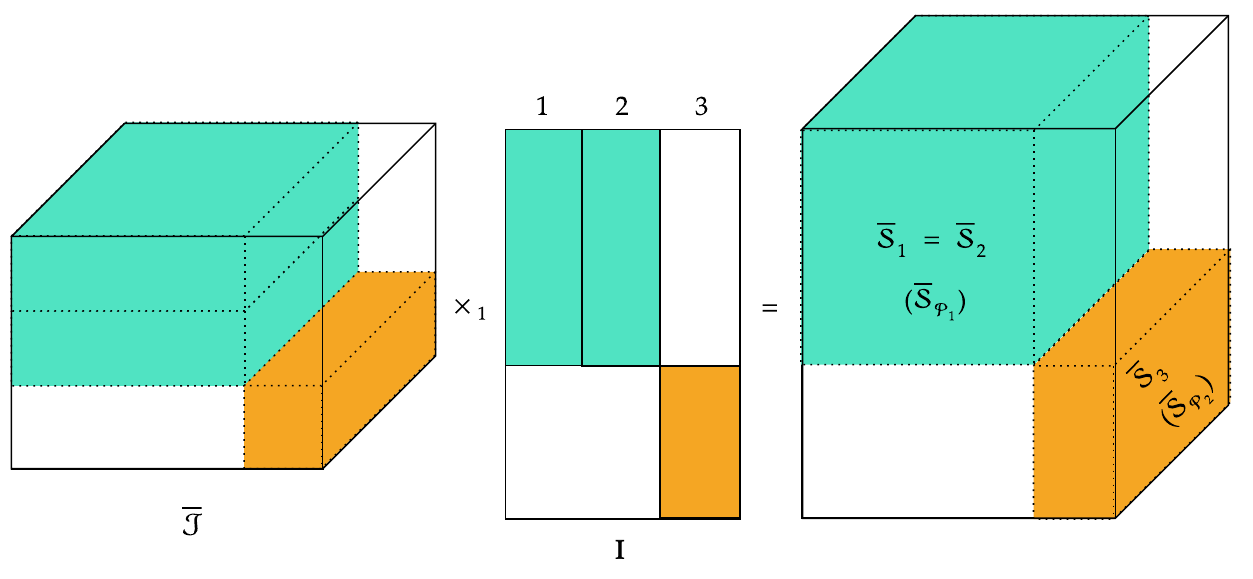}
\caption{This figure illustrates three different rank-one contribution supports $\TensorS_{1},\TensorS_{2},\TensorS_{3}$, where the first two fall into the same equivalence class $\TensorS_{\mathcal{P}_1}  = \TensorS_{1} = \TensorS_{2}$, while $\TensorS_{\mathcal{P}_2}  = \TensorS_{3}$.}
\label{fig:difference}
\end{figure}

We then indicate how each equivalence class of tiles covers the product tensor, composed of the union of the representative rank-one supports.
\begin{defi} \label{Def2c}
For every set $\mathcal{C}' \subseteq \mathcal{C}$ of equivalence class, let us define the \emph{union of the representative rank-one supports} 
\begin{align}
\label{eq:union_rankones}
\mathcal{S}_{\mathcal{C}'}\triangleq \bigcup_{\mathcal{P} \in \mathcal{C}'} \TensorS_{\mathcal{P}} 
\end{align}
to be the point-wise logical OR of the corresponding $\TensorS_{\mathcal{P}}$.
\end{defi}

In the above definition (cf.~\eqref{eq:union_rankones}), $\mathcal{S}_{\mathcal{C}'}$ is simply the area of the product tensor covered by all the tiles $\mathcal{P}\in\mathcal{C}'$.

\ahmad{
We conclude Part I by introducing the maximum rank of a representative rank-one support of an equivalence class $\mathcal{C}$, which will be used to determine the number of servers and derive the upper bound in the proof of Theorem~\ref{Achievability-Converse}.
}
\begin{defi}[\!\!\cite{le2023spurious}] \label{maxrank}
The \emph{maximum rank of a representative rank-one support of class} $\mathcal{P}\in \mathcal{C}$ is obtained as
\begin{align}
   r_{\mathcal{P}} \triangleq \min (|\mathcal{C}_{\mathcal{P}}|, |\mathcal{R}_{\mathcal{P}}|)\, . \label{max-rank}
\end{align}
\end{defi}      
The above simply says that for the case of $\mathbf{I} = \supp(\mathbf{D})$ and $\TensorJ = \supp(\TensorE)$, then the part of tensor $\TensorE\times_1 \mathbf{D}$ covered by tile $\TensorS_\mathcal{P}$ can have rank which is at most $r_{\mathcal{P}}$.

With the above in place, we proceed \ahmad{with Part II of the proof of Theorem~\ref{Achievability-Converse}} to describe the method used to design the matrix $\mathbf{D}$ and the tensor $\TensorE$ for our sparse tensor factorization $\TensorF=\TensorE\times_1 \mathbf{D}$.

\textbf{Part II (Construction of $\mathbf{D},\, \TensorE$).}
We next detail the four-step procedure for deriving the decoding matrix $\mathbf{D}$ and the encoding tensor $\TensorE$.

\paragraph{Step 1: Designing the sizes of the tiles for the $K\times P_1\times \hdots \times P_L$ product tensor $\TensorE\times_1 \mathbf{D}$}

The family of all tiles is initially specified by the set of equivalence classes $\mathcal{C}$ (cf. Definition~\ref{Def2}), defined as 
\begin{align}\label{eq:eq-classes}
    \mathcal{C}\;\triangleq\; \Big\{ \mathcal{P}_{i,\j,r}\;\subseteq\;[K]\times [P_1]\times\cdots\times[P_L] \Big\}
\end{align}
where $\mathcal{P}_{i,\j,r}\;\triangleq\;\mathcal{C}_{\mathcal{P}_{i,\j,r}} \times \mathcal{R}_{\mathcal{P}_{i,\j,r}}$ 
represents a high-dimensional \emph{tile}, which is initially indexed by $(i,\j,r)$ and identified by $\mathcal{C}_{\mathcal{P}_{i,\j,r}},\mathcal{R}_{\mathcal{P}_{i,\j,r}}$,
where 
\begin{align}
    \mathcal{C}_{\mathcal{P}_{i,\j,r}} \;\triangleq\; \big[\,1+(i-1)\Delta:\min(i\Delta,K)\,\big] \subseteq [K]\, ,\ \
    i \in \Big[ \Big\lceil \tfrac{K}{\Delta} \Big\rceil \Big] 
\end{align}
represents the indices of the corresponding columns of $\mathbf{I}$ (cf. Definition~\ref{Def2b}),
$\j\triangleq (j_1,\hdots,j_L)$ with each $j_\ell$ indexing the $\ell^{th}$ basis subfunction in tensor $\TensorF$,
and 
\begin{align}
\mathcal{R}_{\mathcal{P}_{i,\j,r}}
\;&\triangleq\; 
\prod\limits_{\ell\in [L]} [1] \cup \Big(\mathbbm{1}(\ell \in \mathcal{Q}_r) \times [1 + (j_{\ell} -1) \Lambda_\ell : \min(j_\ell \Lambda_\ell,P_{\ell})]\Big)
\;\subseteq\;[P_1]\times\cdots\times[P_L]\, ,\nonumber\\ 
\forall j_\ell &\in \Big[ \Big\lceil \tfrac{P_\ell}{\Lambda_\ell} \Big\rceil \Big], \, r\in \Big[\binom{L}{\Gamma} \Big]
\end{align} 
demonstrates the set of indices of the corresponding rows of $\TensorJ$ (cf. Definition~\ref{Def2b}), wherein $\mathcal{Q}_r\triangleq \{\ell_1,\hdots,\ell_\Gamma\}\subseteq[L]$ denotes a choice of $\Gamma$ active coordinates among all $L$ dimensions, i.e., $\Gamma$ basis subfunctions with the exponent terms $\{p_\ell-1>0\}_{\ell\in \mathcal{Q}_r}$.

The set of equivalence classes $\mathcal{C}$ in~\eqref{eq:eq-classes}, depending on the dimensions of $\TensorF$, include the equivalence classes $\mathcal{C}_s,\, s\in [4]$, which follow from Definition \ref{Def2} for all possible scenarios where $\Delta \mid K$, $\Lambda_\ell \mid P_\ell$, $\Delta \nmid K$, and $\Lambda_\ell \nmid P_\ell$. 
We note that for the cases $\Delta \mid K$ and $\Lambda_\ell \mid P_\ell$, we divide each $(\Gamma+1)$-dimensional space corresponding to each $\mathcal{Q}_r,\, r\in [{L\choose{\Gamma}}]$ to $\frac{K}{\Delta}\times \prod\limits_{\ell\in \mathcal{Q}_r} \frac{P_\ell}{\Lambda_\ell}$ tiles each with dimensions $\Delta\times \Lambda_{\ell_1}\times\hdots\times\Lambda_{\ell_\Gamma}$. A similar procedure holds for the residual space for the general cases $\Delta\nmid K$ and $\Lambda_\ell \nmid P_\ell$.

The number of representative rank-one supports for tiles $\mathcal{P}$ in each equivalence class is thus
\begin{align}\label{eqclass-c}
    |\mathcal{C}_1|&=\lfloor \frac{K}{\Delta} \rfloor \sum\limits_{r=1}^{L\choose{\Gamma}} \prod\limits_{\ell\in \mathcal{Q}_r} \lfloor \frac{P_\ell}{\Lambda_\ell} \rfloor \, ,
    \quad |\mathcal{C}_2|=\sum\limits_{r=1}^{L\choose{\Gamma}} \prod\limits_{\ell\in \mathcal{Q}_r} \lfloor \frac{P_\ell}{\Lambda_\ell} \rfloor \, ,\nonumber\\
    |\mathcal{C}_3|&=\lfloor \frac{K}{\Delta} \rfloor \sum\limits_{r=1}^{L\choose{\Gamma}} \prod\limits_{\ell\in \mathcal{Q}_r:\ \Lambda_\ell | P_\ell} \lfloor \frac{P_\ell}{\Lambda_\ell} \rfloor \times\prod\limits_{\ell'\in \mathcal{Q}_r:\ \Lambda_{\ell'} \nmid P_{\ell'}} \lceil \frac{P_{\ell'}}{\Lambda_{\ell'}} \rceil \, ,\nonumber\\
    |\mathcal{C}_4|&=\sum\limits_{r=1}^{L\choose{\Gamma}} \prod\limits_{\ell\in \mathcal{Q}_r:\ \Lambda_\ell | P_\ell} \lfloor \frac{P_\ell}{\Lambda_\ell} \rfloor \times \prod\limits_{\ell'\in \mathcal{Q}_r:\ \Lambda_{\ell'} \nmid P_{\ell'}} \lceil \frac{P_{\ell'}}{\Lambda_{\ell'}} \rceil \, .
\end{align}

The above information will be essential in enumerating our equivalence classes and associating each such class with a collection of servers.

The next step is to fill the tiles in the product tensor, using the constitutive component columns and rows, as follows.
\paragraph{Step 2: Filling tiles in $\TensorE \times_1 \mathbf{D}$ as a function of $\TensorF$}
Recall that we have a tile $\TensorS_{\mathcal{P}}(\mathcal{R}_{\mathcal{P}}, \mathcal{C}_{\mathcal{P}})$ corresponding to the non-zero elements of $\TensorS_{\mathcal{P}}$. This tile is ``empty'' in the sense that all non-zero entries of $\TensorS_{\mathcal{P}}$ are equal to $1$.

To avoid assigning any entry of $\TensorF$ to more than one tile, we fix a strict total ascending order $\prec$ on the tiles $\{\mathcal{P}\}$, e.g., lexicographic on $(i,\j,\mathcal{Q}_r)$, and maintain a binary mask $\bar{\mathcal{M}}$ (same size as $\TensorF$), initialized to zero. We process tiles in the order $\prec$, and for each tile $\mathcal{P}$, we \emph{zero–force} the previously owned positions as
\begin{align}
\TensorS_{\mathcal{P}}^{\circ}\ \triangleq\ \TensorS_{\mathcal{P}}\odot(\bar{\bf 1}-\bar{\mathcal{M}})\, 
\end{align}
where $\bar{\bf 1}$ denotes an all-ones tensor of appropriate dimensions.

Consequently, all entries already assigned by earlier tiles are set to zero. If $\TensorS_{\mathcal{P}}^{\circ}=\bar{\bf 0}$, with $\bar{\bf 0}$ denoting an all-zeros tensor of appropriate dimensions, we skip $\mathcal{P}$. Otherwise, we define the induced index sets
\begin{align}
\mathcal{C}_{\mathcal{P}}^{\circ}\!&\triangleq\!\{k:\exists\,\pu\ \text{s.t.}\ (k,\pu)\!\in\!\Supp(\TensorS_{\mathcal{P}}^{\circ})\}\, , \, \forall\pu \in \prod_{\ell\in [L]} [P_\ell]\, ,\nonumber\\
\mathcal{R}_{\mathcal{P}}^{\circ}\!&\triangleq\!\{\pu:\exists\,k\ \text{s.t.}\ (k,\pu)\!\in\!\Supp(\TensorS_{\mathcal{P}}^{\circ})\}\, , \, \forall k\in [K] 
\end{align}
to form the cropped subtensor, factorized as
\begin{align}
\widehat{\TensorF}_{\mathcal{P}}\ \triangleq\ \big(\TensorF\odot \TensorS_{\mathcal{P}}^{\circ}\big)\big(\mathcal{R}_{\mathcal{P}}^{\circ},\,\mathcal{C}_{\mathcal{P}}^{\circ}\big)\, .
\end{align}
 
To decompose the high-dimensional tiles in this problem, we consider the multilinear SVD \ahmad{(MLSVD)} approach, detailed in Appendix~\ref{MLSVD}, which is based on the matrix SVD on mode-1 unfolding of $\Gamma$-dimensional tiles $\TensorS_\mathcal{P}$, denoted as $\TensorS_{\mathcal{P}_{(1)}}$.
We then use the rank properties of the matrix unfolding of a tensor, where we have the rank budget 
\begin{align}
r_{\mathcal{P}}^{\circ}=\rank(\TensorS^{\circ}_\mathcal{P})=\rank(\TensorS^{\circ}_{\mathcal{P}_{(1)}})=\min\big(|\mathcal{C}_{\mathcal{P}}^{\circ}|,\,|\mathcal{R}_{\mathcal{P}}^{\circ}|\big)\, .
\end{align}

The maximum rank of each representative rank-one support for such tiles, i.e., of each tile $\mathcal{P}$ of $\TensorE \times_1 \mathbf{D}$, from~\eqref{max-rank} and Definition~\ref{maxrank}, therefore takes the form
\begin{align} \label{max-rank-formula-P'}
    r_{\mathcal{P}}^{\circ}
    =
    \begin{cases}
    \min \big(\Delta, |\mathcal{R}^\circ_{\mathcal{P}}| \big)\, ,\qquad \mathcal{P}\in \mathcal{C}_1\bigcup\mathcal{C}_3\, ,\\
      \min \big(\mod(K,\Delta), |\mathcal{R}^\circ_{\mathcal{P}}| \big)\, ,\qquad  \mathcal{P}\in \mathcal{C}_2\bigcup\mathcal{C}_4
    \end{cases}
\end{align}
where
\begin{align}
|\mathcal{R}^\circ_{\mathcal{P}}|=
   \begin{cases}
       \prod\limits_{\ell\in \mathcal{Q}_r} \Lambda_\ell\, , & \mathcal{P}\in \mathcal{C}_1\bigcup\mathcal{C}_2\, ,\\
       \prod\limits_{\underset{\Lambda_{\ell} \vert P_{\ell}}{\ell\in \mathcal{Q}_r :}} \Lambda_{\ell} \prod\limits_{\underset{\Lambda_{\ell'} \nmid P_{\ell'}}{\ell'\in \mathcal{Q}_r :}} \mod(P_{\ell'},\Lambda_{\ell'})\, , & \mathcal{P}\in \mathcal{C}_3\bigcup\mathcal{C}_4\, .
   \end{cases} 
\end{align}

We then compute the \ahmad{MLSVD} of $\widehat{\TensorF}_{\mathcal{P}}$ as
\begin{align}
\widehat{\TensorF}_{\mathcal{P}} \;=\; \TensorR_{\mathcal{P}} \times_{1} \mathbf{L}_{\mathcal{P}}
\end{align}
with $\mathbf{L}_{\mathcal{P}}\in\mathbb{R}^{|\mathcal{C}_{\mathcal{P}}^{\circ}|\times r_{\mathcal{P}}^{\circ}}$ and $\TensorR_{\mathcal{P}}\in\mathbb{R}^{r_{\mathcal{P}}^{\circ}\times |\mathcal{R}_{\mathcal{P}}^{\circ}(1)|\times\cdots\times|\mathcal{R}_{\mathcal{P}}^{\circ}(L)|}$. The columns of $\mathbf{D}$ reserved for tile $\mathcal{P}$ (one column per rank-$1$ component) are filled with $\mathbf{L}_{\mathcal{P}}$ on the rows indexed by $\mathcal{C}_{\mathcal{P}}^{\circ}$ and zeros elsewhere. The matching frontal slices of $\bar{\mathcal{E}}$ are filled with $\mathcal{R}^{\circ}_{\mathcal{P}}$ on the indices $\mathcal{R}_{\mathcal{P}}$ and zeros elsewhere. Finally, we update the mask by
\begin{align}
\bar{\mathcal{M}}\ \leftarrow\ \bar{\mathcal{M}}\ \vee\ \Supp(\TensorS_{\mathcal{P}}^{\circ})\, .
\end{align}

By construction, the owned supports $\{\Supp(\TensorS_{\mathcal{P}}^{\circ})\}_{\mathcal{P}}$ are pairwise disjoint, hence no entry of $\TensorF$ is assigned twice. Therefore, the later tiles automatically see zeros on intersections. 

In the third step, we create and fill the non-zero tiles in the decoding matrix $\mathbf{D}$ and the encoding tensor $\TensorE$, as follows.
\paragraph{Step 3: Creating and filling the non-zero tiles in  $\mathbf{D}$ and  $\TensorE$}
This step starts with MLSVD (as described in Appendix~\ref{MLSVD}) of the cropped tile $\TensorF_\mathcal{P}$, where this decomposition takes the form
\begin{align}
      \TensorF_{\mathcal{P}} = \TensorE_{\mathcal{P}}\times_1\mathbf{D}_{\mathcal{P}}\label{sub-SVD-o}
  \end{align}
  where $\mathbf{D}_{\mathcal{P}} \in \mathbb{R}^{|\mathcal{C}^{\circ}_{\mathcal{P}}| \times r^{\circ}_{\mathcal{P}}}$ and $\TensorE_{\mathcal{P}} \in \mathbb{R}^{r^{\circ}_{\mathcal{P}} \times |\mathcal{R}^{\circ}_{\mathcal{P}}|} $. In particular, $\TensorF_{\mathcal{P}}, \mathbf{D}_{\mathcal{P}}$, and $\TensorE_{\mathcal{P}}$ are  associated to  $\TensorT$, $ \mathbf{U}^{(1)}$, and $\TensorS$ in all complete SVD decomposition of \eqref{Tensor_SVD}. Naturally, $\rank(\TensorF_{\mathcal{P}}) \leq r^{\circ}_{\mathcal{P}}$.

Let 
\begin{align}\label{eq:tileEnumeration}
\bigcup^{4}_{i=1} \mathcal{C}_i\triangleq \{ \mathcal{P}_1, \mathcal{P}_2, \hdots, \mathcal{P}_{m}\}, \: m \in \mathbb{N}
\end{align}
describe the enumeration we give to each tile. Then, the position that each cropped tile takes inside $\mathbf{D}$, is given by 
\begin{align}\label{eq:tilePositionInD_T1}
\mathcal{C}^{\circ}_{\mathcal{P}_j},\;
\Big[\, \sum_{i=1}^{j-1} r^{\circ}_{\mathcal{P}_i} + 1 \,:\, \sum_{i=1}^{j} r^{\circ}_{\mathcal{P}_i} \,\Big],
\quad \forall\, \mathcal{P}_j \in \bigcup_{s=1}^{4} \mathcal{C}_s
\end{align}
and the position of each cropped tile in $\TensorE$ is given by 
\begin{align}\label{eq:tilePositionInE_T1}
\Big[\, \sum_{i=1}^{j-1} r^{\circ}_{\mathcal{P}_i} + 1 \,:\, \sum_{i=1}^{j} r^{\circ}_{\mathcal{P}_i} \,\Big],\;
\mathcal{R}^{\circ}_{\mathcal{P}_j},
\quad \forall\, \mathcal{P}_j \in \bigcup_{s=1}^{4} \mathcal{C}_s\, .
\end{align}

In particular, these yield
\begin{align}
    \mathbf{D}\left(\mathcal{C}^{\circ}_{\mathcal{P}_j}, \Big[\sum^{j-1}_{i=1} r^{\circ}_{\mathcal{P}_i} +1,\sum^{j}_{i=1} r^{\circ}_{\mathcal{P}_i} \Big]\right) = \mathbf{D}_{\mathcal{P}_j}
    \label{Decoding-Encoding1-o}
\end{align}
and 
\begin{align}
    \TensorE\left(\Big[\sum^{j-1}_{i=1} r^{\circ}_{\mathcal{P}_i} +1,\sum^{j}_{i=1} r^{\circ}_{\mathcal{P}_i} \Big], \mathcal{R}^{\circ}_{\mathcal{P}_j}\right)=\TensorE_{\mathcal{P}_j}\label{Decoding-Encoding2}
\end{align}
while naturally the remaining non-assigned elements of $\mathbf{D}$ and $\TensorE$ are zero.

We are now ready to upper bound $N$.

\paragraph{Step 4: An upper bound to the number of servers}
We now proceed to bound the number of rank-one contribution supports (corresponding to the number of servers), and we do so under our previously stated assumption of disjoint supports (cf. Definition~\ref{tensor-disjointSupportAssumption}), which is equivalent to the disjoint-tiles assumption via the following lemma.
 \begin{lem} \label{Disjoint-Support-Assumption: Equivalence}
For matrix $\mathbf{D}$ and tensor $\TensorE$, the representative supports $\{\TensorS_{\mathcal{P}_i}\}_{i=1}^m$ of $\TensorE\times_1\mathbf{D}$ are disjoint (i.e., $\TensorS_{\mathcal{P}_i} \cap \TensorS_{\mathcal{P}_j} = \bar{\bf 0}, \ j\neq i $) if and only if $\mathbf{D}$ and $\TensorE$ accept the \emph{disjoint support assumption}. 
 \end{lem}

\begin{proof}[Proof of Lemma~\ref{Disjoint-Support-Assumption: Equivalence} ]
\begin{itemize}
    \item Disjoint support assumption $\implies$ disjoint representative supports:
    Assume that $\mathbf{D}\in \mathbb{R}^{K\times N}$ and 
$\TensorE \in \mathbb{R}^{N\times P_1\times \hdots \times P_{L}}$
satisfy the \emph{disjoint support assumption} of Definition~\ref{tensor-disjointSupportAssumption}. 
Then, for any $i,i'\in [N]$, either the supports coincide, i.e.,
\[
\supp(\mathbf{D}(:,i)) = \supp(\mathbf{D}(:,i'))\ \ \text{and} \ \ 
\supp(\TensorE(i,:,\hdots,:)) = \supp(\TensorE(i',:,\hdots,:))
\]
or they are disjoint.
Consider now the induced supports of the mode-1 product
$\TensorE \times_1 \mathbf{D}$.  
By construction, 
either 
\begin{align}\label{eq:lem_DSP_1}
    \supp(\TensorE(i,:,\hdots,:)\times_{1} \mathbf{D}(:,i))=
\supp(\TensorE(i',:,\hdots,:)\times_{1} \mathbf{D}(:,i'))
\end{align}
or the two supports are disjoint.
Therefore, the representative supports 
$\{\TensorS_{\mathcal{P}_i}\}_{i=1}^m$ are pairwise disjoint,
which yields the disjoint representative support equivalence classes
in Definition~\ref{Def2}.
\item Disjoint representative supports $\implies$ disjoint support assumption:
Now assume that $\TensorE\times_1\mathbf{D}$ satisfies the disjoint representative support assumption.  
Let $\mathcal{C}=\{\mathcal{P}_1,\ldots,\mathcal{P}_m\}$ denote the induced equivalence classes.
By the definition of these classes, for any two representatives
$\mathcal{P},\mathcal{P}'\in\mathcal{C}$, their
supports are either identical or disjoint.
This property propagates back to the column supports of $\mathbf{D}$ 
and the row supports of subtensors of $\TensorE$. 
Hence, for all $i,i'\in[N]$, we have that
\begin{align}\label{eq:lem_DSP_21}
    \supp(\mathbf{D}(:,i)) = \supp(\mathbf{D}(:,i'))
\ \ \text{or} \ \
\supp(\mathbf{D}(:,i)) \cap \supp(\mathbf{D}(:,i'))=\emptyset\, ,
\end{align}
and similarly,
\begin{align}\label{eq:lem_DSP_22}
\supp(\TensorE(i,:,\hdots,:)) 
=
\supp(\TensorE(i',:,\hdots,:))
\ \ \text{or} \ \
\supp(\TensorE(i,:,\hdots,:)) 
\cap 
\supp(\TensorE(i',:,\hdots,:))
=\emptyset\, .
\end{align}
Combining these observations in~\ahmad{\eqref{eq:lem_DSP_1}, \eqref{eq:lem_DSP_21}, and \eqref{eq:lem_DSP_22}} shows that the induced supports of the mode-1 products, described as
$\supp(\TensorE(i,:,\hdots,:)\times_1 \mathbf{D}(:,i))$,
are either identical or disjoint for any distinct pair $i,i'$, 
which is precisely the disjoint support assumption of 
Definition~\ref{tensor-disjointSupportAssumption}.\\
This completes the proof of Lemma~\ref{Disjoint-Support-Assumption: Equivalence}.\qedhere
\end{itemize}
\end{proof}

Completing the proof of Lemma~\ref{Disjoint-Support-Assumption: Equivalence}, we \ahmad{next} use the equivalence of the disjoint support assumption and the disjoint representative supports to finalize the proof of Theorem~\ref{Achievability-Converse}. To that end, we recall
from Section~\ref{Formulating} (see also~\eqref{EncodingMatrix}--\eqref{DecodingMatrix}) that each server $n\in[N]$ corresponds to one column of the decoding matrix $\mathbf{D}$ and the associated row of the encoding tensor $\TensorE$. To express $N$ in terms of the scheme parameters, consider the tiles $\mathcal{P}\in\mathcal{C}$ and the associated submatrices $\mathbf{D}_{\mathcal{P}}$ defined in~\eqref{sub-SVD-o}. From~\eqref{Decoding-Encoding1-o}, these submatrices collectively span $\sum_{i=1}^{m} r^\circ_{\mathcal{P}_i}$ columns. 

The total number of required servers is then obtained as $N\leq\sum_{\mathcal{P}} r_{\mathcal{P}}^{\circ}$ while the design constraints $(\Gamma,\Delta,\{\Lambda_{\ell}\}_{\ell\in [L]})$ are satisfied at each server. Finally, as one can readily verify, the above design corresponds to 
\begin{align}\label{result-2-f}
    N_{} &\leq \sum_{s \in [4]} \sum_{\mathcal{P} \in \mathcal{C}_s} r^{\circ}_{\mathcal{P}}\\ 
     &=\lfloor \frac{K}{\Delta} \rfloor \sum\limits_{r=1}^{L\choose{\Gamma}} \min \Big(\Delta, \prod\limits_{\ell\in \mathcal{Q}_r} \Lambda_\ell\Big)\times \prod\limits_{\ell\in \mathcal{Q}_r} \lfloor \frac{P_\ell}{\Lambda_\ell} \rfloor \nonumber\\
     &+\sum\limits_{r=1}^{L\choose{\Gamma}} \min \Big(\mod(K,\Delta), \prod\limits_{\ell\in \mathcal{Q}_r} \Lambda_\ell \Big)\times \prod\limits_{\ell\in \mathcal{Q}_r} \lfloor \frac{P_\ell}{\Lambda_\ell} \rfloor \nonumber\\
     &+\lfloor \frac{K}{\Delta} \rfloor \sum\limits_{r=1}^{L\choose{\Gamma}} \min \Big( \underset{\ell, \ell' \in \mathcal{Q}_r :\ \Lambda_{\ell} \vert P_{\ell},\ \Lambda_{\ell'} \nmid P_{\ell'}}{\Delta, \prod_{\ell} \Lambda_{\ell} \prod_{\ell'} \mod(P_{\ell'},\Lambda_{\ell'})} \Big) \times\prod\limits_{\ell\in \mathcal{Q}_r:\ \Lambda_\ell | P_\ell} \lfloor \frac{P_\ell}{\Lambda_\ell} \rfloor \times \prod\limits_{\ell'\in \mathcal{Q}_r:\ \Lambda_{\ell'} \nmid P_{\ell'}} \lceil \frac{P_{\ell'}}{\Lambda_{\ell'}} \rceil \nonumber\\
     &+\sum\limits_{r=1}^{L\choose{\Gamma}} \min \Big( \underset{\hspace{1.4cm}\ell, \ell' \in \mathcal{Q}_r :\ \Lambda_{\ell} \vert P_{\ell},\ \Lambda_{\ell'} \nmid P_{\ell'}}{\mod(K,\Delta), \prod_{\ell} \Lambda_{\ell} \prod_{\ell'} \mod(P_{\ell'},\Lambda_{\ell'})} \Big)
     \times\prod\limits_{\ell\in \mathcal{Q}_r:\ \Lambda_\ell | P_\ell} \lfloor \frac{P_\ell}{\Lambda_\ell} \rfloor \times \prod\limits_{\ell'\in \mathcal{Q}_r:\ \Lambda_{\ell'} \nmid P_{\ell'}} \lceil \frac{P_{\ell'}}{\Lambda_{\ell'}} \rceil \nonumber\, .
\end{align}

Letting $\Delta\mid K$, $\Lambda_\ell=\Lambda,\, P_\ell=P,\,\ell\in [L]$, and $\Lambda \mid P$, the upper bound on $N$ in \eqref{result-2-f} is simplified to \eqref{achiv-o},
\ahmad{which} completes the proof of Theorem~\ref{Achievability-Converse}.
\end{proof}

\ahmad{
We next present basic examples of multi-user non-linearly decomposable problems and illustrate the interplay among system parameters based on Theorem~\ref{Achievability-Converse}.
}

\begin{figure*}[t!]
    \centering
    \includegraphics[width=0.85\textwidth]{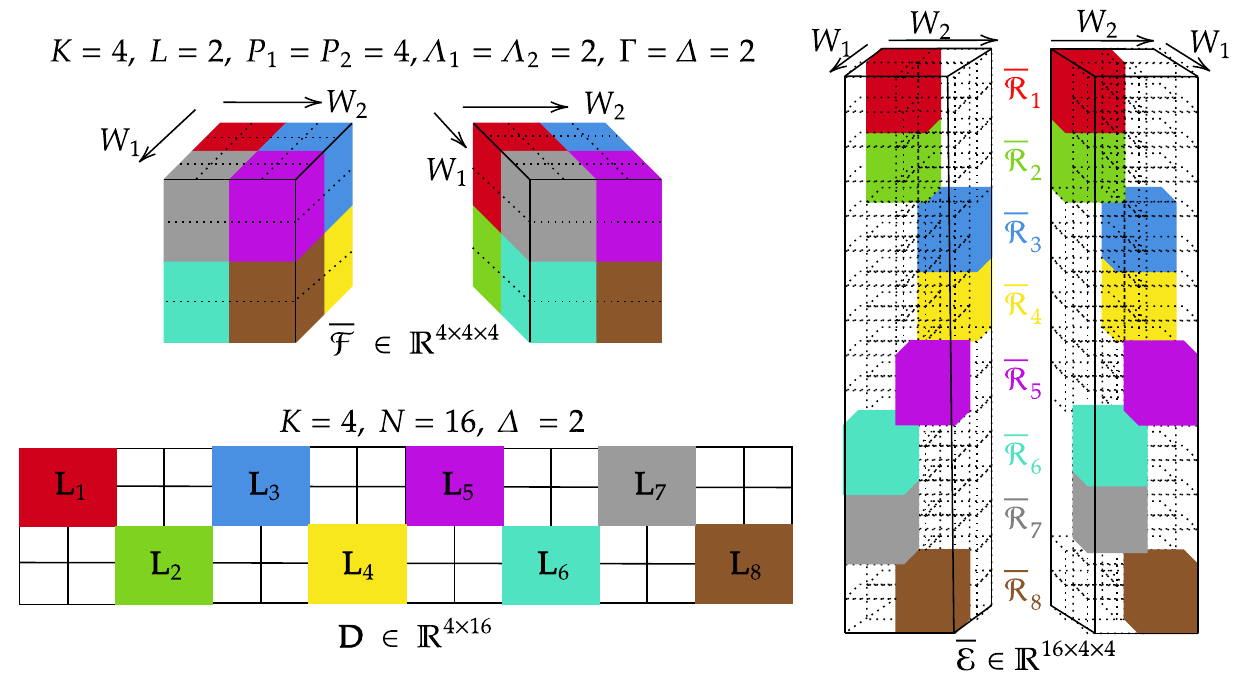}
    \caption{Corresponding to Example~\ref{single-shot-example-simple} \ahmad{(Case $\mathrm{I}$)}, this figure illustrates the partitioning of $\mathcal{\bar F}$ into $8$ tiles of size $ (2\times 2\times 2)$, and the sparse tiling of $\mathbf{D}$ and $\mathcal{\bar E}$ with tiles $\mathbf{L}_j$ and $\bar{\mathcal{R}}_j$, respectively, resulting in the full tiling of $\TensorF = \bar{\mathcal{E}} \times_{1} \mathbf{D}$.
    }
    \label{ex1}
\end{figure*}

\begin{ex}\label{single-shot-example-simple}
Consider the $(K=4,N,L=2,\Gamma=2, \{P_{\ell}=4,\Lambda_{\ell}=2\}_{\ell\in [L]})$ setting, with $N$ servers tasked with computing non-linear functions of $L = 2$ basis subfunctions 
\begin{align} 
\label{eq:example_4_users_4_exponents}
    F_{k}(.) = \sum\limits_{\substack{(p_1,p_2) \in [4]\times[4]}} 
    c_{k,\pu}\ W_1^{p_1-1} W_2^{p_2-1},\quad k\in[4]\, .
\end{align}

The coefficients $c_{k,\pu}$ are described by tensor $\TensorF$, as demonstrated in Figure~\ref{ex1}. 
Given the design constraints, we aim to guarantee lossless reconstruction of (\ref{eq:example_4_users_4_exponents}). 

\ahmad{
\textbf{Case $\mathrm{I}$: $\Delta=2$ .}
}
we directly conclude from~\eqref{achiv-o} that we need $N \leq 16$ servers. To tackle this challenge of reconstruction, we need to construct 
\begin{enumerate}
    \item The $(N\times P_1\times P_2) = (16\times 4\times 4)$ computing tensor $\mathcal{\bar{E}}$, specifying the computational tasks of each server.
    \item The $(K \times N) = (4 \times 16)$ communication matrix $\mathbf{D}$, determining the server-user connections.
\end{enumerate}


These originate from the decomposition of $(K\times P_1\times P_2 )=(4\times 4\times 4)$ tensor $\TensorF$ (cf.~\eqref{Demand-Tensor-1}) as $\TensorF = \bar{\mathcal{E}}\times_{1} \mathbf{D}$, representing the requested functions. The solution is then as follows.
\begin{enumerate}
    \item Initially, we partition $\mathcal{\bar F}$ into $\frac{K}{\Delta} \frac{P_1}{\Lambda_1} \frac{P_2}{\Lambda_2} 
    = 2\cdot 2\cdot 2 = 8$ disjoint $\Delta \times \Lambda_1 \times \Lambda_2$ subtensors 
    $\bar{\mathcal{S}}_j\in 
    \mathbb{R}^{2 \times 2 \times 2},\, j\in [8]$, as depicted in Figure~\ref{ex1}. 
    
    \item  Next, using the standard tensor decomposition form (cf. Appendix~\ref{MLSVD}),
    we decompose each $\bar{\mathcal{S}}_j$ as $\bar{\mathcal{S}}_j =\bar{\mathcal{R}}_j \times_1 \mathbf{L}_j$, where $\bar{\mathcal{R}}_j \in \mathbb{R}^{2\times 2\times 2}, \mathbf{L}_j \in \mathbb{R}^{2\times 2}$ for all $j \in [8]$, noting that such full decomposition is feasible since the maximum rank of each $\bar{\mathcal{S}}_j$ is $\min(\Delta, \Lambda_1\Lambda_2)=2$\,. 
    
    \item Finally, we construct $\mathbf{D} \in \mathbb{R}^{4\times 16}$ and $\mathcal{\bar E}\in \mathbb{R}^{16\times 4\times 4}$ by tiling them with $\mathbf{L}_j$ and $\bar{\mathcal{R}}_j$, respectively, as in Figure~\ref{ex1}.  
    \end{enumerate}
\end{ex}

\begin{figure}[t!]
\centering
\includegraphics[scale=0.7]{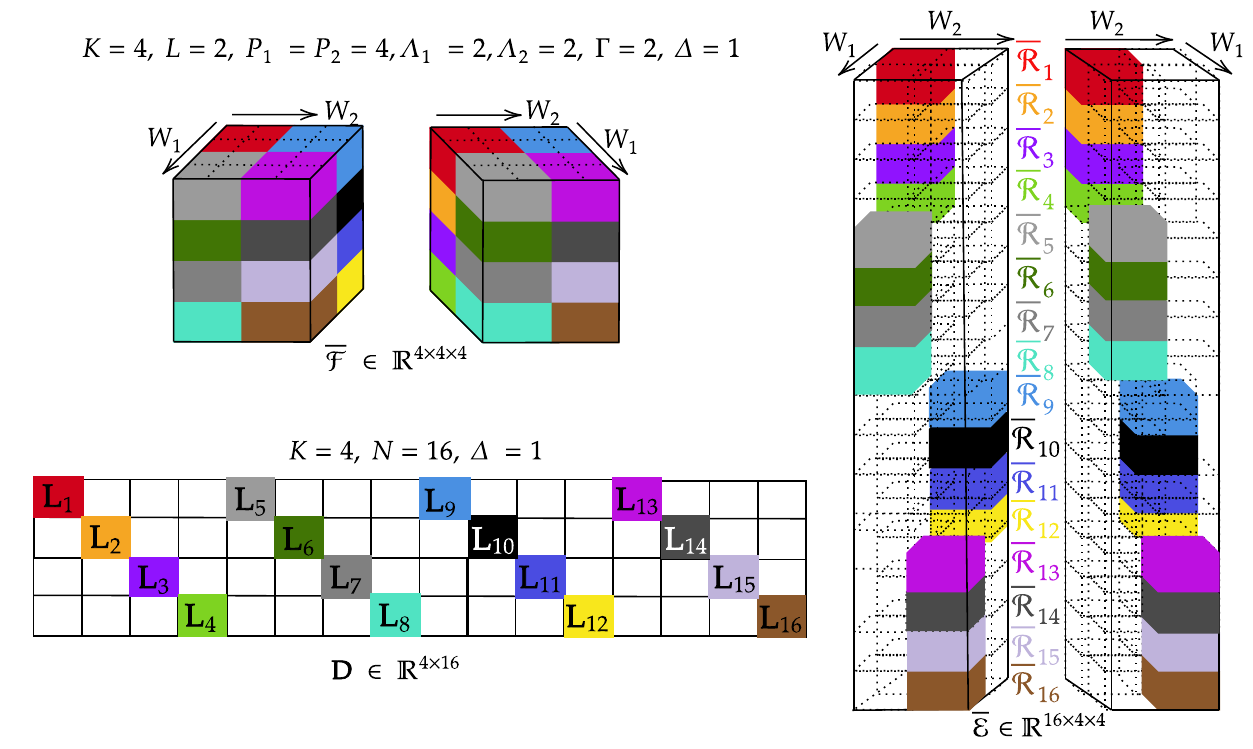}
\caption{Pertaining to Example~\ref{single-shot-example-simple} \ahmad{(Case $\mathrm{II}$)} with 
$N=16$ servers, the new tessellation pattern allows for a reduced $\Delta =1$ reflecting a reduction from $\delta = 1/2$ \ahmad{in Case $\mathrm{I}$} to $\delta = 1/4$.}
\label{ADD1}
\end{figure}

Example~\ref{single-shot-example-simple} provides a glimpse 
of the general principle behind creating our 
scheme. In brief, 
we begin by splitting our $K\times P_1\times P_2$ tensor $\TensorF$ into $\frac{K}{\Delta}\frac{P_1}{\Lambda_1}\frac{P_2}{\Lambda_2}$ subtensors of size $\Delta \times \Lambda_1 \times \Lambda_2$. We decompose these subtensors 
into submatrices $\{\mathbf{L}_j\}_{j\in[8]}$ and into subtensors $\{\bar{\mathcal{R}}_j\}_{j \in [8]}$ that form tiles of $\mathbf{D}$ and $\TensorE$, respectively. The tile placement must respect the sparsity constraints ($\Gamma,\Delta, \Lambda_1, \Lambda_2$) and must yield $\TensorE \times_1 \mathbf{D} = \TensorF$. 
Regarding the required number of servers, the general rule 
is that $N$ is the number of subtensors, multiplied by the rank of each subtensor. 
Since we had $8$ subtensors, each of rank $2$, we need $N=16$ servers.
In contrast, it is easy to show that the linearized approach, employing sparse matrix factorization of~\cite{Khalesi2025Tessellated}, would entail $L_M=\prod_{\ell\in [L]} P_\ell = 16$ basis subfunctions and approximately double the number of servers, for the same $\Gamma$ and $\Delta$.

At the same time, some reductions in $\gamma,\delta$ may \ahmad{arise naturally}. 
\ahmad{In particular, multiple tessellation patterns can yield} the same required value of $N$, 
\ahmad{but variations in tile size and placement may result in different values of} $\Gamma$ and $\Delta$. To illustrate \ahmad{this}, consider the following case.

\ahmad{
\textbf{Case $\mathrm{II}$: $\Delta=1$ .}
In the same lossless computing setting, 
we consider again $N=16$ servers, $K = 4$ users, $L = 2$ subfunctions.
We also maintain a computational cost of $\Gamma = 2$, but now we see that the tessellation pattern in Figure~\ref{ADD1} allows for a reduced $\Delta=1$ corresponding to $\delta = 1/4$. 
}

\begin{figure}[t!]
    \centering
    \includegraphics[scale=0.7]{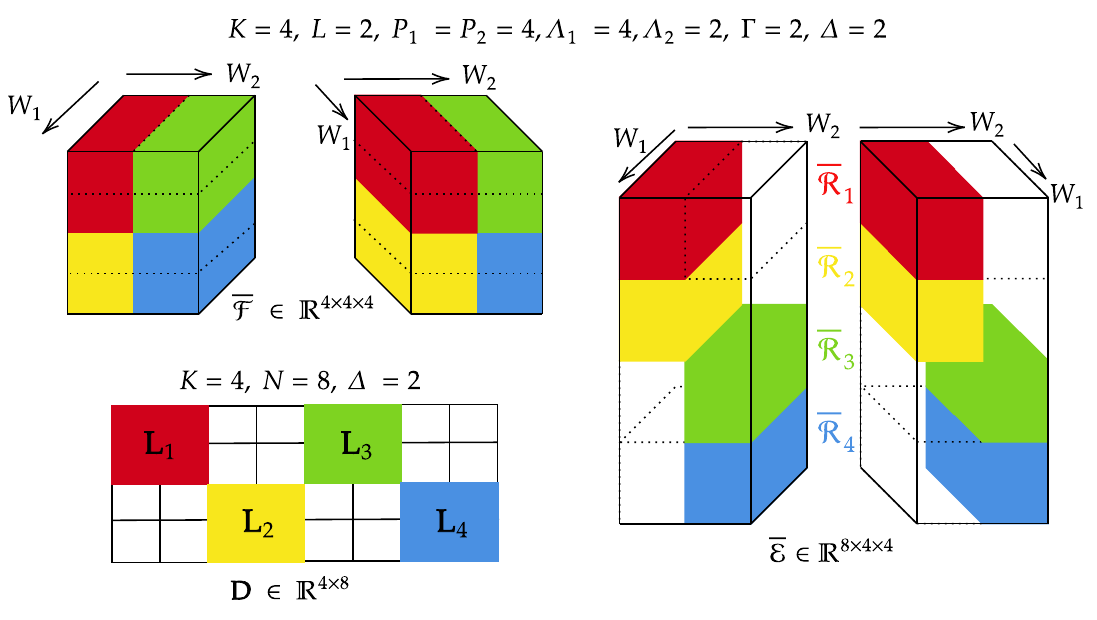}
    \caption{Corresponding to Example~\ref{single-shot-example-simple} with $\Lambda_1 = 4$,
    this figure illustrates the partitioning of $\mathcal{\bar F}$ into 4 tiles of size $ (\Delta \times \Lambda_1\times \Lambda_2) = (2\times 4\times 2)$, and also illustrates the sparse tiling of $\mathbf{D}$ and $\mathcal{\bar E}$ with tiles $\mathbf{L}_j$ and $\bar{\mathcal{R}}_j$ respectively, resulting in the full tiling of $\TensorF = \bar{\mathcal{E}} \times_{1} \mathbf{D}$.}
    \label{ex2}
\end{figure}

On the other hand, once we reduce the computation and communication capabilities of each server, more servers may be required. Considering different multiplicative terms for the subfunctions could also significantly change the number of required servers.
\ahmad{
For example, for the same setting as in Example~\ref{single-shot-example-simple}, again for $K = 4,L = 2, P_1=P_2=4$, and again for $\Delta = 2$ (corresponding to $\delta = \frac{1}{2}$) and $\Gamma =2$ (corresponding to $\gamma = \frac{1}{2}$), if we aim to evaluate different degrees of subfunctions corresponding to $\Lambda_1=4, \Lambda_2=2$, then the required number of servers $N$ is now $8$ (cf.~\eqref{achiv-o}), 
and the corresponding tessellation pattern is presented in Figure~\ref{ex2}, where we see $4$ subtensors of rank $2$.  
}



We next \ahmad{show} in Proposition~\ref{comparison1} that, in the matrix factorization approach of TDC \cite{Khalesi2025Tessellated}, \ahmad{using} a potentially large dimension $L$ to accommodate
\ahmad{more} basis subfunctions
\ahmad{increases the}
total computational load per server. In contrast, our proposed model 
\ahmad{improves computational efficiency by using}
fewer basis subfunctions, each with a bounded range of degrees, captured by the multiplication cost. Our comparison assumes that the time to compute a basis subfunction is the same as the time to compute one power of its output.
\begin{prop}\label{comparison1}
\ahmad{
Consider a lossless $\big(K,N,L,\Gamma,\Delta, \{P_\ell, \Lambda_\ell\}_{\ell\in [L]}\big)$ distributed computing system with $P_\ell=P$ and $\Lambda_\ell=\Lambda$ for all $\ell\in L$, where $P>2$. Under the tensor factorization approach proposed in this work, with computation cost $\Gamma_T = L$ and multiplication cost $\Lambda$, the total computation cost is upper bounded by
\begin{align}
NL(\Lambda+1)\, ,
\end{align}
which scales linearly with $L$. In contrast, under the matrix factorization approach in~\cite{Khalesi2025Tessellated}, with per-server computation cost $\Gamma_M$, the total computation cost is upper bounded by
\begin{align}
N \left( \Gamma_M + \left(\tfrac{P(P-1)}{2}\right)^L \right),
\end{align}
which grows exponentially with $L$.
}
\end{prop}
\begin{proof}
The overall computational cost for the non-linearly separable distributed computing setting proposed in this work can be \ahmad{upper bounded} as
\[
N \big(\Gamma_T + \Lambda_1 + \hdots + \Lambda_L\big)\\
\overset{(a)}{\le} N \big(L(\Lambda+1)\big)
\]
because we have $N$ servers each computing a basis subfunction with a computation cost $\Gamma_T$ as well as the powers of all $L$ basis subfunctions each with a multiplication cost $\Lambda_\ell$ and $(a)$ holds because of the assumptions $\Gamma_T=L$ and $\Lambda_\ell=\Lambda,\, \ell\in [L]$.  

The total computation cost of the matrix factorization approach in \cite{Khalesi2025Tessellated} is \ahmad{upper bounded} as
\begin{align*}
    N \big(\Gamma_M + \sum\limits_{\underline{\mathbf{p}}\in [2:P_1]\times\hdots\times[2:P_L]} p_1\times\hdots\times p_L \big) &\overset{(b)}{=} N \big(\Gamma_M + \sum\limits_{p_1\in [2:P]} p_1 \times \hdots \times \sum\limits_{p_L\in [2:P]} p_L\big)\\
    &\overset{(c)}{=} N \Big(\Gamma_M + \big(\frac{P(P-1)}{2}\big)^L\Big)
\end{align*}
because we have $N$ servers each computing a basis subfunction with a computation cost $\Gamma_M$ as well as the number of multiplications for powers of each basis subfunction $\ell\in [L]$, ranging from $2$ to $P_\ell$, $(b)$ follows from some algebraic manipulations, 
and $(c)$ follows from $1+\hdots+n=\frac{n(n+1)}{2}$, assuming $n=P-1$.
\end{proof}

In the multi-user non-linearly separable distributed computing setting, the coordinator node must split the global computational demand tensor into smaller tensor blocks (subtensors) that can be assigned to different servers.  
These tensor partitions arise naturally from the need to distribute the computation across the servers while respecting the support constraints imposed by the exponent vectors of the requested basis subfunctions.  
However, if the partitioning is not carefully designed, the resulting subtensors may inherit large combinatorial overlaps across multiple dimensions, which in turn leads to substantial rank inflation and consequently requires more servers to compute all subtasks without error.  
Therefore, designing tensor partitions that minimize the resulting \emph{sum rank} is central, as the rank of each subtensor directly determines the computation and communication loads \ahmad{at} each server. We will address this by introducing a novel combinatorial reduction algorithm based on matching in a bipartite graph.

Next, we address the overlap of the induced subtensors (tile closures), which leads to ambiguity in the assignment of admissible index tuples and may result in inefficient designs. 
To resolve this issue, we introduce a capacitated tuple-to-tile assignment algorithm that enforces a disjoint decomposition by assigning each tuple to exactly one feasible tile. This construction yields a valid achievable scheme and improves the achievable rate of the proposed approach in Theorem~\ref{Achievability-Converse}.

\subsection{\ahmad{Refined System Rate via a Combinatorial Reduction Algorithm}}\label{subsec:refined_rate}
We here present a tighter achievable rate result for the proposed lossless distributed computing setting of $(K,N,L,\Gamma,\Delta, \{P_{\ell},\Lambda_{\ell}\}_{\ell\in [L]})$, 
\ahmad{focusing on the simplified case where}
$\delta, \lambda_\ell \in \mathbb{N}$.
\ahmad{In particular, we upper bound}
the required number of servers $N$, which is the common dimension between $\TensorE,\, \mathbf{D}$, \ahmad{and directly impacts} our sparse tensor factorization approach. 
Particularly, we consider a mode-1 matrix unfolding
of tensor $\TensorF$ and use its rank properties to elaborate the number of required servers as the sum of the total number of tiles with different maximum representative rank-one supports. 
Since tiles may have combinatorial intersections, we introduce a combinatorial reduction approach based on a bipartite graph matching and a recursive assignment 
\ahmad{maps each intersecting index tuple to a tile cell.}
This procedure results in tiles with a lower sum rank compared to the tensor factorization approach~(cf. Theorem \ref{Achievability-Converse}).

\begin{theo}[Achievable scheme via capacitated tile assignment]\label{Achievability-new}
Consider the lossless distributed computing system with parameters $(K,N,L,\Gamma,\Delta, \{P_{\ell},\Lambda_{\ell}\}_{\ell\in [L]})$ under the homogeneous setting $\{P_{\ell}=P,\Lambda_{\ell}=\Lambda\}_{\ell\in [L]}$ and $(\Delta \mid K,\, \Lambda \mid P)$.
\ahmad{
Then, there exists a feasible assignment of admissible tuples to tiles, constructed as described in Algorithm~\ref{Alg:GreedyAssignment}, which achieves 
rate $R = K/N$, where the required number of servers satisfies}
\begin{align}\label{achiv}
N
\;\le\;
\frac{K}{\Delta}
\sum_{\beta\in [n_{\mathcal{P}}]}
\min \big(\Delta,\lvert \mathcal{R}^*_{\mathcal{P}_\beta} \rvert\big)
\end{align}
where $\mathcal{R}^*_{\mathcal{P}_\beta}$ denotes the assigned \ahmad{disjoint} row-index set of tile $\mathcal{P}_\beta$ induced by the output of Algorithm~\ref{Alg:GreedyAssignment}.
\end{theo}

\begin{proof}
The proof follows from the same parts and steps as the proof of Theorem \ref{Achievability-Converse}.
However, we have further details in {Part II (Construction of $\mathbf{D},\, \TensorE$)}, which we will explain here. 

We next detail the four-step procedure for deriving $\mathbf{D}$ and $\TensorE$.
\paragraph{Step 1: Designing the sizes of the tiles for the $K\times P_1\times \hdots \times P_L$ product tensor $\TensorE\times_1 \mathbf{D}$}

The family of all tiles is initially specified by the same set of equivalence classes $\mathcal{C}$ as in~\eqref{eq:eq-classes}.

Despite having $L$ basis subfunctions and therefore a $L$-dimensional space for each user's demand, indexed by the tuple $\pu$, the computation cost $\Gamma$ restricts each server to compute at most $\Gamma$ basis subfunctions and their multiplicative combinations. This induces $(\Gamma+1)$-dimensional tiles, denoted by $\mathcal{P}_{i,\j,r}$, corresponding to $\Gamma$ active exponent dimensions and one user dimension.

Each tile $\mathcal{P}_{i,\j,r}$ is geometrically composed of $\Lambda^\Gamma$ unit cells, corresponding to exponent tuples with exactly $\Gamma$ active coordinates. However, for the purpose of assignment, each tile is associated with a larger \emph{row-index set} (closure), denoted by $\mathcal{R}_{\mathcal{P}_{i,\j,r}}$, which includes all tuples whose active support is contained in the tile support and whose active values lie within the selected windows.

The row-index set of each initial tile has a cardinality $\big|\mathcal{R}_{\mathcal{P}_{i,\j,r}}\big| = \prod\limits_{\ell\in\mathcal{Q}_r}\Lambda_\ell$, describing the number of unit cells of size $1\times 1$ in each initial tile. For simplicity, we assume $\{\Lambda_\ell=\Lambda, P_\ell=P\}_{\ell\in [L]}$ and $\Lambda\mid P$, resulting in $\big|\mathcal{R}_{\mathcal{P}_{i,\j,r}}\big|=\Lambda^\Gamma$.
\ahmad{
The total number of initial tiles of various shapes and sizes, whose construction is detailed in the proof of Theorem~\ref{Achievability-Converse}, Part II, a) Step 1, corresponding to each user demand~\footnote{Considering all $K$ user demands, $\frac{K}{\Delta} n_{\mathcal{P}}\min(\Delta,\Lambda^\Gamma)$ recovers the result in Theorem~\ref{Achievability-Converse}.}
within the set of equivalence classes $\mathcal{C}_1$, is given by
}
\begin{align}\label{eq:n_tile}
    n_{\mathcal{P}}\triangleq \binom{L}{\Gamma} \big(\frac{P}{\Lambda}\big)^\Gamma
\end{align}
\ahmad{which} follows from the fact that each tile $\mathcal{P}_{i,\j,r}$ is indexed by $(i,\j,r)$, where $i$ labels the column index, $\j$ denotes the $L$-dimensional tuple of row index set, and $\mathcal{Q}_r$ specifies the $r^{th}$ tile pattern or configuration of active dimensions. 



\ahmad{
Tile closures $\mathcal{R}_{\mathcal{P}_{i,\j,r}}$ may overlap due to two distinct origins, detailed below.
\begin{enumerate}
    \item Combinatorial overlaps: different $\Gamma$-subsets $\mathcal Q_r \subseteq [L]$ may contain the same lower-dimensional support, resulting in combinatorial overlaps across different support classes $\mathcal Q_r$,
    \item Geometric overlaps: within a fixed support class, different window selections may also contain the same tuple due to inactive coordinates taking value $1$.
\end{enumerate}
}


Consequently, a given tuple $\pu$ may belong to multiple row-index sets $\mathcal{R}_{\mathcal{P}_{i,\j,r}}$, and thus must be assigned uniquely to one tile to ensure a disjoint support structure for the MLSVD decomposition.
Because of these two forms of overlap, naive layered counting leads to overcounting. The feasibility graph and the induced unique assignment are therefore essential for obtaining a valid disjoint tile decomposition.

The maximum rank of each representative support, based on the matrix unfolding of the tiles, from~\eqref{max-rank} and Definition~\ref{maxrank} (assuming $\Delta \mid K$), is given by
\begin{align}\label{rank-class}
    r_{\mathcal{P}}=\min\,(\Delta, |\mathcal{R}_{\mathcal{P}_{i,\j,r}}|)\, .
\end{align}

We aim to assign the $L$-dimensional index tuples to the $\Gamma$-dimensional tiles in each user's demand tensor through a combinatorial reduction. The resulting assigned row-index sets are denoted by $\mathcal{R}^*_{\mathcal{P}_{i,\j,r}}$, and the induced number of required servers $N$ is given by the corresponding sum rank of the resulting subtensors.

To obtain an upper bound on $N$, we construct feasible tessellation patterns by formulating the assignment as a combinatorial problem based on bipartite \ahmad{graph} matching with capacities.

\ahmad{We next introduce the set of admissible index tuples, which is required in the bipartite graph construction in Definition~\ref{defi:feasibility_graph}.}

Let $\mathcal{U}$ denote the set of all admissible index tuples $\pu$ with at most $\Gamma$ active coordinates, i.e.,
\[
\mathcal{U} \triangleq
\Big\{\, \pu \in [P]^L  \;\Big|\;
\big|\supp\big(\mathbf{I}({\underline{\mathbf{p}}})\big)\big| \leq \Gamma \,\Big\},
\]
where $\mathbf{I}({\underline{\mathbf{p}}}) \triangleq \big[\mathbbm{1}(p_1>1)\ \ldots\ \mathbbm{1}(p_L>1)\big]$.
The total number of admissible tuples is
\begin{align}\label{eq:n_u}
    n_{\mathcal{U}} = |\mathcal{U}| = \sum_{d=0}^{\Gamma} \binom{L}{d} (P-1)^d.
\end{align}

\begin{defi}[Tuple--tile feasibility graph with capacities]\label{defi:feasibility_graph}
Let $\{\mathcal{P}_\beta\}_{\beta\in[n_{\mathcal P}]}$ denote the set of candidate tiles, each associated with a row-index (closure) set $\mathcal{R}_{\mathcal{P}_\beta}$.

We define the bipartite graph $G(\mathcal{U} \cup \mathcal{V}, \mathcal{E})$ as follows:

\begin{itemize}
    \item \textbf{Left vertex set:} $\mathcal{U} = \{u_\alpha\}_{\alpha\in[n_{\mathcal U}]}$, where each $u_\alpha$ corresponds to an admissible tuple $\pu_\alpha$.

    \item \textbf{Right vertex set:} $\mathcal{V} = \{\mathcal{V}_\beta\}_{\beta\in[n_{\mathcal P}]}$, where each supernode $\mathcal{V}_\beta$ corresponds to a tile $\mathcal{P}_\beta$.

    \item \textbf{Edge set:} An edge exists if and only if
    \[
    (u_\alpha,\mathcal{V}_\beta) \in \mathcal{E}
    \quad \Longleftrightarrow \quad
    \pu_\alpha \in \mathcal{R}_{\mathcal{P}_\beta}\, .
    \]

    \item \textbf{Capacity:} Each tile $\mathcal{V}_\beta$ has capacity
    \[
    \kappa_\beta \triangleq |\mathcal{R}_{\mathcal{P}_\beta}|\, ,
    \]
    representing the maximum number of index tuples that can be assigned to that tile.
\end{itemize}
\end{defi}

We seek a disjoint assignment mapping each tuple $u_\alpha \in \mathcal{U}$ to exactly one feasible tile $\mathcal{V}_\beta$ such that the capacity constraints are satisfied.

Let $\mathcal{M}$ denote the resulting assignment. The induced assigned row-index set of tile $\mathcal{P}_\beta$ is
\[
\mathcal{R}_{\mathcal{P}_\beta}^*
\triangleq
\left\{
\pu_\alpha \mid (u_\alpha,\mathcal{V}_\beta)\in\mathcal M
\right\}\, ,
\]
and the corresponding rank contribution is
\[
r^*_{\mathcal{P}_\beta}
=
\min\,\bigl(\Delta,|\mathcal{R}_{\mathcal{P}_\beta}^*|\bigr)\, .
\]

Thus, the total number of servers induced by an assignment $\mathcal{M}$ is
\[
N_{\mathrm{alg}}
=
\sum_{\beta} r^*_{\mathcal{P}_\beta}
\]
\ahmad{which} coincides with the sum rank of the resulting tile decomposition. While minimizing the number of servers is equivalent to minimizing this sum rank over all feasible assignments, the proposed algorithm focuses on constructing a feasible assignment and then evaluating the induced server count. \ahmad{Accordingly, it provides an achievable scheme rather than minimizing the sum rank objective, which is computationally challenging.}

This problem naturally corresponds to a capacitated bipartite assignment (or flow) problem. At a high level, we iteratively construct a feasible assignment by combining maximum flow computations with a simple greedy fallback step.

We adopt an iterative algorithm \ahmad{(Algorithm~\ref{Alg:GreedyAssignment})} that alternates between global matching and local assignment, \ahmad{as summarized below.}


\begin{algorithm}[h!]
\fontsize{12}{12}\selectfont
\DontPrintSemicolon
\caption{Capacitated Flow-Based Tuple-to-Tile Assignment}
\label{Alg:GreedyAssignment}

\KwInput{Graph $G(\mathcal{U},\mathcal{V},\mathcal{E})$ with capacities $\kappa_\beta$}

Initialize:
$\mathcal{M}\leftarrow\emptyset$,
$\mathcal{U}_{\mathrm{rem}}\leftarrow\mathcal{U}$,
remaining capacities $\kappa_\beta$\;

\While{$\mathcal{U}_{\mathrm{rem}}\neq\emptyset$}{

    Compute a maximum feasible assignment (flow) on the residual graph\;

    \If{all remaining tuples are assigned}{
        update $\mathcal{M}$ and \textbf{break}\;
    }

    Select a tuple $u_\alpha \in \mathcal{U}_{\mathrm{rem}}$\;

    Assign $u_\alpha$ to any feasible tile $\mathcal{V}_\beta$ with remaining capacity\;

    Update $\mathcal{M}$, $\mathcal{U}_{\mathrm{rem}}$, and capacities\;
}

\KwOutput{Assignment $\mathcal{M}$}
\end{algorithm}

\ahmad{We next describe how Algorithm~\ref{Alg:GreedyAssignment} attains global feasibility.}

\begin{remark}[Resolution of infeasibility via local assignment]
When a global feasible assignment does not exist, the corresponding capacitated bipartite graph exhibits an imbalance between a subset of tuples and the available capacity of their neighboring tiles. This can be interpreted as a generalized violation of Hall-type conditions\ahmad{\cite{hopcroft1973n}} in the capacitated setting.

The local assignment step resolves this \ahmad{issue} by assigning at least one tuple to a feasible tile, thereby reducing the size of the remaining instance and modifying the residual capacity structure.
Thus, \ahmad{Algorithm~\ref{Alg:GreedyAssignment}} progressively eliminates infeasible configurations and eventually restores global feasibility.
\end{remark}

\ahmad{We next address the feasibility-first nature of the assignment provided by Algorithm~\ref{Alg:GreedyAssignment}.}

\begin{remark}[Feasibility-first nature of the algorithm]\label{rem:rank}
\ahmad{Algorithm~\ref{Alg:GreedyAssignment} provides a feasible assignment.}
Its objective is to \ahmad{map} every admissible tuple to exactly one tile while respecting capacity constraints. The final number of servers is then computed from the induced tile loads via
\begin{align}\label{optrank}
    r^*_{\mathcal{P}_\beta}=\min(\Delta,|\mathcal{R}^*_{\mathcal{P}_\beta}|).
\end{align}
\end{remark}

\ahmad{
Next, we discuss the effect of tile capacity and threshold $\Delta$ on the performance of Algorithm~\ref{Alg:GreedyAssignment}.}

\begin{remark}[Effect of tile capacity and threshold $\Delta$]\label{rem:cap}
The relation between the tile capacity $\kappa_\beta$ and the threshold $\Delta$ influences the resulting server count.

When $\kappa_\beta$ is large relative to $\Delta$, a tile can absorb many tuples while contributing at most $\Delta$ to the sum rank. In this regime, consolidating assignments within fewer tiles can significantly reduce the number of servers.

In contrast, when $\kappa_\beta < \Delta$, the contribution of a tile grows approximately linearly with the number of assigned tuples, and the benefits of consolidation are reduced.

These observations highlight the role of tile geometry in the achievable performance, despite \ahmad{Algorithm~\ref{Alg:GreedyAssignment}} is primarily designed for feasibility rather than explicit sum rank minimization.
\end{remark}

The above information \ahmad{provided in Remarks~\ref{rem:rank} and~\ref{rem:cap}} will be essential in enumerating our equivalence class of tiles and associating each such class with a collection of servers.

\ahmad{After presenting Algorithm~\ref{Alg:GreedyAssignment}, our next step}
is to fill the tiles in the product tensor, using the constitutive component columns and rows, as follows.
\paragraph{Step 2: Filling tiles in $\TensorE \times_1 \mathbf{D}$ as a function of $\TensorF$}

Having established the set of equivalence classes and the corresponding (position of the) tiles of $\TensorF$ by establishing Algorithm~\ref{Alg:GreedyAssignment}, we proceed to fill (and crop) the tiles in $\TensorF$ as follows:
  \begin{align}
      \TensorF_\mathcal{P} \triangleq (\TensorF \odot  \TensorS_{\mathcal{P}}) (\mathcal{R}^*_{\mathcal{P}}, \mathcal{C}_{\mathcal{P}})\, , \quad \forall \mathcal{P} \in \mathcal{C}
      \, . \label{Formation-of-submatrices}
  \end{align}
  
The first step $\TensorF \odot \TensorS_{\mathcal{P}}$ simply fills up $\TensorS_{\mathcal{P}}$ with the corresponding entries of $\TensorF$, and the second step $(\TensorF \odot \TensorS_{\mathcal{P}}) (\mathcal{R}^*_{\mathcal{P}}, \mathcal{C}_{\mathcal{P}})$ crops these\footnote{We here see a small distinction between the previously uncropped tiles, and the tiles that are cropped here. These cropped tiles will be the outcomes of an \ahmad{MLSVD} decomposition of subtensors of $\TensorF$, yielding our SVD-based factorization.}.

\paragraph{Step 3: Creating and filling the non-zero tiles in  $\mathbf{D}$ and  $\TensorE$}
This step starts with the \ahmad{MLSVD} decomposition (described in \ahmad{Appendix}~\ref{MLSVD}) of the cropped tile $\TensorF_\mathcal{P}$ where this decomposition takes the form
\begin{align}
  \TensorF_{\mathcal{P}} = \TensorE_{\mathcal{P}}\times_1\mathbf{D}_{\mathcal{P}} \label{sub-SVD}
\end{align}
where $\mathbf{D}_{\mathcal{P}} \in  \mathbb{R}^{|\mathcal{C}_{\mathcal{P}}| \times r^*_{\mathcal{P}}}, \TensorE_{\mathcal{P}} \in \mathbb{R}^{r^*_{\mathcal{P}} \times |\mathcal{R}^*_{\mathcal{P}}|} $. In particular, $\TensorF, \mathbf{D}$, and $\TensorE$ are  associated to  $\TensorT$, $ \mathbf{U}^{(1)}$, and $\TensorS$ in all complete \ahmad{MLSVD} decomposition of \eqref{Tensor_SVD}. Naturally, $\rank(\TensorF_{\mathcal{P}}) \leq r^*_{\mathcal{P}}$.
Let \begin{align}\label{eq:tileEnumeration}
\mathcal{C} \triangleq \{ \mathcal{P}_1, \mathcal{P}_2, \hdots, \mathcal{P}_{m}\}, \: m \in \mathbb{N}\end{align}
describe the enumeration we give to each tile. 
Then the position that each cropped tile takes inside $\mathbf{D}$, is given by 
\begin{align}\label{eq:tilePositionInD}
    \mathcal{C}_{\mathcal{P}_j},\:  [\sum^{j-1}_{i=1}  r^*_{\mathcal{P}_i} 
+1: \sum^{j}_{i=1} r^*_{\mathcal{P}_i}]\, ,\quad \forall \mathcal{P}_j \in \mathcal{C} 
\end{align} 
and the position of each cropped tile in $\TensorE$ is given by 
\begin{align} \label{eq:tilePositionInE}
[\sum^{j-1}_{i=1} r^*_{\mathcal{P}_i} 
+1: \sum^{j}_{i=1} r^*_{\mathcal{P}_i}],\: \mathcal{R}^*_{\mathcal{P}_j}\, , \quad \forall \mathcal{P}_j \in \mathcal{C}
\end{align}

This yields
\begin{align}
\mathbf{D}\Big(\mathcal{C}_{\mathcal{P}_j},\:  [\sum^{j-1}_{i=1} r^*_{\mathcal{P}_i}+1: \sum^{j}_{i=1} r^*_{\mathcal{P}_i}]\Big)= \mathbf{D}_{\mathcal{P}_j}\, ,\quad \forall \mathcal{P}_j \in \mathcal{C}
    \label{Decoding-Encoding1}
\end{align}
and 
\begin{align}
\TensorE\Big([\sum^{j-1}_{i=1} r^*_{\mathcal{P}_i}
+1: \sum^{j}_{i=1} r^*_{\mathcal{P}_i}],\: \mathcal{R}^*_{\mathcal{P}_j}\Big)=\TensorE_{\mathcal{P}_j}\, ,\quad \forall \mathcal{P}_j \in \mathcal{C}
    \label{Decoding-Encoding2}
\end{align}
while naturally the remaining non-assigned elements of $\mathbf{D}$ and $\TensorE$ are zero.

\paragraph{Step 4: An upper bound to the number of servers}
We now proceed to bound the number of rank-one contribution supports (corresponding to the number of servers), and we do so under our previously stated assumption of disjoint supports (cf. Definition~\ref{tensor-disjointSupportAssumption}), which is equivalent to the disjoint-tiles assumption \ahmad{(see Lemma~\ref{Disjoint-Support-Assumption: Equivalence}).}

We recall from Section~\ref{Formulating} (see also~\eqref{EncodingMatrix}--\eqref{DecodingMatrix}) that each server $n\in[N]$ corresponds to one column of the decoding matrix $\mathbf{D}$ and the associated row of the encoding tensor $\TensorE$. To express $N$ in terms of the \ahmad{system} parameters, consider the tiles $\mathcal{P}\in\mathcal{C}$ and the associated submatrices $\mathbf{D}_{\mathcal{P}}$ defined in~\eqref{sub-SVD}. From~\eqref{Decoding-Encoding1}, these submatrices collectively span $\sum_{i=1}^{m} r^*_{\mathcal{P}_i}$ columns. 

Therefore, from~\eqref{optrank}, the upper bound on the number of servers is obtained as
\begin{align}
N&\le
\sum\limits_{\mathcal{P}\in \mathcal{C}} r_{\mathcal{P}}^{*}=\frac{K}{\Delta} \sum_{\mathcal{\beta}\in [n_{\mathcal{P}}]}
\min \big(\Delta,\lvert \mathcal{R}^*_{\mathcal{P}_\beta} \rvert\big)
\end{align}
which corresponds to~\eqref{achiv}. \qedhere
\end{proof}

We next provide a comprehensive example to illustrate how the system parameters determine the tile structure, how overlaps arise among admissible tuples, and how Algorithm~\ref{Alg:GreedyAssignment} resolves these overlaps through a unique tuple-to-tile assignment.

\begin{figure*}[t!]
    \centering
    
    \begin{subfigure}{0.45\textwidth}
        \centering
        \includegraphics[width=\linewidth]{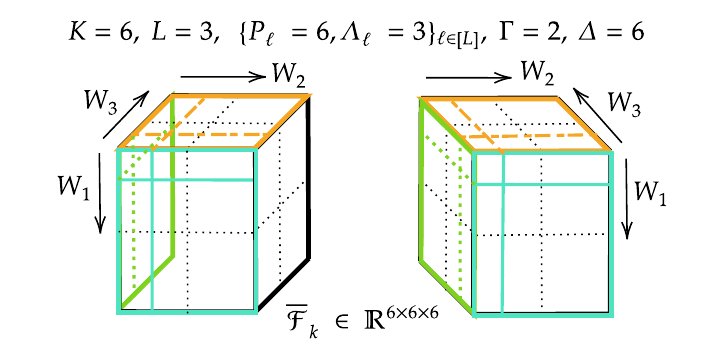}
        \caption{}
        \label{fig:ex_L3_merge_a}
    \end{subfigure}
    \hfill
    \begin{subfigure}{0.45\textwidth}
        \centering
        \includegraphics[width=\linewidth]{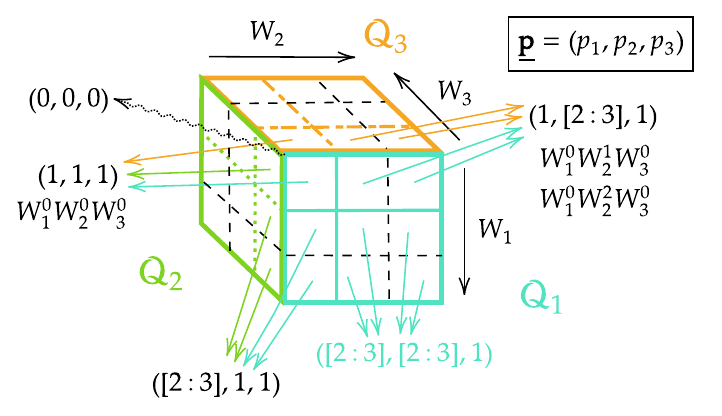}
        \caption{}
        \label{fig:ex_L3_merge_b}
    \end{subfigure}

    \vspace{0.4em}

    \begin{subfigure}{0.45\textwidth}
        \centering
        \includegraphics[width=\linewidth]{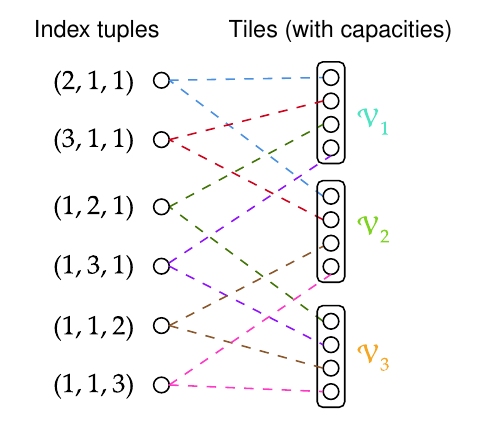}
        \caption{}
        \label{fig:ex_L3_merge_c}
    \end{subfigure}
    \hfill
    \begin{subfigure}{0.54\textwidth}
        \centering
        \includegraphics[width=\textwidth]{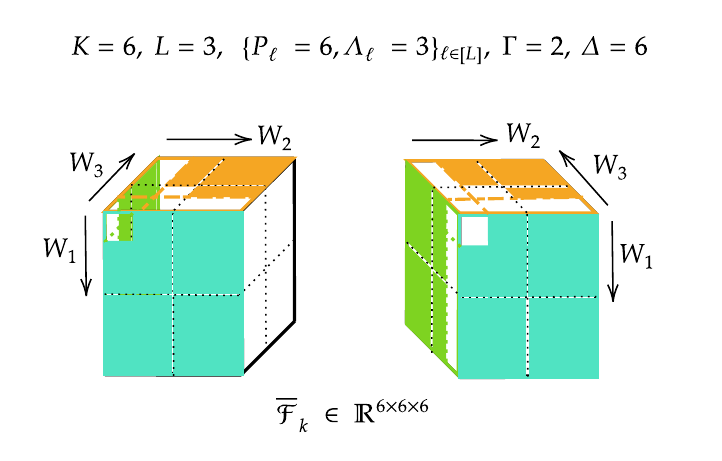}
        \caption{}
        \label{fig:ex_L3_merge_d}
    \end{subfigure}

    \caption{Illustration of Example~\ref{ex:L3merged}. 
    (a) Initial construction of $\Gamma$-dimensional tiles in $\TensorF_1$ and their combinatorial overlaps. 
    (b) Corresponding index-tuple representation, where a single admissible tuple may belong to multiple tile closures. 
    (c) Bipartite feasibility graph between admissible tuples and candidate tiles; each tile node represents a tile with capacity (illustrated by stacked slots corresponding to its closure size), and edges indicate feasible assignments based on closure membership. 
    (d) Final disjoint tile assignment after applying Algorithm~\ref{Alg:GreedyAssignment}. The assignment shown is illustrative and represents one feasible outcome of the algorithm. The assignment induces disjoint sets $\mathcal{R}_{\mathcal{P}}^*$, enabling the MLSVD construction. Tiles with identical colors correspond to the same active subset $\mathcal{Q}_r\subseteq [L]$ of cardinality $\Gamma$. }

    \label{fig:ex_L3_merged}
\end{figure*}

\begin{ex}[Detailed \ahmad{tuple-to-tile assignment} example] 
\label{ex:L3merged}
Consider a distributed computing setting with the parameters $(K=6,L=3,\{P_\ell=6,\Lambda_\ell=3\}_{\ell\in[L]},\Gamma=2,\Delta=6)$. To illustrate the representative supports, the induced overlaps, and the role of Algorithm~\ref{Alg:GreedyAssignment}, we focus on the demanded tensor of one user, e.g., $1$, denoted by $\TensorF_1$, shown in Figure~\ref{fig:ex_L3_merged}.

Following the protocol in the proof of Theorem~\ref{Achievability-new}, the $\Gamma$-subsets of basis subfunctions are
\[
\mathcal Q_1=\{1,2\},\qquad
\mathcal Q_2=\{1,3\},\qquad
\mathcal Q_3=\{2,3\}.
\]
Since $\Gamma=2$, every candidate tile is associated with exactly one of these three support classes.

We detail the solution in the following steps.
\begin{parletters}
\paragraph{Step 1: Admissible tuples}
Each index tuple has the form $\pu=(p_1,p_2,p_3)\in[6]^3$,
and must satisfy the sparsity constraint $\big|\supp(\mathbf I(\pu))\big|\le 2$,
that is, at most two coordinates may exceed $1$. Hence, the admissible tuple set is
\[
\mathcal U
=
\Big\{
\pu\in[6]^3:\big|\supp(\mathbf I(\pu))\big|\le2
\Big\}
\]
with cardinality $|\mathcal U|
=
\sum_{d=0}^2 \binom{3}{d}(6-1)^d
=
1+3\cdot5+3\cdot25
=
91$.

\paragraph{Step 2: Window structure}
Since $P_\ell=6$ and $\Lambda_\ell=3$, the active values $\{2,3,4,5,6\}$ are grouped into the windows
\[
\mathcal W^{(1)}=\{2,3,4\},
\qquad
\mathcal W^{(2)}=\{5,6\}.
\]
Thus, for each support class $\mathcal Q_r$, the active exponent coordinates are partitioned into window choices, which define the candidate tiles.

\paragraph{Step 3: Tile cells versus closures}
Consider the support class $\mathcal Q_1=\{1,2\}$ and the window pair $(\mathcal W^{(1)},\mathcal W^{(2)})$. The corresponding geometric tile cells are the tuples with exactly two active coordinates,
\[
\mathcal P(\mathcal Q_1,\mathcal W^{(1)},\mathcal W^{(2)})
=
\Big\{
(p_1,p_2,p_3)\in[6]^3:
p_1\in\mathcal W^{(1)},\;
p_2\in\mathcal W^{(2)},\;
p_3=1
\Big\}.
\]
These are the atomic \ahmad{$\Gamma=2$}-dimensional cells displayed geometrically in Figure~\ref{fig:ex_L3_merged} (a).

However, for assignment purposes, each tile is associated with its larger row-index set, or closure,
\[
\mathcal R_{\mathcal P(\mathcal Q_1,\mathcal W^{(1)},\mathcal W^{(2)})}
=
\Big\{
(p_1,p_2,p_3)\in[6]^3:
p_1\in \{1\}\cup\mathcal W^{(1)},\;
p_2\in \{1\}\cup\mathcal W^{(2)},\;
p_3=1
\Big\}.
\]
Therefore,
\[
\mathcal R_{\mathcal P}
=
\{1,2,3,4\}\times\{1,5,6\}\times\{1\},
\]
and
\[
|\mathcal R_{\mathcal P}|=(1+3)(1+2)=12.
\]
This distinction is crucial: the geometric tile contains only fully active tuples, whereas the closure contains all admissible tuples whose support is contained in the tile support and whose active entries lie in the selected windows.

\paragraph{Step 4: Why overlaps arise}
As shown in Figure~\ref{fig:ex_L3_merged} (b), overlaps arise for two reasons.

First, a lower-support tuple may be contained in multiple support classes. For instance, the tuple $(2,1,1)$
has support $\{1\}$, and hence belongs to closures associated with both $\mathcal Q_1=\{1,2\}$ and $\mathcal Q_2=\{1,3\}$.

Second, even within a fixed support class, an inactive coordinate equal to $1$ is compatible with several window choices. For example,
\[
(2,1,1)\in
\mathcal R_{\mathcal P(\mathcal Q_1,\mathcal W^{(1)},\mathcal W^{(1)})}
\cap
\mathcal R_{\mathcal P(\mathcal Q_1,\mathcal W^{(1)},\mathcal W^{(2)})}\, .
\]

Therefore, lower-support tuples may overlap both across support classes and across window selections.
In contrast, fully active tuples are rigid. For example, the tuple $(2,5,1)$
has support $\{1,2\}$, with $2\in\mathcal W^{(1)},\ 5\in\mathcal W^{(2)}$,
so both its support class and its window pair are uniquely determined. Hence, it belongs to exactly one candidate tile.

\paragraph{Step 5: Feasibility graph and capacities}
To resolve the ambiguity, we construct the tuple--tile feasibility graph of Definition~\ref{defi:feasibility_graph}. The left vertices correspond to admissible tuples in $\mathcal U$, while the right vertices correspond to candidate tiles. An edge is drawn whenever the tuple belongs to the closure of the tile. Figure~\ref{fig:ex_L3_merged} (c) illustrates the resulting bipartite graph for the intersecting tiles.

Each tile node is equipped with a capacity equal to the size of its closure. In the present homogeneous setting, this capacity depends on the selected windows. For instance, for the tile above, we have $\kappa_\beta
=
|\mathcal R_{\mathcal P(\mathcal Q_1,\mathcal W^{(1)},\mathcal W^{(2)})}|
=
12$\,.

Algorithm~\ref{Alg:GreedyAssignment} then assigns each tuple to exactly one feasible tile while respecting these capacities.

\paragraph{Step 6: Induced disjoint assignment}
The output of Algorithm~\ref{Alg:GreedyAssignment} is a collection of assigned sets
$\mathcal R^*_{\mathcal P_\beta}\subseteq \mathcal R_{\mathcal P_\beta}$
such that each admissible tuple belongs to exactly one assigned set. Consequently, the resulting supports are disjoint. A feasible tessellation pattern for $\TensorF_1$ is illustrated in Figure~\ref{fig:ex_L3_merged} (d).


\ahmad{This disjointness condition is exactly what is needed to apply the MLSVD construction described in Appendix~\ref{MLSVD}.}

\paragraph{Step 7: Induced sum rank and server count}
Once the assignment is fixed, the contribution of each tile is $r^*_{\mathcal P_\beta}
=
\min\bigl(\Delta,|\mathcal R^*_{\mathcal P_\beta}|\bigr)$.
Hence, the per-user sum rank is $\sum_{\beta\in[n_{\mathcal P}]}
\min\bigl(\Delta,|\mathcal R^*_{\mathcal P_\beta}|\bigr)$,
and the total number of required servers is obtained by scaling this quantity by $K/\Delta$, as in Theorem~\ref{Achievability-new}.

For this example, applying Algorithm~\ref{Alg:GreedyAssignment} yields a feasible disjoint assignment whose induced tile loads have sum rank
\[
\sum_{\beta}\min\bigl(\Delta,|\mathcal R_{\mathcal P_\beta}^*|\bigr)=66\, .
\]

Since $K=\Delta=6$, the scaling factor $K/\Delta$ is equal to $1$, and thus the total number of required servers is $N=66$.
\end{parletters}

For comparison, the default tensor factorization \ahmad{of} Theorem~\ref{Achievability-Converse} gives
\[
\frac{K}{\Delta}\binom{L}{\Gamma}\Big(\frac{P}{\Lambda}\Big)^\Gamma\min(\Delta,\Lambda^\Gamma)
=
72\, .
\]

\ahmad{Consequently, the assignment-based construction achieves a reduction of $6$ servers in this example.}

\end{ex}

We next examine the computational complexity of the proposed assignment algorithm (cf. Algorithm~\ref{Alg:GreedyAssignment}) for tiling designs.

\begin{remark}\label{rem:complexity}
The complexity of the proposed algorithm is dominated by the construction of the tuple--tile feasibility graph and the repeated solution of capacitated bipartite assignment problems via maximum flow.

The bipartite graph consists of $n_{\mathcal U}$ left vertices (admissible tuples) and $n_{\mathcal P}$ right vertices (candidate tiles). Each edge corresponds to a feasible assignment, i.e., a tuple belonging to the closure of a tile. Hence, the total number of edges is
\[
|\mathcal E|
=
\sum_{\alpha\in[n_{\mathcal U}]}
\bigl|\{\beta:\pu_\alpha \in \mathcal R_{\mathcal P_\beta}\}\bigr|.
\]

Each iteration of the algorithm solves a capacitated bipartite assignment problem, which can be formulated as a maximum flow instance. Using standard reductions (e.g., node-splitting to handle capacities~\cite{cormen2022introduction}), this can be solved in
\[
\mathcal{O}\!\big(|\mathcal E|\sqrt{|\mathcal U|+|\mathcal V|}\big),
\quad \text{where} \quad |\mathcal U|=n_{\mathcal U}\quad \text{and} \quad |\mathcal V|=n_{\mathcal P}
\]
by classical algorithms for bipartite matching, such as the Hopcroft--Karp algorithm~\cite{hopcroft1973n}.

In practice, the algorithm typically converges in a small number of iterations. When the global assignment is feasible, a single flow computation suffices. Otherwise, the algorithm performs a sequence of residual updates, each assigning at least one tuple and strictly reducing the number of remaining tuples.
The total complexity is therefore bounded by $\mathcal{O}\!\big(N_{\mathrm{iter}} \cdot |\mathcal E|\sqrt{n_{\mathcal U}+n_{\mathcal P}}\big)$,
where $N_{\mathrm{iter}} \le n_{\mathcal U}$. While this bound is linear in the number of iterations in the worst case, empirical behavior is significantly more favorable, with $N_{\mathrm{iter}}$ typically remaining small.

In regimes where tile capacities are sufficiently large relative to the number of admissible tuples, the global assignment succeeds in a single iteration, yielding overall complexity $\mathcal{O}\!\big(|\mathcal E|\sqrt{n_{\mathcal U}+n_{\mathcal P}}\big)$.
In contrast, in highly constrained regimes, multiple iterations may be required. However, each iteration strictly reduces the number of remaining tuples, guaranteeing termination in at most $n_{\mathcal U}$ steps.

Finally, in the homogeneous setting~\ahmad{with $\{P_\ell=P,\, \Lambda_\ell=\Lambda\}_{\ell\in [L]}$}, the number of edges satisfies $|\mathcal E| = \mathcal{O}(n_{\mathcal U}\Lambda^\Gamma)$
since each tuple is compatible with multiple tiles across support configurations and sliding windows.
\end{remark}

\begin{figure}[t!]
\centering
\includegraphics[width=0.8\textwidth]{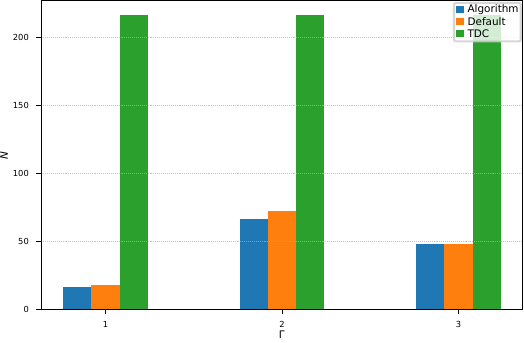}
\caption{Comparison of the required number of servers \(N\) for the lossless distributed computing setting \((K=6,L=3,\Gamma,\Delta=6,\{P_\ell=6,\Lambda_\ell=3\}_{\ell\in[L]})\). The bars compare the proposed algorithmic achievable scheme, the default tensor factorization scheme, and the TDC baseline as functions of the computation cost \(\Gamma\).}
\label{fig:L3}
\end{figure}

\begin{figure}[t!]
\centering
\includegraphics[width=0.8\columnwidth]{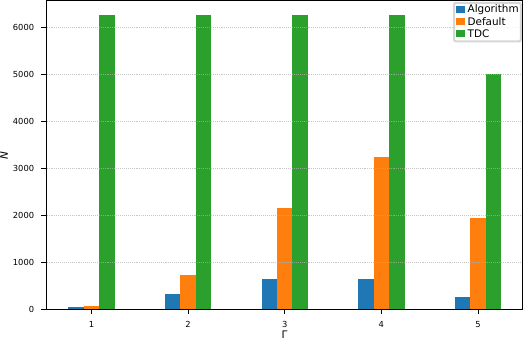}
\caption{Comparison of the required number of servers \(N\) for the lossless distributed computing setting \((K=8,L=5,\Gamma,\Delta=4,\{P_\ell=5,\Lambda_\ell=2\}_{\ell\in[L]})\). The bars compare the proposed algorithmic achievable scheme, the default tensor factorization scheme, and the TDC baseline with respect to the computation cost \(\Gamma\).}
\label{fig:L5}
\end{figure}

\section{Numerical Analysis and Evaluation of Algorithm~\ref{Alg:GreedyAssignment}}\label{numerical}

In this section, we provide numerical evaluations of the required number of servers for different system configurations using Algorithm~\ref{Alg:GreedyAssignment}, and compare the results with the default tensor factorization scheme of Theorem~\ref{Achievability-Converse} and the sparse matrix factorization of the TDC scheme~\cite{Khalesi2025Tessellated}.

We consider homogeneous lossless systems with parameters $(K,\Delta,L,\Gamma,P,\Lambda)$, where $K=\Delta$ is considered in the first part of the simulation so that the total number of required servers coincides with the induced sum rank per block. The performance is evaluated as a function of the computation cost $\Gamma$ for two representative cases $L=3$ and $L=5$. The corresponding results are illustrated in Figure~\ref{fig:L3} and Figure~\ref{fig:L5}. In particular, the gap between the proposed scheme and the default tensor factorization increases with $L$, highlighting the growing importance of assignment in higher-dimensional settings.

We next discuss the results shown in Figures~\ref{fig:L3} and~\ref{fig:L5}, which illustrate the required number of servers $N$ as a function of the computation cost $\Gamma$ under the lossless setting. In both figures, we consider homogeneous parameters with $P_\ell = P$ and $\Lambda_\ell = \Lambda$ for all $\ell\in [L]$, and we compare the proposed assignment-based scheme (Algorithm~\ref{Alg:GreedyAssignment}) with the default tensor factorization scheme of Theorem~\ref{Achievability-Converse} and the TDC baseline~\cite{Khalesi2025Tessellated}. We next detail the comparisons.

\textbf{Moderate dimensionality ($L=3$)\,.}
In this regime,~\ahmad{as shown in Figure~\ref{fig:L3}}, the number of admissible tuples grows relatively slowly with $\Gamma$, and all schemes exhibit a moderate increase in $N$ as $\Gamma$ increases. However, the proposed assignment-based scheme consistently outperforms the default tensor factorization scheme of Theorem~\ref{Achievability-Converse} by eliminating overlap among subtensors. This gain is particularly visible for intermediate values of $\Gamma$, where multiple tiles share common tuples. By enforcing a disjoint assignment, Algorithm~\ref{Alg:GreedyAssignment} avoids redundant counting and achieves a smaller sum rank of the resulting subtensors.
In contrast, the TDC scheme~\cite{Khalesi2025Tessellated} yields significantly larger values of $N$ across all $\Gamma$ since it does not exploit the multiplicative structure of the underlying functions and instead treats the problem with much larger dimensionality to capture all multiplicative terms, leading to inefficient use of computational resources.

\textbf{Higher dimensionality ($L=5$)\,.}
The impact of the assignment step becomes more pronounced as the dimension increases,~\ahmad{as illustrated in Figure~\ref{fig:L5}}. In this setting, the number of admissible tuples grows combinatorially with $\Gamma$, which leads to a rapid increase in the required number of servers for the default tensor factorization scheme of Theorem~\ref{Achievability-Converse}. This growth is due to the increasing overlap between high-dimensional subtensors, which are treated independently in the default construction in Theorem~\ref{Achievability-Converse}.
The proposed assignment-based scheme mitigates this effect by consolidating overlapping tuples into a reduced set of tiles. As a result, the growth of $N$ is significantly limited compared to the default scheme, especially for $\Gamma=3$ and $\Gamma=4$, where overlap is most pronounced.
Interestingly, for $\Gamma=L$, a slight reduction in $N$ can be observed. This is due to the fact that higher-dimensional tiles allow more efficient packing of tuples, reducing the number of active tiles required. Such behavior is not captured by the default scheme of Theorem~\ref{Achievability-Converse}, which does not perform any global assignment. Furthermore,~\ahmad{similarly as in}  Figure~\ref{fig:L3}, the TDC baseline~\cite{Khalesi2025Tessellated} remains significantly above both tensor-based schemes and shows limited sensitivity to $\Gamma$, reflecting its failure to leverage the underlying tensor structure.

\section{Discussion and Conclusions}\label{Conclusion}

In this paper, we investigated the problem of lossless multi-user distributed computing \ahmad{with $K$ users and $N$ servers} for real-valued multivariate polynomials through the lens of structured tensor representations. \ahmad{By transforming the problem into a sparse tensor factorization, we devised novel achievability scheme, and characterized the achievable system rate \( R = \frac{K}{N} \), as described in Theorem~\ref{Achievability-Converse}, reflecting how the system parameters, including computation cost \(\Gamma\), communication cost \(\Delta\), and the multiplication cost \(\Lambda\) interact jointly, and how exploiting this interplay reduces the number of required servers \(N\). 
Furthermore, we introduced a combinatorial reduction algorithm (Algorithm~\ref{Alg:GreedyAssignment}) in Theorem~\ref{Achievability-new}, which results in a tighter achievable rate result compared to Theorem~\ref{Achievability-Converse} by resolving the redundant computations across the servers.
} 

From a performance perspective, the proposed scheme offers clear advantages over existing approaches. Compared to the TDC scheme~\cite{Khalesi2025Tessellated}, it benefits from explicitly exploiting the multiplicative structure of the problem. Relative to the default tensor factorization approach, it avoids the inefficiencies caused by overlapping components by enforcing a consistent assignment. As demonstrated in the numerical results, this leads to a substantial reduction in the number of required servers.


Several directions remain open as future work. One natural extension is to consider the lossy setting, where approximate recovery is allowed, and new tradeoffs emerge between accuracy and \ahmad{design constraints (i.e., $\Gamma$, $\Delta$, and $\Lambda$).}
Another important direction is to establish converse bounds that precisely characterize the fundamental limits of this problem. 
\appendix
The appendix is structured as follows. Appendix~\ref{tensor_basic} includes the basic tensor concepts and definitions. Next, Appendix~\ref{MLSVD} provides a primer on multilinear SVD for tensor decomposition, which is used in the Proof of Theorem~\ref{Achievability-new}. Finally, we present all the details for the multi-shot scenario with $T>1$ shots, which \ahmad{generalizes the results derived for} the single-shot scenario, in Appendix~\ref{sec:multishot}.

\subsection{Basic Tensor Definitions}\label{tensor_basic}
We first introduce a tensor in the following definition.
\begin{defi}
\emph{An order-N tensor}, $\mathcal{\bar{X}} \in \mathbb{R}^{I_1 \times \cdots \times I_N}$, is a multi-way array with $N$ modes, with the $n$-th mode of dimensionality $I_n$, for $n\in [N]$. Special cases of tensors include matrices as order-2 tensors (e.g., $\mathbf{X} \in \mathbb{R}^{I_1 \times I_2}$), vectors as order-1 tensors (e.g., $\textbf{x} \in \mathbb{R}^{I_1}$), and scalars as order-0 tensors (e.g., ${x} \in \mathbb{R}$).
\end{defi}
 
We next define the matrix unfolding of a tensor.

\begin{defi}\label{def:mode-n-unfold}
    A \emph{mode-n unfolding} of a tensor is the procedure of mapping the elements from a
    multidimensional array to a two-dimensional array (matrix). Conventionally, such a procedure is associated with stacking mode-$n$ fibers (modal vectors) as column vectors of the resulting matrix.
    For instance, mode-$1$ unfolding of $\mathcal{\bar{X}} \in \mathbb{R}^{I_1 \times I_2 \times
    \cdots \times I_N}$ is represented as $\mathcal{\bar{X}}_{(1)} \in \mathbb{R}^{I_1 \times I_2 I_3
    \cdots I_N}$, and given by 
    \begin{equation}
        \mathcal{\bar{X}}_{(1)}\big(i_1,\overline{i_2 i_3 \ldots i_N} \big) = \mathcal{\bar{X}}(i_1,i_2,\ldots, i_N)\, .
    \end{equation}
    Note that the overlined subscripts refer to linear indexing (or
    Little-Endian) \cite{Dolgov2014}, is given by
    \begin{align}\label{eq:mode-n-unfold}
        \overline{i_1 i_2 \dots i_N}
            &= 1 + \sum_{n=1}^N (i_n - 1) \prod_{k=1}^{n-1}I_{k} \\
            &= 1 + i_1 + (i_2 - 1)I_1 + \cdots + (i_N-1)I_1 \ldots I_{N-1}. \nonumber
    \end{align}
\end{defi}

We proceed to define tensor folding of a vector as follows.
\begin{defi}
     Any given vector $\mathbf{x} \in \mathbb{R}^{I_1 I_2 \ldots I_N}$ can be
    \emph{folded} into an $N$-th order tensor, $\mathcal{\bar{X}} \in \mathbb{R}^{I_1 \times I_2 \times \cdots \times I_N}$, with the relation between their entries defined by
    \begin{equation}\label{eq:folding}
        \mathcal{\bar{X}}(i_1, i_2, \dots, i_N) = \mathbf{x}(i), \quad \forall i_n \in [I_n]
    \end{equation}
    where $i=1+\sum_{n=1}^{N}(i_n-1)\prod_{k=1}^{n-1}I_k$.
\end{defi}

We next formally define the stacking operation in tensors.
\begin{defi}
Consider grouping $J$ order-$N$ tensor samples $\mathcal{\bar{X}}_j \in \mathbb{R}^{I_1 \times \cdots
\times I_N},\, j\in [J]$, so as to form an order-$(N + 1)$ data tensor, $\mathcal{\bar{Y}} \in \mathbb{R}^{I_1
\times \cdots \times I_N \times J}$. This \emph{stacking} operation is
denoted by
\begin{equation}\label{eq:stacking_app}
    \mathcal{\bar{Y}} = \textit{stack}_{N+1}\left({\mathcal{\bar{X}}_1, \ldots, \mathcal{\bar{X}}_J}\right)\, .
\end{equation}

In other words, the combined tensor samples introduce another dimension, the $(N+1)$-th mode of 
$\mathcal{\bar{Y}}$, such that its mode-$(N+1)$ unfolding $\mathcal{\bar{Y}}_{(N+1)}\in\; \mathbb{R}^{(I_1 I_2 \cdots I_N)\times J}$ is
\begin{align}
    \mathcal{\bar{Y}}_{(N+1)}\big(\overline{i_1i_2 i_3 \ldots i_N}, j \big) \;=\; 
    \begin{bmatrix}
        \mathbf{x}_1,
        \hdots,
        \mathbf{x}_J
    \end{bmatrix}
\end{align}
where $\mathbf{x}_j\in \mathbb{R}^{I_1 I_2 \cdots I_N}$ denotes the vectorization of 
the tensor $\mathcal{\bar{X}}_j$, obtained by stacking all its entries into a single column vector in 
Little-Endian order consistent with~\eqref{eq:mode-n-unfold}.

\end{defi}

We next define the mode-n product for tensors.

\begin{defi}
    The \emph{mode-n product} takes as input an order-$N$ tensor, $\mathcal{\bar{X}} \in \mathbb{R}^{I_1\times I_2 \times \cdots \times I_N}$, and a matrix $\mathbf{A} \in \mathbb{R}^{J \times I_n}$, to
    produce another tensor, $\mathcal{\bar{Y}}$, of the same order as the original tensor $\mathcal{\bar{X}}$. The operation is denoted by
    \begin{equation}
        \mathcal{\bar{Y}} = \mathcal{\bar{X}} \times_n \mathbf{A} 
    \end{equation}
    where $\mathcal{\bar{Y}} \in \mathbb{R}^{I_1 \times \cdots \times I_{n-1} \times J \times I_{n+1} \times\cdots \times I_N}$. The mode-$n$ product is comprised of 3 consecutive steps:   
    \begin{equation}   \begin{aligned}\label{eq:mode-n-product_app}
        \mathcal{\bar{X}} \rightarrow \mathcal{\bar{X}}_{(n)}\, , \quad 
        \mathcal{\bar{Y}}_{(n)} = \A \mathcal{\bar{X}}_{(n)}\, , \quad  
        \mathcal{\bar{Y}}_{(n)} \rightarrow \mathcal{\bar{Y}}\ \, .
    \end{aligned}
    \end{equation}
\end{defi}

To see the transition to the sparse tensor factorization, we next introduce the tensor contraction product.
\begin{defi}\cite{kolda2009tensor}
\emph{Tensor Contraction Product} (TCP) is at the core of tensor decompositions, an operation similar to the mode-$n$ product, but the arguments of which are multidimensional arrays that can be of a different order. For instance, given an $N$-th order tensor $\mathcal{\bar{X}} \in \mathbb{R}^{I_1\times \cdots \times I_N}$ and an $M$-th order tensor $\mathcal{\bar{Y}}\in \mathbb{R}^{J_1\times \cdots \times J_M} $, with common modes $I_n = J_m$, then, their $(n,m)$-contraction denoted by $\times^m_n$,  yields a third tensor $\mathcal{\bar{Z}}\in \mathbb{R}^{I_1 \times \cdots \times I_{n-1} \times I_{n+1}  \times \cdots \times I_N \times J_1 \times \cdots \times J_{m-1} \times J_{m+1}  \times \cdots \times J_M}$ of order $(N+M-2)$, $\mathcal{\bar{Z}}=\mathcal{\bar{X}} \times_n^m \mathcal{\bar{Y}}$ with the entries
\begin{multline}\label{eq:cont}
    \bar{\mathcal{Z}}(i_1,\dots,i_{n-1}, i_{n+1}, \dots, i_N, j_1, \dots, j_{m-1}, j_{m+1}, \dots, j_M)\\
    =\sum_{i_n\in [I_n]} \bar{\mathcal{X}}(i_1,\dots, i_{n-1}, i_n, i_{n+1},\dots, i_N) \\
    \times \bar{\mathcal{Y}}(j_1, \dots, j_{m-1}, i_n, j_{m+1},\dots, j_M) \, . 
\end{multline}
\end{defi}
The overwhelming indexing associated with the TCP operation in \eqref{eq:cont} becomes unmanageable for larger tensor networks, whereby multiple TCPs are carried out across a large number of tensors. Manipulation of such expressions is prone to errors and prohibitive to the manipulation of higher-order tensors. 

We next generalize the TCP operation for the cases where there is more than one common mode between two tensors. Recall that this operation is used in~\eqref{Functions}.


\begin{defi}\label{GTCP}
\emph{Generalized TCP} is an operation similar to TCP, but it contracts a whole \emph{ordered block of modes} (a multi-index) between two tensors. For instance, given an $N$-th order tensor $\mathcal{\bar X}\in\mathbb{R}^{I_1\times\cdots\times I_N}$ and an $M$-th order tensor $\mathcal{\bar Y}\in\mathbb{R}^{J_1\times\cdots\times J_M}$, assume the tails match pairwise, i.e.,
$(I_n,\dots,I_N)=(J_m,\dots,J_M)$ and $N-n+1 = M-m+1$.
Then, their $([[n:N]],[[m:M]])$-contraction, denoted by $\times^{[[m:M]]}_{[[n:N]]}$, yields a third tensor $\mathcal{\bar Z}\in\; \mathbb{R}^{I_1 \times \cdots \times I_{n-1} \times J_1 \times \cdots \times J_{m-1}}$, where
\[
\mathcal{\bar Z} \;=\; \mathcal{\bar X}\times^{[[m:M]]}_{[[n:N]]}\mathcal{\bar Y}
\]
of order $(n+m-2)$, with entries
\begin{multline}
\bar{\mathcal{Z}}(i_1,\dots,i_{n-1},\,j_1,\dots,j_{m-1})\\
=\sum_{\underline{\mathbf{i}}\in\mathcal I} \bar{\mathcal{X}}(i_1,\dots,i_{n-1},\,i_n,\dots,i_N)\\
\times\bar{\mathcal{Y}}(j_1,\dots,j_{m-1},\,i_n,\dots,i_N)
\end{multline}
where $\mathcal N\triangleq [[n:N]]$ denotes the range of contracted modes, $\underline{\mathbf{i}}\triangleq (i_n)_{n\in\mathcal N}=(i_n,\dots,i_N)$ refers to their corresponding indices, and $\mathcal I\triangleq \prod_{n\in\mathcal N}[I_n]$. The sum ranges over all $\underline{\mathbf{i}}\in\mathcal I$ and contracts $N-n+1$ paired modes (from $n$ to $N$), resulting in an order-$(n+m-2)$ tensor $\mathcal{\bar Z}$.
\end{defi}


We next provide a brief \ahmad{review} of the fundamental concepts related to SVD decompositions for high-dimensional data, which will be useful for proceeding with the achievable scheme and the Proof of Theorem~\ref{Achievability-new}, presented in Section~\ref{Results}.

\subsection{A Primer on Multilinear SVD} \label{MLSVD} 
To prove our main result (cf. Theorem~\ref{Achievability-new}), we need the notion of multilinear SVDs. 
The multilinear SVD (MLSVD) extends the concept of the matrix SVD into the multilinear domain~\cite{de2000}. This decomposition provides a powerful tool for analyzing tensors and obtaining low-rank multilinear approximations. 

For matrices, the SVD is well-known and expressed as 
\begin{align}
    {\bf M} = {\bf U}{\bf \Sigma}{\bf V}^{\intercal}
\end{align}
where ${\bf M}\in \mathbb{R}^{J_1\times J_2}$ is an arbitrary real-valued matrix, ${\bf \Sigma}\in \mathbb{R}^{I_1\times I_2}$ is a diagonal matrix with the entries $\sigma_1 \geq \sigma_2 \geq \hdots \geq \sigma_{\min(I_1,I_2)} \geq 0$ in descending order, and ${\bf U}\in \mathbb{R}^{J_1\times I_1} $ and ${\bf V}\in \mathbb{R}^{J_2\times I_2}$ are orthogonal matrices. 

Using mode-$n$ tensor-matrix products, we have:
\begin{align}
{\bf M} = {\bf \Sigma}\times_1 {\bf U}\times_2 {\bf V}^{\intercal}\, .
\end{align}

The MLSVD generalizes this decomposition to higher-order tensors. In the literature, e.g.,~\cite{de2000}, it is also referred to as the higher-order SVD or Tucker decomposition, though ``Tucker decomposition" has evolved into a broader term. The MLSVD of a $N$-th order tensor is represented as
\begin{align}
\TensorT = \TensorS \times_1 {\bf U}^{(1)} \times_2 {\bf U}^{(2)}\hdots \times_N {\bf U}^{(N)}
\label{Tensor_SVD}
\end{align}
where $\TensorT\in \mathbb{R}^{J_1\times J_2\times\hdots\times J_N}$, $\TensorS\in \mathbb{R}^{I_1\times I_2\times\hdots\times I_N}$, and ${\bf U}^{(n)}\in \mathbb{R}^{J_n\times I_n},\, n\in [N]$. 
Similar to the matrix case, where ${\bf U}$ and ${\bf V}$ serve as orthonormal bases for the column and row spaces, the MLSVD computes $N$ orthonormal mode matrices ${\bf U}^{(n)} \in \mathbb{R}^{I_n \times J_n},\, n \in [N]$, each spanning the subspace of mode-$n$ vectors. These mode matrices are directly analogous to the singular vector bases in the matrix SVD. Concretely, the MLSVD unfolds the tensor along each mode, applies the matrix SVD to the resulting unfolding, and collects the mode-$n$ singular vectors to form the factor matrices, yielding an orthogonal decomposition of the tensor into a core tensor and its mode matrices.
The subtensors $\TensorS_{i_n=\alpha}$ of $\TensorS$ are obtained by fixing the $n$-th index to $\alpha$ with the following properties:
\begin{itemize}
    \item \emph{All-orthogonality\footnote{Arrays with a scalar product of zero are considered orthogonal.}}: Subtensors $\TensorS_{i_n=\alpha}$ and $\TensorS_{i_n=\beta}$ are orthogonal for all possible $n, \alpha, \beta$ if
    \begin{align}
        \langle\TensorS_{i_n=\alpha},\TensorS_{i_n=\beta}\rangle=0 \ \ \text{when} \ \ \alpha\neq \beta 
    \end{align}
    where the scalar product of two tensors $\TensorT,\TensorS\in \mathbb{R}^{I_1\times\hdots\times I_N}$ is defined as
    \begin{align}
        \langle \TensorT, \TensorS \rangle \triangleq \sum\limits_{i_1}\hdots \sum\limits_{i_N} t_{i_1\hdots i_N} s_{i_1\hdots i_N}
    \end{align}
    where $t$ and $s$ represent the elements of tensors $\TensorT$ and $\TensorS$, respectively.

    \item \emph{Ordering}: 
    \begin{align}
        \sigma^{(n)}_1 \geq \sigma^{(n)}_2 \geq \hdots \geq \sigma^{(n)}_{I_n} \geq 0 
    \end{align}
where symbols $\sigma^{(n)}_i$ represent the $n$-mode singular values of $\TensorS$, and are equal to the Frobenius-norms $\lVert \TensorS_{i_n=i}\rVert,\, i\in [I_n]$, where the Frobenius-norm of tensor $\TensorT$ is described by
\begin{align}
    \lVert \TensorT\rVert \triangleq \langle \TensorT, \TensorT \rangle\, .
\end{align}
\end{itemize}

The mode-$n$ rank (or $n$-rank) of a tensor is defined using the matrix-based methods. Specifically, it is the rank of the mode-$n$ unfolding of the tensor, i.e., the dimension 
of the subspace spanned by its mode-$n$ vectors. The mode-$n$ vectors of $\TensorT$ are precisely the column vectors of its mode-$n$ matrix unfolding $\TensorT_{(n)}$. We thus have: 
\begin{align}
    \rank_n(\TensorT)=\rank(\TensorT_{(n)})\, .
\end{align}



\subsection{Multi-Shot Scenario with $T>1$}\label{sec:multishot}
In this part of the Appendix, \ahmad{we study the multi-shot scenario with $T$ shots in which each server can compute the assigned tasks and transmit to the users during $T$ shots, allowing for a reduced number of required servers, compared to the single-shot scenario.} 
We follow the same procedure as in the single-shot scenario and highlight the differences in the following.

\subsubsection{System Model}
In the multi-shot scenario with $T$ shots, we will have $e_{n,\pu,t}$ and $z_{n,t}$ instead of $e_{n,\pu}$ and $z_{n}$ in \eqref{EncodedFiles}, \eqref{DecedFiles}, respectively, as well as $d_{k,n,t}$ instead of $d_{k,n}$ in \eqref{DecedFiles}, allowing for utilizing the servers by operating both computation and communication phases across multiple shots $t\in [T]$.
These yield 
\begin{align}
\label{EncodedFiles-multishot}
  z_{n,t}\triangleq \sum\nolimits_{\pu \in \mathcal{P}_{n}} e_{n,\pu,t} \prod\nolimits_{\ell\in [L]} W_\ell^{p_{\ell}-1},\quad n\in [N]
  \, ,\, t\in [T]
\end{align} 
and
\begin{align}
    F'_{k} \triangleq \sum\nolimits_{n \in [N]} d_{k,n,t} z_{n,t}\, . \label{DecedFiles-multishot}
\end{align}

The system will be parametrized by $(K,N,T,L,\Gamma,\Delta, \{P_{\ell},\Lambda_{\ell}\}_{\ell\in [L]})$ for the multi-shot scenario with $T$ shots. Therefore, for $T$ shots, \eqref{eq:N_TDC}, from~\cite[Theorem~1]{Khalesi2025Tessellated}, will take the form
\begin{align}\label{eq:N_TDC-multishot}
    N=\frac{K}{\Delta} \frac{L'}{\Gamma} \frac{\min(\Delta,\Gamma)}{T}\, .
\end{align}

\subsubsection{Problem Formulation in Tensor Form}\label{problem-multishot}

For the multi-shot scenario with $T>1$, we have the following differences in problem formulation details.


In the communication phase, 
from~\eqref{EncodedFiles-multishot}, 
the transmission from server $n$ during time $t\in [T]$ takes the form $z_{n,t} = \bar{\mathscr{e}}_{n,t} \times^{[[1:L]]}_{[[1:L]]} \bar{\mathcal{W}},\, z_{n,t}\in \mathbb{R}$, 
where for all $\pu \in \prod_{\ell \in [L]} [P_{\ell}]$,  the encoding coefficients $e_{n,\pu,t}$  
yield a tensor of the form
\begin{align}
    \bar{\mathscr{e}}_{n,t}(\pu) & \triangleq e_{n,\underline{\mathbf{p}},t},\ \bar{\mathscr{e}}_{n,t} \in \mathbb{R}^{P_1 \times \hdots \times P_L},\: n \in [N],\: t\in [T],\: \pu\in \mathcal{P}_n 
    \label{encoding-tensor-per-shot}
\end{align}
thus allowing us to represent all the per-shot encoding tensors in a general tensor of the form
\begin{align}
    \bar{\mathcal{E}}_{n}(\pu) &\triangleq \textit{stack}_{1}(\bar{\mathscr{e}}_{n,1},\hdots, 
        \bar{\mathscr{e}}_{n,T})\in \mathbb{R}^{T \times P_1\times \hdots \times P_L},\: n \in [N] \label{encoding-tensor-multi-shot}
\end{align}
where $\textit{stack}_{1}(\bar{\mathscr{e}}_{n,1},\hdots,\bar{\mathscr{e}}_{n,T})$ forms an order-$(L+1)$ tensor $\TensorE_n\in\mathbb{R}^{T\times P_1\times \hdots\times P_L}$ 
composed by stacking the $T$ tensors $\{\bar{\mathscr{e}}_{n,t}\}_{t\in[T]}$ (each of order $L$) along a new first mode.
thus yielding the per-server transmission vector $\z_{n}\triangleq [z_{n,1},\hdots, z_{n,T}]^ \intercal$, obtained as
\begin{align}
\z_{n} 
= \TensorE_n \times^{[[1:L]]}_{[[2:L+1]]} \bar{\mathcal{W}},\ \z_{n}\in \mathbb{R}^{T},\: n \in [N] \label{EncodedCashedData-multi-shot}
\end{align}
and the overall transmitted vector $\z \triangleq [\z_{1}^{\intercal},\z^{\intercal}_{2},\hdots, \z^{\intercal}_{N}]^ \intercal \in \mathbb{R}^{NT}$,
taking the form
\begin{align}
    \z =[\TensorE_{1},\hdots, \TensorE^{}_{N}]_1 \times^{[[1:L]]}_{[[2:L+1]]} \bar{\mathcal{W}}  
    \label{EncodedCashedData-1-multi-shot}
\end{align}
where $[\TensorE_{1},\hdots, \TensorE^{}_{N}]_1$ forms an order-$(L+1)$ tensor $\TensorE\in\mathbb{R}^{NT\times P_1\times \hdots\times P_L}$ 
composed by concatenating the $N$ tensors $\{\TensorE_n\}_{n\in[N]}$ (each of order $(L+1)$) along their first mode.



In the decoding phase, similarly,
each retrieved function output takes the form $F'_k= \mathbf{d}_{k}^{\intercal}\mathbf{z}\in\mathbb{R}$, 
thus resulting in the vector of all outputs taking the form 
\begin{align}
\label{eq:received_functions}
\f' \triangleq [F'_{1},F'_{2},\hdots, F'_{K}]= [\mathbf{d}_1,\mathbf{d}_2,\hdots,\mathbf{d}_K]^{\intercal} \mathbf{z} \in\mathbb{R}^K
\end{align}
where the decoding coefficients $d_{k,n,t}$ from~\eqref{DecedFiles-multishot} lead to
\begin{align}
\mathbf{d}_{k,n} &\triangleq [d_{k,n,1},\hdots, d_{k,n,T}]^ \intercal \in \mathbb{R}^{T},\: k \in [K]\, ,\: n \in [N]\, , \label{decoding-vectors-per-shot-1-multi-shot}\\
        \mathbf{d}_k &\triangleq [\mathbf{d}_{k,1}^{\intercal},\hdots, \mathbf{d}_{k,N}^{\intercal}]^ \intercal \in \mathbb{R}^{N T},\: k \in [K]\, . \label{decoding-vectors-multi-shot}
\end{align}

The tensors of encoding coefficients in~\eqref{encoding-tensor-multi-shot}, and the vector of decoding coefficients in~\eqref{decoding-vectors-multi-shot} allow us to form the respective tensors 
\begin{align}
        \TensorE &\triangleq [\TensorE_{1},\hdots, \TensorE^{}_{N}]_1 \in \mathbb{R}^{NT \times P_1\times P_2 \times \hdots \times P_{L}},\label{EncodingMatrix-multi-shot}\\
    \mathbf{D} &\triangleq [\mathbf{d}_1, \hdots , \mathbf{d}_K]^{\intercal} \in \mathbb{R}^{K \times NT}\, . \label{DecodingMatrix-multi-shot}
\end{align}
 

%

In terms of the corresponding connection to the sparsity of $\mathbf{D}$ and $\TensorE $, we recall from~\eqref{eq:Gamma}, our metric $\Gamma$, which directly from~\eqref{encoding-tensor-multi-shot} and~\eqref{EncodingMatrix-multi-shot}, implies the computation constraint as 
\[
\max_{n \in [N]} \sum_{\ell\in [L]} \Bigl| \mathbbm{1} \Big(\cup^{T}_{t=1} \supp\big(\bar{\mathcal{E}}((n-1)T +t, \underbrace{:,\hdots,:}_{\ell-1\,\,\text{terms}},p_\ell, 
 \underbrace{:, \hdots,:}_{L -\ell\,\,\text{terms}})\big) \neq \emptyset \Big) \Bigr| \leq \Gamma\, ,\quad p_\ell\in [2:P_{\ell}]\, .
\]

Furthermore, from~\eqref{eq:Delta}, we recall $\Delta$, which from~\eqref{decoding-vectors-multi-shot} and~\eqref{DecodingMatrix-multi-shot}, implies a communication constraint 
\[
\max_{n \in [N]} \bigl|\cup^{T}_{t=1}\supp\big(\mathbf{D}(:,(n-1)T +t)\big) \bigr| \leq \Delta \, .
\]

Finally from~\eqref{eq:Lambda}, we recall $\Lambda_{\ell}$, which from~\eqref{encoding-tensor-multi-shot} and 
\eqref{EncodingMatrix-multi-shot}, suggests
the  multiplication constraints 
 \begin{align*}     
 \bigl| \cup^{T}_{t=1}\  \|\bar{\mathcal{E}}\big((n-1)T +t,p_1,p_2, \hdots, p_{\ell-1}, :,p_{\ell +1}, \hdots, p_L \big)\|_0 \bigr| &\leq \Lambda_\ell\, ,\quad \ell\in [L],\, p_\ell \in [P_\ell]\, .
\end{align*}

\subsubsection{Main Results}

\paragraph*{Default Tensor Factorization}
For the multi-shot scenario with $T$ shots, Theorem~\ref{Achievability-Converse} will reform as follows.
\begin{theo} 
\label{Achievability-Converse-multishot}
The achievable rate of the lossless $(K,N,T,L,\Gamma,\Delta, \{P_{\ell},\Lambda_{\ell}\}_{\ell\in [L]})$ distributed computing system, under $\{P_{\ell}=P,\Lambda_{\ell}=\Lambda\}_{\ell\in [L]}$ and $(\Delta | K$, $\Lambda | P)$, takes the form $R = K/N$, where 
\begin{align}\label{achiv-o-multi}
N&\leq \frac{K}{\Delta}\, {L\choose{\Gamma}}\, \Big\lceil\frac{\min(\Delta, \Lambda^\Gamma)}{T}\Big\rceil\,
\big(\frac{P}{\Lambda}\big)^\Gamma\, .
\end{align}
\end{theo}

\begin{proof}
    The proof follows from the same steps as in the proof of Theorem~\ref{Achievability-new-multishot}, detailed next, though with a difference in the rank of each tile, which is simply $\Big\lceil\frac{\min(\Delta, \Lambda^\Gamma)}{T}\Big\rceil$.
\end{proof}

\paragraph*{Assignment-based Algorithmic Tensor Factorization}
For the multi-shot scenario with $T$ shots, instead of Theorem~\ref{Achievability-new}, we will have the following theorem.
\begin{theo} 
\label{Achievability-new-multishot}
The achievable rate of the lossless $(K,N,T,L,\Gamma,\Delta, \{P_{\ell},\Lambda_{\ell}\}_{\ell\in [L]})$ distributed computing system, under $\{P_{\ell}=P,\Lambda_{\ell}=\Lambda\}_{\ell\in [L]}$ and $(\Delta | K$, $\Lambda | P)$, takes the form $R = K/N$, where 
\begin{align}\label{achiv-multishot}
N\le
\frac{K}{\Delta} \sum_{\mathcal{\beta}\in [n_{\mathcal{P}}]}
\left\lceil
  \frac{\min \big(\Delta,\lvert \mathcal{R}^*_{\mathcal{P}_\beta} \rvert\big)}{T}
\right\rceil
\end{align}
where $\mathcal{R}^*_{\mathcal{P}_\beta}$ is the row index set of $\beta^{th}$ designated tile, resulting from Algorithm~\ref{Alg:GreedyAssignment} in the proof.

\end{theo}

\begin{proof}
The proof follows from the same parts and steps as the proof of Theorem \ref{Achievability-new}. However, in Part I, the notion of rank-one contribution support and the related concepts will have the new dimension $n\in [NT]$, where $T$ appears as the number of shots, allowing for operating both computation and communication phases across multiple shots.

Part I and Steps a) and b) in Part II, including Algorithm~\ref{Alg:GreedyAssignment} 
will remain the same, but Steps c) and d) in Part II will 
\ahmad{incorporate the number of shots $T$ to design tiles} in $\mathbf{D}$ and  $\TensorE$ and enumerating the number of required servers, which we detail next. 

\begin{parletters}[start=3]
\paragraph{Step 3: Creating and filling the non-zero tiles in  $\mathbf{D}$ and  $\TensorE$}

For $T=1$, each tile $\mathcal{P}_i$ of rank $r^*_{\mathcal{P}_i}$ is assigned to $r^*_{\mathcal{P}_i}$ distinct servers, each contributing one rank dimension. For $T>1$, a server may contribute up to $T$ rank dimensions within the same tile. Since $r^*_{\mathcal{P}_i}$ may not be divisible by $T$, at most one server per tile may be partially utilized. Associating such a server with additional tiles may violate the communication, computation, or multiplication constraints. Therefore, we allocate $\Big\lceil \frac{r^*_{\mathcal{P}}}{T} \Big\rceil$
servers to each equivalence class of tiles.
Therefore, the position that each cropped tile takes inside $\mathbf{D}$, is given by 
\begin{align}\label{eq:tilePositionInD-multishot}
    \mathcal{C}_{\mathcal{P}_j},\:  [\sum^{j-1}_{i=1} T \lceil \frac{r^*_{\mathcal{P}_i}}{T} \rceil
+1: \sum^{j}_{i=1} T \lceil \frac{r^*_{\mathcal{P}_i}}{T} \rceil]\, ,\quad \forall \mathcal{P}_j \in \mathcal{C} 
\end{align} and the position of each cropped tile in $\TensorE$ is given by 
\begin{align} \label{eq:tilePositionInE-multishot}
[\sum^{j-1}_{i=1} T \lceil \frac{r^*_{\mathcal{P}_i}}{T} \rceil
+1: \sum^{j}_{i=1} T \lceil \frac{r^*_{\mathcal{P}_i}}{T} \rceil],\: \mathcal{R}^*_{\mathcal{P}_j}\, , \quad \forall \mathcal{P}_j \in \mathcal{C}
\end{align}

This yields
\begin{align}
\mathbf{D}\Big(\mathcal{C}_{\mathcal{P}_j},\:  [\sum^{j-1}_{i=1} T \lceil \frac{r^*_{\mathcal{P}_i}}{T} \rceil+1: \sum^{j}_{i=1} T \lceil \frac{r^*_{\mathcal{P}_i}}{T} \rceil]\Big)= \mathbf{D}_{\mathcal{P}_j}\, ,\quad \forall \mathcal{P}_j \in \mathcal{C}
    \label{Decoding-Encoding1-multishot}
\end{align}
and 
\begin{align}
\TensorE\Big([\sum^{j-1}_{i=1} T \lceil \frac{r^*_{\mathcal{P}_i}}{T} \rceil
+1: \sum^{j}_{i=1} T \lceil \frac{r^*_{\mathcal{P}_i}}{T} \rceil],\: \mathcal{R}^*_{\mathcal{P}_j}\Big)=\TensorE_{\mathcal{P}_j}\, ,\quad \forall \mathcal{P}_j \in \mathcal{C}
    \label{Decoding-Encoding2-multishot}
\end{align}
while naturally the remaining non-assigned elements of $\mathbf{D}$ and $\TensorE$ are zero.

\paragraph{Step 4: An upper bound to the number of servers}
We now proceed to bound the number of rank-one contribution supports in $T$ shots (corresponding to the number of servers), and we do so under our previously stated assumption of disjoint supports by substituting $N$ with $NT$ in Definition \ref{tensor-disjointSupportAssumption},
which is equivalent to the disjoint-tiles assumption again by substituting $N$ with $NT$ in Lemma \ref{Disjoint-Support-Assumption: Equivalence}.
We then recall from Appendix~\ref{problem-multishot} (see also~\eqref{EncodingMatrix-multi-shot}--\eqref{DecodingMatrix-multi-shot}) that each server $n\in[NT]$ corresponds to one column of the decoding matrix $\mathbf{D}$ and the associated row of the encoding tensor $\TensorE$. To express $N$ in terms of the scheme parameters, consider the tiles $\mathcal{P}\in\mathcal{C}$ and the associated submatrices $\mathbf{D}_{\mathcal{P}}$ defined in~\eqref{sub-SVD}. From~\eqref{Decoding-Encoding1-multishot}, these submatrices collectively span $\sum_{i=1}^{m} r^*_{\mathcal{P}_i}$ columns, which results in $NT\le \sum_{i=1}^{m} r^*_{\mathcal{P}_i}$. 

Finally, from~\eqref{optrank} and considering the multi-shot scenario with $T$ transmissions, the upper bound on the number of servers is expressed as 
\begin{align}
N&\leq\sum\limits_{\mathcal{P}\in \mathcal{C}} \lceil\frac{r_{\mathcal{P}}^{*}}{T}\rceil=\frac{K}{\Delta} \sum_{\mathcal{\beta}\in [n_{\mathcal{P}}]}
\left\lceil
  \frac{\min \big(\Delta,\lvert \mathcal{R}^*_{\mathcal{P}_\beta} \rvert\big)}{T}
\right\rceil
\end{align}
which completes the proof for Theorem~\ref{Achievability-new-multishot}.\qedhere
\end{parletters}

\end{proof}



\bibliographystyle{IEEEtran}
\bibliography{refs}

@STRING{tit = {IEEE Trans. Inf. Theory}}

@STRING{AIStats = {Proc., Int. Conf. Art. Intell. Stat. (AISTATS)}}

@STRING{icml = {Proc., Int. Conf. Mach. Learn. (ICML)}}

@STRING{iclr = {Proc., Int. Conf. Learn. Represent. (ICLR)}}

@STRING{isit = {Proc., IEEE Int. Symp. Inf. Theory (ISIT)}}

@STRING{itw = {Proc., Inf. Theory Wksh. (ITW)}}

@STRING{neurips = {Proc., Adv. Neural Inf. Process. Syst. (NeurIPS)}}

@STRING{spawc = {Proc., IEEE Int. Wksh. Signal Process. Adv. Wireless Commun. (SPAWC)}}

@STRING{preprint = {Preprint}}

@article{verbraeken2020survey,
  title={A survey on distributed machine learning},
  author={Verbraeken, Joost and Wolting, Matthijs and Katzy, Jonathan and Kloppenburg, Jeroen and Verbelen, Tim and Rellermeyer, Jan S},
  journal={ACM Comput. Surv.},
  volume={53},
  number={2},
  pages={1--33},
  year={2020}
}

@article{wan2022secure,
  author={Wan, Kai and Sun, Hua and Ji, Mingyue and Caire, Giuseppe},
  journal=tit, 
  title={Distributed Linearly Separable Computation}, 
  year={2022},
  volume={68},
  number={2},
  pages={1259-1278},
  doi={10.1109/TIT.2021.3127910}
}

@article{jia2021capacity,
  title={On the capacity of secure distributed batch matrix multiplication},
  author={Jia, Zhuqing and Jafar, Syed Ali},
  journal=tit,
  volume={67},
  number={11},
  pages={7420--7437},
  year={2021}
}

@article{dean2008mapreduce,
  title={{MapReduce}: Simplified data processing on large clusters},
  author={Dean, Jeffrey and Ghemawat, Sanjay},
  journal={Commun. ACM},
  volume={51},
  number={1},
  pages={107--113},
  year={2008}
}

@inproceedings{yu2019lagrange,
  title={Lagrange coded computing: Optimal design for resiliency, security, and privacy},
  author={Yu, Qian and Li, Songze and Raviv, Netanel and Kalan, Seyed Mohammadreza Mousavi and Soltanolkotabi, Mahdi and Avestimehr, Salman A.},
  booktitle={Int. Conf. Artif. Intell. Stat. (AISTATS)},
  pages={1215--1225},
  year={2019},
  organization={PMLR}
}

@article{raviv2020gradient,
  title={Gradient coding from cyclic {MDS} codes and expander graphs},
  author={Raviv, Netanel and Tamo, Itzhak and Tandon, Rashish and Dimakis, Alexandros G.},
  journal=tit,
  volume={66},
  number={12},
  pages={7475--7489},
  year={2020}
}

@inproceedings{tandon2017gradient,
  title={Gradient coding: Avoiding stragglers in distributed learning},
  author={Tandon, Rashish and Lei, Qi and Dimakis, Alexandros G. and Karampatziakis, Nikos},
  booktitle=icml,
  pages={3368--3376},
  year={2017},
  month    = aug,
  address   = {Sydney, Australia},
  organization={PMLR}
}

@article{dutta2016short,
  title={Short-Dot: Computing large linear transforms distributedly using coded short dot products},
  author={Dutta, Sanghamitra and Cadambe, Viveck and Grover, Pulkit},
  journal=neurips,
  volume={29},
  year={2016},
month     = Dec,
  address   = {Barcelona, Spain},
}

@inproceedings{ramamoorthy2019universally,
  author    = {Ramamoorthy, Aditya and Tang, Li and Vontobel, Pascal O.},
  title     = {Universally decodable matrices for distributed matrix-vector multiplication},
  booktitle = isit,
  year      = {2019},
  pages     = {1777--1781},
  month     = jul,
  address   = {Paris, France}
}

@article{wang2018fundamental,
  title={Fundamental limits of coded linear transform},
  author={Wang, Sinong and Liu, Jiashang and Shroff, Ness and Yang, Pengyu},
  journal={arXiv preprint arXiv:1804.09791},
  year={2018}
}

@inproceedings{ye2018communication,
  title        = {Communication-computation efficient gradient coding},
  author       = {Ye, Min and Abbe, Emmanuel},
  booktitle    = icml,
  pages        = {5610--5619},
  year         = {2018},
  organization = {PMLR},
  month        = jul,
  address      = {Stockholm, Sweden}
}

@article{le2023spurious,
  title={Spurious valleys, {NP}-hardness, and tractability of sparse matrix factorization with fixed support},
  author={Le, Quoc-Tung and Riccietti, Elisa and Gribonval, Rémi},
  journal={SIAM J. Matrix Anal. Appl.},
  volume={44},
  number={2},
  pages={503--529},
  year={2023}
}

@INPROCEEDINGS{Brunero1,
  author={Brunero, Federico and Wan, Kai and Caire, Giuseppe and Elia, Petros},
  booktitle=itw, 
  title={Coded Distributed Computing for Sparse Functions With Structured Support}, 
  year={2023},
  month = Apr,
  address = {Saint-Malo, France},
  pages={474--479},
  doi={10.1109/ITW55543.2023.10160235}
}

@inproceedings{AVEST1,
  title     = {Securing secure aggregation: Mitigating multi-round privacy leakage in federated learning},
  author    = {So, Jinhyun and Ali, Ramy E. and G{\"u}ler, Ba{\c{s}}ak and Jiao, Jiantao and Avestimehr, A. Salman},
  booktitle = {Proc. AAAI Conf. Artif. Intell.},
  volume    = {37},
  pages     = {9864--9873},
  year      = {2023},
  month     = feb,
  address   = {Washington, DC, USA}
}

@ARTICLE{Malak,
  author={Malak, Derya and Médard, Muriel},
  journal={IEEE Trans. Signal Process.}, 
  title={A Distributed Computationally Aware Quantizer Design via Hyper Binning}, 
  year={2023},
  volume={71},
  pages={76--91},
  doi={10.1109/TSP.2023.3238888}
}

@ARTICLE{Charalambides25,
  author={Charalambides, Neophytos and Mahdavifar, Hessam and Hero, Alfred O.},
  journal=tit, 
  title={Generalized Fractional Repetition Codes for Binary Coded Computations}, 
  year={2025},
  volume={71},
  number={3},
  pages={2170-2194},
  doi={10.1109/TIT.2025.3529680}}

@ARTICLE{Vithana23tit,
  author={Vithana, Sajani and Ulukus, Sennur},
  journal=tit, 
  title={Private Read-Update-Write With Controllable Information Leakage for Storage-Efficient Federated Learning With Top-$r$ Sparsification}, 
  year={2023},
  volume={70},
  number={5},
  pages={3669--3692},
  doi={10.1109/TIT.2023.3339450}
}

@article{ordentlich2025quant,
  author    = {Or Ordentlich and Yury Polyanskiy},
  title     = {Optimal Quantization for Matrix Multiplication},
  journal   = {arXiv preprint arXiv:2410.13780},
  year      = {2025},
  publisher = {arXiv},
  doi       = {10.48550/arXiv.2410.13780}
}

@inproceedings{tanha24,
  author    = {Tanha, Ahmad and Malak, Derya},
  title     = {The Influence of Placement on Transmission in Distributed Computing of {B}oolean Functions},
  booktitle = spawc,
  address   = {Lucca, Italy},
  month     = Sep,
  year      = {2024}
}

@article{de2000,
  author  = {De Lathauwer, Lieven and De Moor, Bart and Vandewalle, Joos},
  title   = {A Multilinear Singular Value Decomposition},
  journal = {SIAM J. Matrix Anal. Appl.},
  volume  = {21},
  number  = {4},
  pages   = {1253--1278},
  year    = {2000},
  doi     = {10.1137/S0895479896305696}
}

@article{Dolgov2014,
  author  = {Dolgov, Sergey V. and Savostyanov, Dmitry V.},
  title   = {Alternating Minimal Energy Methods for Linear Systems in Higher Dimensions},
  journal = {SIAM J. Sci. Comput.},
  volume  = {36},
  number  = {5},
  pages   = {A2248--A2271},
  year    = {2014},
  doi     = {10.1137/140953289}
}

@ARTICLE{Yan22,
  author   = {Yan, Qifa and Yang, Sheng and Wigger, Michèle},
  journal  = tit, 
  title    = {Storage-Computation-Communication Tradeoff in Distributed Computing: Fundamental Limits and Complexity}, 
  year     = {2022},
  volume   = {68},
  number   = {8},
  pages    = {5496--5512},
  doi      = {10.1109/TIT.2022.3158828}
}

@inproceedings{Jayakumar2020Multiplicative,
title={Multiplicative Interactions and Where to Find Them},
author={Siddhant M. Jayakumar and Wojciech M. Czarnecki and Jacob Menick and Jonathan Schwarz and Jack Rae and Simon Osindero and Yee Whye Teh and Tim Harley and Razvan Pascanu},
booktitle = iclr,
year={2020},
month     = Apr,
address   = {Addis Ababa, Ethiopia}
}

@inproceedings{sefidgaran2023mdl,
  author    = {Milad Sefidgaran and Abdellatif Zaidi and Piotr Krasnowski},
  title     = {Minimum Description Length and Generalization Guarantees for Representation Learning},
  booktitle = neurips,
  year      = {2023}
}

@article{kavian2025heterogeneity,
  author    = {Masoud Kavian and Romain Chor and Milad Sefidgaran and Abdellatif Zaidi},
  title     = {Heterogeneity Matters even More in Distributed Learning: Study from Generalization Perspective},
  journal   = {arXiv preprint arXiv:2503.01598},
  year      = {2025},
  publisher = {arXiv},
  doi       = {10.48550/arXiv.2503.01598}
}

@inproceedings{haziza2025sparsity,
  author    = {Haziza, Daniel and Steiner, Benoit and Yang, Edward and Sirotenko, Mikhail and Jacob, Benoit and Vaswani, Ashish},
  title     = {2:4 Sparsity for Activation Functions in {LLM}s},
  booktitle = iclr,
  year      = {2025},
  month     = may,
  address   = {Singapore}
}

@article{hassani2025generalized,
  author={Ali Hassani and Fengzhe Zhou and Aditya Kane and Jiannan Huang and Chieh-Yun Chen and Min Shi and Steven Walton and Markus Hoehnerbach and Vijay Thakkar and Michael Isaev and Qinsheng Zhang and Bing Xu and Haicheng Wu and Wen-mei Hwu and Ming-Yu Liu and Humphrey Shi},
  title     = {Generalized Neighborhood Attention: Multi-dimensional Sparse Attention at the Speed of Light},
  journal   = {arXiv preprint arXiv:2504.16922},
  year      = {2025},
  publisher = {arXiv},
  doi       = {10.48550/arXiv.2504.16922}
}

@article{Khalesi2025Tessellated,
  author    = {Khalesi, Ali and Elia, Petros},
  title     = {Tessellated distributed computing},
  journal   = tit,
  volume    = {71},
  number    = {6},
  pages     = {4754--4784},
  year      = {2025}
}

@inproceedings{zaharia2010spark,
  author    = {Zaharia, Matei and Chowdhury, Mosharaf and Franklin, Michael J. and Shenker, Scott and Stoica, Ion},
  title     = {Spark: Cluster computing with working sets},
  booktitle = {2nd USENIX Wksh. Hot Topics Cloud Comput. (HotCloud)},
  year      = {2010},
  month     = aug,
  address   = {Sydney, Australia}
}

@misc{Khalesi25Tensor,
  author = {Khalesi, Ali and  Tanha, Ahmad and  Malak, Derya and  Elia, Petros},
  title = {Tessellated distributed computing of non-linearly separable functions},
  howpublished = {Recent Results, Wksh. Distrib. Comput. Optim. Learn (WDCL)},
  year = {2025},
  month=sep,
  address={Munich, Germany}
}

@article{maheri2026universal,
      title={Universal and Asymptotically Optimal Data and Task Allocation in Distributed Computing}, 
      author={Javad Maheri and K. K. Krishnan Namboodiri and Petros Elia},
      year={2026},
      journal={arXiv preprint 2601.05873}
}

@article{reisizadeh2021coded,
  title={Coded computing for distributed graph analytics},
  author={Reisizadeh, Amirhossein and Prakash, Gauri and Mokhtari, Aryan and Hassibi, Babak and Pedarsani, Ramtin},
  journal={IEEE Transactions on Information Theory},
  year={2021},
  volume={67},
  number={4},
  pages={2379--2396}
}

@article{vavasis2009complexity,
  title={On the complexity of nonnegative matrix factorization},
  author={Vavasis, Stephen A},
  journal={SIAM Journal on Optimization},
  volume={20},
  number={3},
  pages={1364--1377},
  year={2009}
}

@article{gillis2014complexity,
  title={The complexity of the two-factor polynomial approximation problem},
  author={Gillis, Nicolas and Glineur, Fran{\c{c}}ois},
  journal={IEEE Transactions on Signal Processing},
  volume={62},
  number={7},
  pages={1578--1589},
  year={2014}
}

@inproceedings{Hanauer24,
author = {Kathrin Hanauer and Martin P. Seybold and Julian Unterweger},
title = {Covering Rectilinear Polygons with Area-Weighted Rectangles},
booktitle = {Proceedings, The Symposium on Algorithm Engineering and Experiments (ALENEX)},
chapter = {},
pages = {157-169},
doi = {10.1137/1.9781611977929.12},
URL = {https://epubs.siam.org/doi/abs/10.1137/1.9781611977929.12},
eprint = {https://epubs.siam.org/doi/pdf/10.1137/1.9781611977929.12},
year={2024},
publisher={SIAM}
}

@article{hillar2013most,
  title={Most tensor problems are {NP}-hard},
  author={Hillar, Christopher J and Lim, Lek-Heng},
  journal={Journal of the ACM},
  volume={60},
  number={6},
  pages={1--39},
  year={2013}
}

@article{khalesi2026non,
  title={Non-Linearly Separable Distributed Computing: A Sparse Tensor Factorization Approach},
  author={Khalesi, Ali and Tanha, Ahmad and Malak, Derya and Elia, Petros},
  journal={arXiv preprint 2601.16171},
  year={2026}
}

@article{hopcroft1973n,
  title={An n\^{}5/2 algorithm for maximum matchings in bipartite graphs},
  author={Hopcroft, John E and Karp, Richard M},
  journal={SIAM Journal on computing},
  volume={2},
  number={4},
  pages={225--231},
  year={1973},
  publisher={SIAM}
}

@INPROCEEDINGS{Moradi25,
  author={Moradi, Parsa and Akbarinodehi, Hanzaleh and Maddah-Ali, Mohammad Ali},
  booktitle=isit, 
  title={General Coded Computing: Adversarial Settings}, 
  year={2025},
  address={Ann Arbor, MI, USA},
  month=june,
  volume={},
  number={},
  pages={1-6},
  doi={10.1109/ISIT63088.2025.11195419}}

@ARTICLE{Abadi25,
  author={Abadi Khooshemehr, Nastaran and Maddah-Ali, Mohammad Ali},
  journal={IEEE Transactions on Information Theory}, 
  title={Vers: Coded Computing System With Distributed Encoding}, 
  year={2025},
  volume={71},
  number={10},
  pages={7609-7625},
  doi={10.1109/TIT.2025.3591523}}

@article{kolda2009tensor,
  title={Tensor Decompositions and Applications},
  author={Kolda, Tamara G. and Bader, Brett W.},
  journal={SIAM Review},
  volume={51},
  number={3},
  pages={455--500},
  year={2009}
}

@book{cormen2022introduction,
  title={Introduction to algorithms},
  author={Cormen, Thomas H and Leiserson, Charles E and Rivest, Ronald L and Stein, Clifford},
  year={2022},
  publisher={MIT press}
}

@inproceedings{tanhaCDMM26,
  author    = {Ahmad Tanha and Mohammad Reza Deylam Salehi and Monolina Dutta and Derya Malak},
  title     = {Cooperative Coded Matrix Multiplication in Secrecy-Constrained Vehicular Networks},
  booktitle = {IEEE 103rd Vehicular Technology Conference (VTC)},
  year      = {2026},
  note      = {to appear}
}

\end{document}